\documentclass[a4paper,nofootinbib,final,floatfix,showkeys,twocolumn]{revtex4}
\usepackage{amsfonts}
\usepackage{amsmath}
\usepackage{amssymb}
\usepackage[usenames,dvips]{color}
\usepackage[english]{babel}
\usepackage[latin1]{inputenc}
\usepackage{amsfonts}
\usepackage{appendix}
\usepackage{epsfig}
\usepackage{graphics,rotating}
\usepackage{dcolumn}
\usepackage{bm}
\usepackage{color}
\usepackage[usenames,dvipsnames,svgnames]{xcolor}
\usepackage[T1]{fontenc}
\usepackage{multirow}
\usepackage{float}
\usepackage{subfigure}
\usepackage{enumitem}
\usepackage[font=scriptsize]{caption}
\usepackage{mwe}
\usepackage{xcolor}
\usepackage{float}

\begin{document}

\title{Spin dynamics of moving bodies in rotating black hole spacetimes}
\author{Bal\'azs Mik\'oczi$^{1,\dag }$}
\author{Zolt\'an Keresztes$^{2,\ddag }$}

\affiliation{$^{1}$ Research Institute for Particle and Nuclear
Physics, Wigner RCP
H-1525 Budapest 114, P.O. Box 49, Hungary \\
$^{2}$ Department of Theoretical Physics, University of Szeged,
Tisza Lajos
krt. 84-86, Szeged 6720, Hungary \\
$^{\dag }${\small E-mail: mikoczi.balazs@wigner.hu\quad }$^{\ddag }
${\small E-mail: zkeresztes@titan.physx.u-szeged.hu}}

\begin{abstract}
The dynamics of spinning test bodies, moving in rotating black hole (Kerr,
Bardeen-like and Hayward-like) spacetimes, are investigated. In Kerr
spacetime, all the spherical, zoom-whirl and unbound orbits are considered
numerically. Along spherical orbits and for high spin, an amplitude
modulation is found in the harmonic evolution of the spin precessional
angular velocity, caused by the spin-curvature coupling. Along the discussed
zoom-whirl and unbound orbits, the test body approaches the center so much
that it passes through the ergosphere. Near and inside the ergosphere, the
variation of the spin direction can be very rapid. The effects of the
spin-curvature coupling is also investigated. The initial values are chosen
such a way, that the body and its spin move in the equatorial plane of the
coordinate space and of the comoving frame, respectively. Hence, a clear
effect of the spin-curvature coupling is observed as the orbit and the spin
vector leave the equatorial plane. Additional effects in the spin
precessional angular velocity and in the evolution of the Boyer-Lindquist
coordinate components of the spin vector is also considered. Finally, in
case of different regular black holes, the spin-curvature coupling
influences differently the orbit and the spin evolutions.
\end{abstract}

\keywords{black hole physics, spinning test particles,
Mathisson-Papapetrou-Dixon equations, Kerr spacetime}
 \maketitle

\section{Introduction}

Both the orbital and the spin dynamics of compact binary systems have a
renewed interest. All observed gravitational waves originated from compact
binary systems composed of black holes or neutron stars (\cite%
{LIGO1,LIGO2,LIGO3,LIGO4,LIGO5NS,LIGO6,LIGO7,LIGO8}). In two cases the spin
of the merging black holes was identified with high significance \cite%
{LIGO2,LIGO8,LIGOspin}. In addition, in a binary system the dominant
supermassive black hole spin precession was identified from VLBI radio data
spanning over 18 years \cite{VLBIKun}.

In the post-Newtonian (PN) approximation the lowest order spin contributions
to the dynamics come from the spin-orbit, spin-spin and quadrupole-monopole
interactions \cite{BOC1,BOC2,Kidder,Poisson}. The spin effects on the orbit
leaded to set up generalized Kepler equations \cite{Wex,KG,KMG,KG2}. The
analytical description of the secular spin dynamics for black holes is given
in Refs. \cite{Racine} and \cite{KGSBS}. Based on the PN description several
interesting spin related behaviours were identified in compact binary
systems, like transitional precession \cite{ACST}, equilibrium
configurations \cite{Schnittman}, spin-flip \cite{spinflip}, spin flip-flop
\cite{flipflop} and wide precession \cite{Gerosa2019}.

The Mathisson--Papapetrou--Dixon (MPD) equations \cite%
{Mathisson,Papapetrou,Dixon1964,Dixon1970,Dixon1976} describe the dynamics
of binaries with significantly different masses more accurately than the PN
approximation in the strong gravitational field regime, where the PN
parameter is not small. The black hole binary systems with small mass ratio
are among the most promising sources for gravitational waves in the
frequency sensitivity range of the planned LISA - Laser Interferometer Space
Antenna \cite{LISA,Jetzer}. In addition, near the central supermassive black
holes in the galaxies many stellar black holes are expected to exist \cite%
{BahcallWolf,BHKM,BHsSagittariusA}.

The MPD equations are not closed, a spin supplementary condition\footnote{%
Both the PN dynamics with spin-orbit coupling and the gravitational
multipole moments depend on the SSC \cite{Kidder,Mikoczi}.} (SSC) is
necessary to choose \cite%
{Frenkel,Pirani,CorPap,NW,Pryce,Tulczyjew,Dixon1964,Ohashi,KyrianSemerak2007}%
, which defines the point at which the four-momentum and the spin are
evaluated. The Hamiltonian formulations in different SSCs are discussed in
Refs. \cite{BRB,LSK,KLLS,DR1,WSL}. A non-spinning body follows a geodesic
trajectory, while a spinning one does not \cite{BOC1974,Semerak1999}.
Spinning bodies governed by the MPD equations were already studied on Kerr
background. Circular orbits in the equatorial plane can be unstable not only
in the radial direction but also in the perpendicular direction to the
equatorial plane due to the spin \cite{SuzukiMaeda1998}. The spin-curvature
effect strengthens with spin and with non-homogeneity of the background
field \cite{Semerak1999}. The MPD equations admit many chaotic solutions,
however, these do not occur in the case of extreme mass ratio binary black
hole systems \cite{Hartl2003b,Hartl2003,WBHAN1,WBHAN}. Analytic studies on
the deviation of the orbits from geodesics due to the presence of a small
spin are presented in Refs. \cite{MashhoonSingh2006,Bini2008,Binispingeoddev}%
. Highly relativistic circular orbits in the equatorial plane occur in much
wider space region for a spinning body than for a non-spinning one \cite%
{Plyatsko2013}. Spin-flip-effects may occur when the magnetic type
components of quadrupole tensor are non-negligible \cite{BiniGeralico2014}.
Corrections due to the electric type components of quadrupole tensor to the
location of innermost stable circular orbit in the equatorial plane and to
the associated motion's frequency were determined in Ref. \cite%
{BiniFayeGeralico2015}. An exact expression for the periastron shift of a
spinning test body moving in the equatorial plane is derived in Ref. \cite%
{HLOPS}. The influences of the affine parameter choice on the constants of
motion in different SSC were also considered \cite{LukesGerakopoulos}.
Frequency domain analysis of motion and spin precession was presented in
Ref. \cite{RVHughes}. The evolution of spinning test particles were
investigated in the $\gamma$ space-time \cite{Toshmatov2019}, in
non-asymptotically flat space-times \cite{Toshmatov2020}, and in wormholes
\cite{Benavides2021}.

Considering geodesic trajectories, the periastron advance can become such
significant in the strong gravitational field regime that the test particle
follows a zoom-whirl orbit \cite{Glampedakis,Glampedakis2002,Glampedakis2005}%
. For non-spinning particles, the topology of these orbits was encoded by a
rational number \cite{Levin1,Levin2}. Numerical relativity confirmed the
existence of zoom-whirl orbits \cite%
{PretoriousKhurana,HLS,Sperhake,GoldBrugmann1,GoldBrugmann2,East}, and they
also occur in the 3 PN dynamics with spin-orbit interaction \cite%
{Levin3,Grossman2}. Here we will present zoom-whirl orbits occurring in the
MPD dynamics for the first time. In addition, these orbits pass over the
ergosphere where the PN approximation fails.\footnote{%
At the ergosphere the value of the PN expansion parameter is typically about
$1/2$.}

Hyperbolic orbits of spinning bodies were analytically studied in both the
PN \cite{VGGJ} and the MPD \cite{BiniHyp} dynamics. Analytic computations in
Ref. \cite{BiniHyp} were carried out for small spin magnitudes, when the
spin is parallel to the central black hole rotation axis and the body moves
in the equatorial plane. In this configuration both the spin magnitude and
direction are conserved, but they have non-negligible influences on the
orbit. In our numerical consideration the spin is not parallel to the black
hole rotation axis. As a consequence, the body's orbit is not confined to
the equatorial plane and the spin direction evolves. In addition, the
closest approach distance is inside the ergosphere where the PN
approximation cannot be used.

Our investigations are not only\ applied in the Kerr spacetime but also in
regular black hole backgrounds. The first spacetime containing a nonrotating
regular black hole was suggested by Bardeen \cite{Bardeen1968}. This metric
was interpreted as the spacetime surrounding a magnetic monopole occurring
in a nonlinear electrodynamics \cite{BeatoGarcia2000}. Another nonrotating
regular black hole was introduced by Hayward \cite{Hayward2005} having
similar interpretation \cite{FanWang2016}. The spacetime family containing
the Bardeen and Hayward cases was generalized for rotating black holes \cite%
{Toshmatov2017} which we will use here\footnote{%
There are discussions (see Refs. \cite{Bronnikov,RodriguesJunior,Toshmatov})
on that the rotating regular black hole spacetimes given in Ref. \cite%
{Toshmatov2017} are not exact solutions of the field equations. However the
spacetime family given analytically differs only perturbatively from the
exact solution \cite{Toshmatov}, therefore it is suitable for consideration
of spinning bodies evolutions.}.

In this paper, we investigate the orbit and spin evolutions of bodies moving
in Kerr, Bardeen-like and Hayward-like spacetimes and governed by the MPD
equations with Frenkel--Mathisson--Pirani (FMP) and Tulczyjew--Dixon (TD)
SSCs. When the covariant derivatives of the spin tensor and the
four-momentum along the integral curve of the centroid determined by the SSC
are small, this system reduces to a geodesic equation with parallel
transported spin discussed in Ref. \cite{Bini2017}. In this sense the
present article can be considered as the generalization of Ref. \cite%
{Bini2017} with non-negligible spin-curvature corrections causing that the
centroid orbit is non-geodesic and the spin is not parallel transported. As
Bini, Geralico and Jantzen pointed out that the spin dynamics can be
described suitably in the comoving Cartesian-like frame obtained by boosting
the Cartesian-like frame associated to the family of static observers (SOs).
This is because SOs do not move with respect to the distant stars. Hence,
the Cartesian-like axes locked to SOs define good reference directions to
which the variation of the spin vector can be compared. Here, we derive the
spin evolution equation in the comoving Cartesian-like frame based on the
MPD system. However, SO does not exist inside the ergosphere of the rotating
black hole, and thus its frame cannot be used for description of the
dynamics when the spinning body passes over this region. Therefore, the spin
dynamics in a Cartesian-like frame obtained by an instantaneous
Lorentz-boost from the frame associated to the zero angular momentum
observer (ZAMO) is also presented, which can be used inside the ergosphere.
The boosted SO and ZAMO frames relate to each other by a spatial rotation
outside the ergosphere. The rotation angle between these boosted frames is
unsignificant far from the rotating black hole.

In Section \ref{EOM}, the MPD equations, the spin supplementary conditions,
the rotating (Kerr, Bardeen-like and Hayward-like) black hole spacetimes and
the frames associated with the families of SOs and ZAMOs are introduced. In
Section \ref{REP} the representations of spin evolution are given. For this
purpose, we introduce two frames by instantaneous Lorentz boosts of SO and
ZAMO frames, which comoves with an observer having an arbitrary four
velocity $U$. The relation between the boosted frames is discussed
(additional expressions are given in Appendix \ref{comFRametraf}). We use
the TD SSC, and $U$ means either the centroid or the zero 3-momentum
observer four velocity. The spin evolution equation is derived in these $U$%
-frames. First, the spin precession is described with respect to the
boosted spherical coordinate triad associated with either the SOs or
ZAMOs. Then, we introduce Cartesian-like triads in the rest spaces
of SOs and ZAMOs. The spin precession with respect to the
corresponding boosted Cartesian-like frames is also derived. The
relations between the spin angular velocities in the boosted SO and
ZAMO frames are discussed. In Section \ref{num}, we apply the
derived spin equations for numeric investigations when the body
moves along spherical-like, zoom-whirl and unbound orbits. In
Subsection \ref{Kerr} the background is the Kerr spacetime, while in
Subsection \ref{reg}, it is one of the rotating regular black hole
spacetimes. In Appendix \ref{para}, the avoidance of paradoxical
behaviour of the MPD equations is checked. Finally, Section
\ref{Conclusion} contains the conclusions.

We use the signature $-+++$, and units where $c=G=1$, with speed of light $c$
and gravitational constant $G$. The bold small Greek indices with or without
prime take values $1$, $2$ and $3$, while the bold capital and the small
Latin indices $0$, $1$, $2$ and $3$. In addition, the following small bold
Latin indices $\mathbf{i}$, $\mathbf{j}$, $\mathbf{k}$ and $\mathbf{i}%
^{\prime }$, $\mathbf{k}^{\prime }$ take values from $\left\{ \mathbf{x},%
\mathbf{y},\mathbf{z}\right\} $. Finally, the bold indices are frame
indices, while the non-bold indices are spacetime coordinate indices.

\section{Equations of motion for spinning bodies in rotating black hole
spacetimes\label{EOM}}

\subsection{MPD equations and SSC\label{MPDSSC}}

In the pole-dipole approximation, the motion of an extended spinning body in
curved spacetime is governed by the MPD equations \cite%
{Mathisson,Papapetrou,Dixon1964,Dixon1970,Dixon1976} which read as
\begin{equation}
\frac{Dp^{a}}{d\tau }\equiv u^{c}\nabla _{c}p^{a}=F^{a},  \label{MPD1}
\end{equation}%
\begin{equation}
\frac{DS^{ab}}{d\tau }\equiv u^{c}\nabla _{c}S^{ab}=p^{a}u^{b}-u^{a}p^{b},
\label{MPD2}
\end{equation}%
with%
\begin{equation}
F^{a}=-\frac{1}{2}R_{~bcd}^{a}u^{b}S^{cd}.  \label{F}
\end{equation}%
Here $\nabla _{c}$ is the covariant derivative, $p^{a}$ and $S^{ab}$ are the
four-momentum and the spin tensor of the moving body, respectively, and $%
R_{~bcd}^{a}$ is the Riemann tensor. Finally, $u^{a}=dx^{a}/d\tau $ is the
four-velocity of the representative point for the extended body at spacetime
coordinate $x^{a}\left( \tau \right) $ with an affine parameter $\tau $.
Note that higher multipoles of the body should occur in the MDP equations
when they are nonvanishing. Here they are taken to be zero.

Choosing the affine parameter $\tau $ as the proper time \cite%
{LSK,Hackmann2014} $u^{a}u_{a}=-1$\footnote{%
Below we derive a condition for the spin magnitude in TD SSC when the proper
time parametrization has a sense.}, Equation (\ref{MPD2}) can be written as%
\begin{equation}
p^{a}=mu^{a}-u_{b}\frac{DS^{ab}}{d\tau },  \label{purel1}
\end{equation}%
where $m=-u_{a}p^{a}$ is the mass in the rest frame of the observer moving
with velocity $u^{a}$. Equation (\ref{purel1}) shows that the momentum $%
p^{a} $ and the kinematic four velocity $u^{a}$ are not proportional to each
other for a spinning body in general.

We note that if the covariant derivatives of the spin tensor and the
four-momentum along the integral curve of $u^{a}$ are small, i.e. the right
hand sides of Equations (\ref{MPD1}) and (\ref{MPD2}) are negligible, $p^{a}$
becomes proportional to $u^{a}$ which satisfies the geodesic equation
because $m$ is a constant. Then introducing a spin four-vector perpendicular
to $u^{a}$ (see Equation (2.5) of Ref. \cite{Bini2018Sch}), it will be
parallel transported along the trajectory. The geodesic equations with
parallel transported spin vector was investigated in Ref. \cite{Bini2017}.

In general, in order to close the MPD equations an SSC is necessary to
choose, which defines the representative point of the extended body referred
as the center of mass or the centroid. There are some proposed SSC, namely
the Frenkel-Mathisson-Pirani \cite{Frenkel,Mathisson,Pirani}, the
Newton-Wigner-Pryce \cite{NW,Pryce}, the Corinaldesi-Papapetrou \cite%
{Papapetrou,CorPap}, and the Tulczyjew-Dixon \cite{Tulczyjew,Dixon1964}. We
will apply the Tulczyjew-Dixon SSC imposing that%
\begin{equation}
p_{a}S^{ab}=0.  \label{TSSC}
\end{equation}%
This SSC yields two constants of motion, the spin magnitude $%
S^{2}=S_{ab}S^{ab}/2$ and the dynamical mass $M=\sqrt{-p^{a}p_{a}}$ (see
Ref. \cite{Semerak1999}). In addition, the TD SSC together with the MPD
equations results in the following velocity-momentum relation \cite%
{TODFELICE1976,Hojman1977,Semerak1999}:%
\begin{equation}
u^{b}=\frac{m}{M^{2}}\left( p^{b}+\frac{4S^{2}}{\eta }v^{b}\right),
\label{u}
\end{equation}
with
\begin{equation}
v^{b}=\frac{S^{ba}R_{aecd}p^{e}S^{cd}}{2S^{2}},
\end{equation}
and
\begin{equation}
\eta =4M^{2}+2\alpha _{R}S^{2},
\end{equation}%
where $\alpha _{R}=R_{aecd}S^{ae}S^{cd}/2S^{2}$. Since $p^{b}$ and $u^{b}$
are not parallels, $u^{b}$ may become spacelike from timelike along an
integral curve. Where the causal character of a curve is changed, it is
known as superluminal bound and has been discussed in different cases (e.g.
Refs.: \cite{Semerak1999,Toshmatov2019,Toshmatov2020,Benavides2021}). The
superluminal motion has no physical meaning, and the timelike condition for $%
u^{b}$ yields a bound for the spin magnitude as%
\begin{equation}
S^{2}<\frac{2M^{3}}{2v-\alpha _{R}M},  \label{spinconstr}
\end{equation}%
where $v=\sqrt{v_{a}v^{a}}$\footnote{%
Note that $v$ and $\alpha _{R}$ do not carry information on the spin
magnitude since $S^{ab}/\sqrt{2}S$ has unit norm.}. When the spin magnitude
obeys this constraint, the proper time parametrization has sense and the
normalization $u^{b}u_{b}=-1$ gives a relation $m^{2}=m^{2}\left(
p^{a},S^{bc}\right) $ as%
\begin{equation}
m^{2}=\frac{M^{4}}{\left( M^{2}-\frac{16S^{4}}{\eta ^{2}}v^{2}\right) }.
\label{msq}
\end{equation}
Since the relation (\ref{u}) can be inverted \cite{Costa2018}, both initial
data sets $\left\{ x^{a},p^{a},S^{ab}\right\} |_{\tau _{in}}$ and $\left\{
x^{a},m,u^{a},S^{ab}\right\} |_{\tau _{in}}$ provide a unique solution of
the MPD equations with TD SSC.

The spin vector being perpendicular to $p^{a}$ is introduced as%
\begin{equation}
S^{a}=-\frac{1}{2M}\eta ^{abcd}p_{b}S_{cd}.  \label{SvecT}
\end{equation}%
Since%
\begin{equation}
S_{a}S^{ab}=0=S_{a}p^{a}=0,  \label{Sconstr}
\end{equation}%
the contraction of Equation (\ref{u}) with $S_{b}$ results in
$S_{b}u^{b}=0$. Finally, the covariant derivative of $S^{a}$ along
the worldline of the
centroid is%
\begin{equation}
\frac{DS^{a}}{d\tau }=\frac{S^{b}F_{b}}{M^{2}}p^{a}.  \label{SevoTD}
\end{equation}%
If $F^{a}$ is negligible, $S^{a}$ is parallel transported along the
worldline of the centroid, and the centroid moves along a geodesic curve.
The latter can be shown from the MPD equations together with (\ref{purel1})
and (\ref{SevoTD}).

Finally, we mention that the MPD equations are valid only for test particles
whose backreaction to the background spacetime curvature are negligible.
Hence, when the spinning body is moving in a spacetime around a black hole
with a mass parameter $\mu $, the dimensionless spin magnitude $S/M\mu $
must be small \cite{SuzukiMaeda1997,Hartl2003}. This is consistent with the
constraint (\ref{spinconstr}), which becomes for the dimensionless spin
magnitude as%
\begin{equation}
\left( \frac{S}{M\mu }\right) ^{2}<\frac{2}{2v-\alpha _{R}M}\frac{M}{\mu },
\end{equation}%
where the mass ratio $M/\mu $ gives a small factor.

\subsection{\textbf{Rotating black hole spacetimes}}

The line element squared describing the considered rotating black hole
spacetimes in Boyer-Lindquist coordinates reads as \cite%
{Kerr1963,Toshmatov2017}
\begin{eqnarray}
ds^{2} &=&-\frac{\Delta -a^{2}\sin ^{2}\theta }{\Sigma }dt^{2}-\frac{2a%
\mathcal{B}\sin ^{2}\theta }{\Sigma }dtd\phi   \notag \\
&&+\frac{\Sigma }{\Delta }dr^{2}+\Sigma d\theta ^{2}+\frac{\mathcal{A}}{%
\Sigma }\sin ^{2}\theta d\phi ^{2},  \label{metric}
\end{eqnarray}%
with%
\begin{eqnarray}
\Sigma  &=&r^{2}+a^{2}\cos ^{2}\theta ~,~  \notag \\
\Delta  &=&r^{2}+a^{2}-2\left[ \mu +\alpha \left( r\right) \right] r~,
\notag \\
\mathcal{B} &=&r^{2}+a^{2}-\Delta ~,~  \notag \\
\mathcal{A} &=&\left( r^{2}+a^{2}\right) ^{2}-\Delta a^{2}\sin ^{2}\theta .
\end{eqnarray}%
In the Kerr spacetime $\alpha \left( r\right) $ vanishes and $\mu $ and $a$
denote the mass and rotation parameters, respectively. The function $\alpha
\left( r\right) $ occurs when a non-linear electromagnetic field is present.
It is given by%
\begin{equation}
\alpha \left( r\right) =\frac{\mu _{em}r^{\gamma }}{\left( r^{\nu
}+q_{m}^{\nu }\right) ^{\gamma /\nu }},  \label{alpha}
\end{equation}%
where $\mu _{em}=q_{m}^{3}/\sigma $ is the electromagnetically induced ADM
mass. Here $\sigma \ $controls the strength of nonlinear electrodynamic
field and carries the dimension of length squared, $q_{m}$\ is related to
the magnetic charge (see Ref. \cite{FanWang2016}), and the powers are ($%
\gamma =3$,$\nu =2$)\ for the Bardeen-like and ($\gamma =3$,$\nu =3$)\ for
the Hayward-like spacetimes.

The stationary limit surfaces and the event horizon (if they exist) are
determined by the solutions of equations $g_{tt}=\Delta -a^{2}\sin
^{2}\theta =0$ and $g^{rr}=\Delta =0$, respectively. The structure of the
spacetime depends on the number of real, positive solutions of these
equations. For the Kerr spacetime $\mu _{em}=0$, then there are two
stationary limit surfaces and event horizons for $a/\mu <1$. The region
which is located outside the outer event horizon but inside the outer
stationary limit surface is called ergosphere. The spacetime is free from
the singularity for $\mu =0$ and $\gamma \geq 3$. The first and the second
panels of Figure 3 in Ref. \cite{Toshmatov2017} indicate the regions in the
parameter space of $a$ and $q=q_{m}/\mu _{em}$ for the Bardeen and the
Hayward subcases, respectively, where the above line element squared
describes a regular black hole.

In the spacetimes having symmetries, constants of motion associated to each
Killing vector $\xi ^{a}$ (which obeys the Killing equation $\nabla _{(a}\xi
_{b)}=0$) emerge \cite{Dixon1970}. Since the rotating black hole spacetimes
have a timelike $\partial _{t}$ and a spatial $\partial _{\phi }$ Killing
vectors due to the staticity and axial symmetry, there are two constants of
motion \cite{Hartl2003}:%
\begin{eqnarray}
E &=&-p_{t}-\frac{1}{2}S^{ab}\partial _{a}g_{bt},~  \notag \\
J_{z} &=&p_{\phi }+\frac{1}{2}S^{ab}\partial _{a}g_{b\phi }.
\end{eqnarray}%
At spatial infinity $E$ means the energy of the spinning body and $J_{z}$ is
the projection of the total momentum to the symmetry axis. These constants
are used for checking the numerical accuracy.

\subsubsection{Static and zero angular momentum observers}

The worldlines of static observers are the integral curves of the Killing
vector field $\partial _{t}$. This family of observers exists outside the
ergosphere, where their frame is given by%
\begin{eqnarray}
e_{\mathbf{0}} &=&u_{\left( SO\right) }=\frac{1}{\sqrt{-g_{tt}}}\partial
_{t},~e_{\mathbf{1}}=\sqrt{\frac{\Delta }{\Sigma }}\partial _{r},~e_{\mathbf{%
2}}=\frac{\partial _{\theta }}{\sqrt{\Sigma }},  \notag \\
e_{\mathbf{3}} &=&-\frac{1}{\sqrt{\Delta }}\left( \frac{a\mathcal{B}\sin
\theta }{\Sigma \sqrt{-g_{tt}}}\partial _{t}-\frac{\sqrt{-g_{tt}}}{\sin
\theta }\partial _{\phi }\right) .  \label{SOdown}
\end{eqnarray}%
The dual basis is obtained as $e_{a}^{\mathbf{A}}=g_{ab}\eta ^{\mathbf{AB}%
}e_{\mathbf{B}}^{b}$, where $\eta ^{\mathbf{AB}}=diag\left( -1,1,1,1\right) $%
.

The orbit of a zero angular momentum observer is orthogonal to the $t=$%
const. hypersurfaces \cite{Bardeen,Semerak93}. The four velocity along this
orbit is%
\begin{equation}
u_{\left( ZAMO\right) }=\sqrt{\frac{\mathcal{A}}{\Sigma \Delta }}\left(
\partial _{t}+\frac{a\mathcal{B}}{\mathcal{A}}\partial _{\phi }\right) ,
\end{equation}%
which corresponds to the 1-form: $-dt/\sqrt{-g^{tt}}$. In contrast to the
SOs, this family of observers also exists inside the ergosphere but outside
the outer event horizon. The frame of the ZAMOs is given by%
\begin{eqnarray}
f_{\mathbf{0}} &=&u_{\left( ZAMO\right) },~f_{\mathbf{1}}=\sqrt{\frac{\Delta
}{\Sigma }}\partial _{r},  \notag \\
~f_{\mathbf{2}} &=&\frac{\partial _{\theta }}{\sqrt{\Sigma }},~f_{\mathbf{3}%
}=\sqrt{\frac{\Sigma }{\mathcal{A}}}\frac{\partial _{\phi }}{\sin \theta },
\label{ZAMOdown}
\end{eqnarray}%
with dual basis: $f_{a}^{\mathbf{A}}=g_{ab}\eta ^{\mathbf{AB}}f_{\mathbf{B}%
}^{b}$.

\section{Representations of spin evolution\label{REP}}

The spin vector (\ref{SvecT}) will be considered in both comoving and zero
3-momentum frames. The definitions of comoving and zero 3-momentum observers
will be introduced in the next subsection. Then the spin evolution equations
will be derived using the boosted spatial spherical and Cartesian-like
triads.

\subsection{Comoving and zero 3-momentum frames}

In the TD SSC, the center of mass is unique and measured in the zero
3-momentum frame with four velocity $p^{a}/M$. On the other hand the four
velocity of the centroid is $u^{a}$. The comoving indicative will refer to
that observer which moves along the centroid worldline. The spin dynamics
will be described in both the zero 3-momentum and the comoving observer's
frames. The velocity of the chosen observer will be denoted by $U$. The
comoving and zero 3-momentum observers' frames will be set up from the
frames of the static and the zero angular momentum observers by an
instantaneous Lorentz-boost knowing $U$ numerically.

The comoving and zero 3-momentum frames (hereafter unanimously referred as $%
U $-frame) obtained from the SO frame are given by%
\begin{equation*}
E_{\mathbf{0}}\left( e,U\right) \equiv U=\Gamma _{\left( S\right) }\left( e_{%
\mathbf{0}}+\mathbf{v}_{\left( S\right) }\right) ,
\end{equation*}%
\begin{equation}
E_{\mathbf{\alpha }}\left( e,U\right) =e_{\mathbf{\alpha }}+\frac{U\cdot e_{%
\mathbf{\alpha }}}{1+\Gamma _{\left( S\right) }}\left( U+u_{\left( SO\right)
}\right) .  \label{BoostedFrame}
\end{equation}%
Here $\mathbf{\alpha }=\left\{ \mathbf{1},\mathbf{2},\mathbf{3}\right\} $, $%
\mathbf{v}_{\left( S\right) }=\Gamma _{\left( S\right) }^{-1}U-u_{\left(
SO\right) }$ is the relative spatial velocity of either the comoving or the
zero 3-momentum observer with respect to the SO frame, which is
perpendicular to $e_{\mathbf{0}}$, and the Lorentz factor is $\Gamma
_{\left( S\right) }=-U\cdot u_{\left( SO\right) }$. The dot denotes the
inner product with respect to the metric $g_{ab}$. The inverse
transformation is given by
\begin{equation*}
e_{\mathbf{0}}=\Gamma _{\left( S\right) }\left( E_{\mathbf{0}}\left(
e,U\right) +\mathbf{w}_{\left( S\right) }\right) ,
\end{equation*}%
\begin{equation}
e_{\mathbf{\alpha }}=E_{\mathbf{\alpha }}\left( e,U\right) +\frac{u_{\left(
SO\right) }\cdot E_{\mathbf{\alpha }}\left( e,U\right) }{1+\Gamma _{\left(
S\right) }}\left( U+u_{\left( SO\right) }\right) ,
\end{equation}%
where%
\begin{equation}
\mathbf{w}_{\left( S\right) }=w_{\left( S\right) }^{\mathbf{\alpha }}E_{%
\mathbf{\alpha }}\left( e,U\right) =\Gamma _{\left( S\right) }^{-1}u_{\left(
SO\right) }-U,  \label{Ws}
\end{equation}%
is the relative spatial velocity of the static observer with respect to the $%
U$-frame.

The corresponding Lorentz-boost from the ZAMO frame reads as%
\begin{equation*}
E_{\mathbf{0}}\left( f,U\right) \equiv U=\Gamma _{\left( Z\right) }\left( f_{%
\mathbf{0}}+\mathbf{v}_{\left( Z\right) }\right) ,
\end{equation*}%
\begin{equation}
E_{\mathbf{\alpha }}\left( f,U\right) =f_{\mathbf{\alpha }}+\frac{U\cdot f_{%
\mathbf{\alpha }}}{1+\Gamma _{\left( Z\right) }}\left( U+u_{\left(
ZAMO\right) }\right) ,
\end{equation}%
with relative spatial velocity $\mathbf{v}_{\left( Z\right) }=\Gamma
_{\left( Z\right) }^{-1}U-u_{\left( ZAMO\right) }$ of the $U$-frame with
respect to the ZAMO frame, and Lorentz factor: $\Gamma _{\left( Z\right)
}=-U\cdot u^{\left( ZAMO\right) }$. The inverse boost transformation is
given by%
\begin{equation*}
f_{\mathbf{0}}=\Gamma _{\left( Z\right) }\left( E_{\mathbf{0}}\left(
f,U\right) +\mathbf{w}_{\left( Z\right) }\right) ,
\end{equation*}%
\begin{equation}
f_{\mathbf{\alpha }}=E_{\mathbf{\alpha }}\left( f,U\right) +\frac{u_{\left(
ZAMO\right) }\cdot E_{\mathbf{\alpha }}\left( f,U\right) }{1+\Gamma _{\left(
Z\right) }}\left( U+u_{\left( ZAMO\right) }\right) ,
\end{equation}%
where%
\begin{equation}
\mathbf{w}_{\left( Z\right) }=w_{\left( Z\right) }^{\mathbf{\alpha }}E_{%
\mathbf{\alpha }}\left( f,U\right) =\Gamma _{\left( Z\right) }^{-1}u_{\left(
ZAMO\right) }-U,  \label{Wz}
\end{equation}%
is the relative spatial velocity of the ZAMO with respect to either the
comoving or the zero 3-momentum frame.

Since $E_{\mathbf{0}}\left( e,U\right) =U=E_{\mathbf{0}}\left( f,U\right) $,
the transformation between the frames $E_{\mathbf{A}}\left( e,U\right) $ and
$E_{\mathbf{A}}\left( f,U\right) $ is a rotation in the rest space of either
the comoving or the zero 3-momentum observer. The rotation axis has the
following non-zero components in both the $E_{\mathbf{A}}\left( e,U\right) $
and the $E_{\mathbf{A}}\left( f,U\right) $ frames:%
\begin{eqnarray}
n^{\mathbf{1}} &=&-\frac{w_{^{\left( Z\right) }}^{\mathbf{2}}}{\sqrt{\left(
w_{^{\left( Z\right) }}^{\mathbf{1}}\right) ^{2}+\left( w_{^{\left( Z\right)
}}^{\mathbf{2}}\right) ^{2}}}  \notag \\
&=&-\frac{w_{^{\left( S\right) }}^{\mathbf{2}}}{\sqrt{\left( w_{^{\left(
S\right) }}^{\mathbf{1}}\right) ^{2}+\left( w_{^{\left( S\right) }}^{\mathbf{%
2}}\right) ^{2}}},  \label{n1}
\end{eqnarray}%
and%
\begin{eqnarray}
n^{\mathbf{2}} &=&\frac{w_{^{\left( Z\right) }}^{\mathbf{1}}}{\sqrt{\left(
w_{^{\left( Z\right) }}^{\mathbf{1}}\right) ^{2}+\left( w_{^{\left( Z\right)
}}^{\mathbf{2}}\right) ^{2}}}  \notag \\
&=&\frac{w_{^{\left( S\right) }}^{\mathbf{1}}}{\sqrt{\left( w_{^{\left(
S\right) }}^{\mathbf{1}}\right) ^{2}+\left( w_{^{\left( S\right) }}^{\mathbf{%
2}}\right) ^{2}}}.  \label{n2}
\end{eqnarray}%
The rotation angle $\Theta $ is determined by%
\begin{eqnarray}
\sin \Theta  &=&\left[ \left( 1-\sqrt{\frac{\Sigma \Delta }{-g_{tt}\mathcal{A%
}}}\right) \frac{\Gamma _{\left( Z\right) }w_{\left( Z\right) }^{\mathbf{3}}%
}{1+\Gamma _{\left( Z\right) }}+\frac{a\mathcal{B}\sin \theta }{\sqrt{%
-g_{tt}\Sigma \mathcal{A}}}\right]   \notag \\
&&\times \frac{\Gamma _{\left( Z\right) }\sqrt{\left( w_{\left( Z\right) }^{%
\mathbf{1}}\right) ^{2}+\left( w_{\left( Z\right) }^{\mathbf{2}}\right) ^{2}}%
}{1+\Gamma _{\left( S\right) }}  \notag \\
&=&-\left[ \left( 1-\sqrt{\frac{\Sigma \Delta }{-g_{tt}\mathcal{A}}}\right)
\frac{\Gamma _{\left( S\right) }w_{\left( S\right) }^{\mathbf{3}}}{1+\Gamma
_{\left( S\right) }}-\frac{a\mathcal{B}\sin \theta }{\sqrt{-g_{tt}\Sigma
\mathcal{A}}}\right]   \notag \\
&&\times \frac{\Gamma _{\left( S\right) }\sqrt{\left( w_{\left( S\right) }^{%
\mathbf{1}}\right) ^{2}+\left( w_{\left( S\right) }^{\mathbf{2}}\right) ^{2}}%
}{1+\Gamma _{\left( Z\right) }},  \label{sinTheta}
\end{eqnarray}%
and%
\begin{eqnarray}
\frac{\cos \Theta -1}{1-\sqrt{\frac{\Sigma \Delta }{-g_{tt}\mathcal{A}}}} &=&%
\frac{\Gamma _{\left( Z\right) }^{2}\left[ \left( w_{\left( Z\right) }^{%
\mathbf{1}}\right) ^{2}+\left( w_{\left( Z\right) }^{\mathbf{2}}\right) ^{2}%
\right] }{\left( 1+\Gamma _{\left( S\right) }\right) \left( 1+\Gamma
_{\left( Z\right) }\right) }  \notag \\
&=&\frac{\Gamma _{\left( S\right) }^{2}\left[ \left( w_{\left( S\right) }^{%
\mathbf{1}}\right) ^{2}+\left( w_{\left( S\right) }^{\mathbf{2}}\right) ^{2}%
\right] }{\left( 1+\Gamma _{\left( S\right) }\right) \left( 1+\Gamma
_{\left( Z\right) }\right) }.  \label{cosTheta}
\end{eqnarray}%
The frame $E_{\mathbf{\alpha }}\left( e,U\right) $ ($E_{\mathbf{\alpha }%
}\left( f,U\right) $) is obtained from $E_{\mathbf{\alpha }}\left(
f,U\right) $ ($E_{\mathbf{\alpha }}\left( e,U\right) $) by a rotation with
the angle $\Theta $ ($-\Theta $) about the axis $\mathbf{n}$. The rotation
angle $\Theta $ exists outside the ergosphere where the terms under the
square roots in Equations (\ref{sinTheta}) and (\ref{cosTheta}) are
positive. The transformation between $E_{\mathbf{\alpha }}\left( e,U\right) $
and $E_{\mathbf{\alpha }}\left( f,U\right) $ in another form is given in
Appendix \ref{comFRametraf}. The above transformation is a special case of
the Wigner-rotation \cite{Wigner1939}, which was discussed recently in Ref.
\cite{Bini2018}. However explicit expressions for the rotation between the
frames which we denote $E_{\mathbf{\alpha }}\left( f,U\right) $ and $E_{%
\mathbf{\alpha }}\left( e,U\right) $ were not presented in \cite{Bini2018}.

\subsection{MPD spin equations in comoving and zero 3-momentum frames}

We investigate two cases related to the chosen $U$-frame: $i)$ $%
U^{a}=p^{a}/M $ when we work in the zero 3-momentum frame; and $ii)$ $%
U^{a}=u^{a}$ which is the four velocity of the center of mass measured in
the zero 3-momentum frame. In all cases the spin vector can be expanded as
\begin{equation}
S=S^{\mathbf{\alpha }}E_{\mathbf{\alpha }},
\end{equation}%
since $S^{\mathbf{0}}=0$. Here, the spatial frame vector $E_{\mathbf{\alpha }%
} $ in the $U$-frame denotes either $E_{\mathbf{\alpha }}\left( e,U\right) $
or $E_{\mathbf{\alpha }}\left( f,U\right) $, which are obtained by boosting
the SO and ZAMO frames, respectively.

The covariant derivative of the spin vector along the integral curve of $u$
is%
\begin{equation}
\frac{DS}{d\tau }=\frac{dS^{\mathbf{\alpha }}}{d\tau }E_{\mathbf{\alpha }%
}+S^{\mathbf{\alpha }}\frac{DE_{\mathbf{\alpha }}}{d\tau }.  \label{SpinEq0}
\end{equation}%
Since the frame vectors are perpendicular to each other, we have%
\begin{equation}
E_{\mathbf{A}}\cdot \frac{DE_{\mathbf{B}}}{d\tau }=-E_{\mathbf{B}}\cdot
\frac{DE_{\mathbf{A}}}{d\tau },  \label{rel1}
\end{equation}%
for $\mathbf{A}\neq \mathbf{B}$, and because of the normalization:%
\begin{equation}
E_{\mathbf{A}}\cdot \frac{DE_{\mathbf{A}}}{d\tau }=0.
\end{equation}%
Therefore the covariant derivatives of the spatial frame vectors along the
integral curve of $u$ can be expressed as
\begin{equation}
\frac{DE_{\mathbf{\alpha }}}{d\tau }=-\left( E_{\mathbf{0}}\cdot \frac{DE_{%
\mathbf{\alpha }}}{d\tau }\right) E_{\mathbf{0}}-\varepsilon _{\mathbf{%
\alpha \beta }}^{~\ \ \mathbf{\gamma }}\Omega ^{\mathbf{\beta }}E_{\mathbf{%
\gamma }},  \label{DEdt}
\end{equation}%
where $\varepsilon _{\mathbf{\alpha \beta }}^{~\ \ \mathbf{\gamma }}$
Levi-Civita symbol in the 3-dimensional Euclidean space, whose frame indices
are raised and lowered by the 3-dimensional Kronecker $\delta $, and the
frame components of the angular velocity are%
\begin{equation}
\Omega ^{\mathbf{\alpha }}=-\frac{1}{2}\varepsilon ^{\mathbf{\alpha \beta
\gamma }}E_{\mathbf{\beta }}\cdot \frac{DE_{\mathbf{\gamma }}}{d\tau }.
\label{OmegaDef}
\end{equation}%
Due to Equation (\ref{rel1}), the first term\footnote{%
Note that this term vanishes when the first order spin corrections,
i.e. the right hand sides of Equations (\ref{MPD1}) and
(\ref{MPD2}), are neglected
(see Ref. \cite{Bini2017}).} in (\ref{DEdt}) can be written as%
\begin{equation}
E_{\mathbf{0}}\cdot \frac{DE_{\mathbf{\alpha }}}{d\tau }=-E_{\mathbf{\alpha }%
}\cdot \mathbf{a},
\end{equation}%
where $\mathbf{a}$ denotes the acceleration $\mathbf{a}=DE_{\mathbf{0}%
}/d\tau $. Now the spin Equation (\ref{SpinEq0}) becomes as%
\begin{equation}
\frac{DS}{d\tau }=\left( \frac{dS^{\mathbf{\alpha }}}{d\tau }+\varepsilon _{~%
\mathbf{\beta \gamma }}^{\mathbf{\alpha }}\Omega ^{\mathbf{\beta }}S^{%
\mathbf{\gamma }}\right) E_{\mathbf{\alpha }}+\left( S\cdot \mathbf{a}%
\right) E_{\mathbf{0}}.  \label{spinevo}
\end{equation}

Finally, we take into account the spin Equation (\ref{SevoTD}). When
considering the spin evolution in the zero 3-momentum frame, we find the
following equation for the spin:%
\begin{equation}
\frac{dS^{\mathbf{\alpha }}}{d\tau }+\varepsilon _{~\mathbf{\beta \gamma }}^{%
\mathbf{\alpha }}\Omega ^{\mathbf{\beta }}S^{\mathbf{\gamma }}=0.
\label{SpinTDp}
\end{equation}

The second case when considering the evolution of $S^{\mathbf{\alpha }}$ in
the comoving frame, requires somewhat longer computation. Equations (\ref%
{SevoTD}) and (\ref{spinevo}) results in%
\begin{equation}
\left( \frac{dS^{\mathbf{\alpha }}}{d\tau }+\varepsilon _{~\mathbf{\beta
\gamma }}^{\mathbf{\alpha }}\Omega ^{\mathbf{\beta }}S^{\mathbf{\gamma }%
}\right) E_{\mathbf{\alpha }}+\mathbf{\Upsilon }=0,  \label{SpinTDu0}
\end{equation}%
with%
\begin{equation}
\mathbf{\Upsilon =}\left[ \left( S\cdot \mathbf{a}\right) u^{\mathbf{A}%
}-\left( S\cdot \mathbf{F}\right) \frac{p^{\mathbf{A}}}{M^{2}}\right] E_{%
\mathbf{A}}.
\end{equation}%
Using Equations (\ref{u}) and (\ref{Sconstr}), a straightforward computation
shows that $u\cdot \mathbf{\Upsilon }=0$. Therefore, $\mathbf{\Upsilon }$
can be expanded as $\mathbf{\Upsilon }=\Upsilon ^{\mathbf{\alpha }}E_{%
\mathbf{\alpha }}$. On the other hand $\mathbf{\Upsilon }$ is perpendicular
to $S$, hence, we can introduce a vector $\mathbf{\omega }$, whose frame
components obey the relation%
\begin{equation}
\varepsilon _{~\mathbf{\beta \gamma }}^{\mathbf{\alpha }}\omega ^{\mathbf{%
\beta }}S^{\mathbf{\gamma }}=\mathbf{\Upsilon ^{\mathbf{\alpha }}.}
\label{omegadef}
\end{equation}%
The vector $\mathbf{\omega }$ is determined ambiguously since its frame
component parallel with $S$ vanishes in the cross product. As a natural
choice, we choose $\mathbf{\omega }$ to be perpendicular to $S$. Using the
definition (\ref{omegadef}), Equation (\ref{SpinTDu0}) reads as%
\begin{equation}
\frac{dS^{\mathbf{\alpha }}}{d\tau }+\varepsilon _{~\mathbf{\beta \gamma }}^{%
\mathbf{\alpha }}\left( \Omega ^{\mathbf{\beta }}+\omega ^{\mathbf{\beta }%
}\right) S^{\mathbf{\gamma }}=0.  \label{SpinTDu}
\end{equation}

The Equations (\ref{SpinTDp}) and (\ref{SpinTDu}) can be considered in
either the $E_{\mathbf{\alpha }}\left( e,U\right) $ or the $E_{\mathbf{%
\alpha }}\left( f,U\right) $ frame. Introducing the notations%
\begin{equation}
k=\left\{ e,f\right\} ,~\Gamma =\left\{ \Gamma _{\left( S\right) },\Gamma
_{\left( Z\right) }\right\} ,
\end{equation}%
the angular velocity components $\Omega ^{\mathbf{\alpha }}\left( k,U\right)
$ can be determined by using%
\begin{eqnarray}
&&E_{\mathbf{\beta }}\left( k,U\right) \cdot \frac{DE_{\mathbf{\alpha }%
}\left( k,U\right) }{d\tau }  \notag \\
&=&k_{\mathbf{\beta }}\cdot \frac{Dk_{\mathbf{\alpha }}}{d\tau }  \notag \\
&&+\frac{1}{1+\Gamma }\left[ \left( U\cdot k_{\mathbf{\alpha }}\right) k_{%
\mathbf{\beta }}-\left( U\cdot k_{\mathbf{\beta }}\right) k_{\mathbf{\alpha }%
}\right] \cdot \frac{Dk_{\mathbf{0}}}{d\tau }  \notag \\
&&+\frac{1}{1+\Gamma }\left[ \left( U\cdot k_{\mathbf{\alpha }}\right) k_{%
\mathbf{\beta }}-\left( U\cdot k_{\mathbf{\beta }}\right) k_{\mathbf{\alpha }%
}\right] \cdot \frac{DU}{d\tau },  \label{Omrels}
\end{eqnarray}%
where $\mathbf{\alpha }\neq \mathbf{\beta }$. This can be computed once $U$
is determined.\footnote{%
We note that when the right hand sides of Equations (\ref{MPD1}) and (\ref%
{MPD2}) are neglected, the centroid moves along a geodesic, thus
$\mathbf{\ \omega }$ and the last term in (\ref{Omrels}) vanish. The
four-velocity $U$ is determined from the geodesic equation, and for
$k=e$, we obtain the same system which was investigated in Ref.
\cite{Bini2017}.}

When both SO and ZAMO frames exist, a rotation about the axis $\mathbf{n}$
defined by Equations (\ref{n1}) and (\ref{n2}) [see also Appendix \ref%
{comFRametraf} for the explicit expressions] relates $E_{\mathbf{\alpha }%
}\left( f,U\right) $ to $E_{\mathbf{\alpha }^{\prime }}\left( e,U\right) $
which can be written as%
\begin{equation}
E_{\mathbf{\beta }}\left( e,U\right) =\mathcal{R}_{~\mathbf{\beta }}^{%
\mathbf{\alpha }^{\prime }}\,E_{\mathbf{\alpha }^{\prime }}\left( f,U\right)
.
\end{equation}%
Here $\mathcal{R}_{~\mathbf{\beta }}^{\mathbf{\alpha }^{\prime }}$ denotes
the components of the corresponding rotation matrix. From the definitions of
$\Omega ^{\mathbf{\alpha }}\left( e,U\right) $ and $\Omega ^{\mathbf{\alpha }%
^{\prime }}\left( f,U\right) $, the following relation between them can be
derived:

\begin{equation}
\mathcal{R}_{~\mathbf{\alpha }}^{\mathbf{\beta }^{\prime }}\Omega ^{\mathbf{%
\alpha }}\left( e,U\right) =\Omega ^{\mathbf{\beta }^{\prime }}\left(
f,U\right) +\mathcal{R}_{~\mathbf{\alpha }}^{\mathbf{\beta }^{\prime
}}\Omega _{\left( \mathcal{R}\right) }^{\mathbf{\alpha }}.
\end{equation}%
Here we have introduced $\Omega _{\left( \mathcal{R}\right) }^{\mathbf{%
\gamma }}$ as%
\begin{equation}
\left( \mathcal{R}^{-1}\right) _{~\mathbf{\nu }^{\prime }}^{\mathbf{\alpha }}%
\frac{d\mathcal{R}_{~\mathbf{\beta }}^{\mathbf{\nu }^{\prime }}}{d\tau }%
=\varepsilon _{~~\mathbf{\gamma \beta }}^{\mathbf{\alpha }}\Omega _{\left(
\mathcal{R}\right) }^{\mathbf{\gamma }},
\end{equation}%
which is the angular velocity of rotation between the frame bases along the
body's trajectory.

\subsubsection{Cartesian-like triads and the characterization of spin
evolution\label{spindesc}}

The evolution of the spin vector can be illustrated suitably by comparison
its direction with Cartesian axes which are fixed with respect to the
distant stars. The static observers are those fiducial observers, whose
frame does not move with respect to the black hole's asymptotic frame \cite%
{MTW}. A static observer sees the same \textquotedblleft
nonrotating\textquotedblright\ sky during the evolution. In this sense the
static observers are preferred fiducial observers in the investigation of
spin dynamics. Following Ref. \cite{Bini2017}, we introduce a spatial
Cartesian-like triad $e_{\mathbf{x}}$, $e_{\mathbf{y}}$ and $e_{\mathbf{z}}$
in the local rest space of the static observer as $e_{\mathbf{\alpha }}=R_{~%
\mathbf{\alpha }}^{\mathbf{i}}e_{\mathbf{i}}$, where $\mathbf{\alpha }%
=\left\{ \mathbf{1},\mathbf{2},\mathbf{3}\right\} $, $\mathbf{i}=\left\{
\mathbf{x},\mathbf{y},\mathbf{z}\right\} $ and $R$ is the same rotation
matrix, which relates the Cartesian and spherical coordinates in the
3-dimensional Euclidean space (see Equation (85) of \cite{Bini2017}). Since
the rotation $R$ and the boost can be interchanged, we have $E_{\mathbf{%
\alpha }}\left( e,U\right) =R_{~\mathbf{\alpha }}^{\mathbf{i}}E_{\mathbf{i}%
}\left( e,U\right) $.

The family of static observers only determines a frame outside the
ergosphere. Therefore, we introduce another Cartesian-like triad $f_{\mathbf{%
x}}$, $f_{\mathbf{y}}$ and $f_{\mathbf{z}}$ in the local rest space of ZAMO
for representation of the spin evolution inside the ergosphere\footnote{%
Note that the frame associated to the ZAMO moves with respect to the distant
stars.} as $f_{\mathbf{\alpha }}=R_{~\mathbf{\alpha }}^{\mathbf{i}}f_{%
\mathbf{i}}$. Then the boost transformation results in $E_{\mathbf{\alpha }%
}\left( f,U\right) =R_{~\mathbf{\alpha }}^{\mathbf{i}}E_{\mathbf{i}}\left(
f,U\right) $.

The Cartesian-like triad components of the spin vector in both the boosted
SO and ZAMO frames are obtained as%
\begin{equation}
S^{\mathbf{i}}=R_{~\mathbf{\alpha }}^{\mathbf{i}}S^{\mathbf{\alpha }},
\label{Cart}
\end{equation}%
which obeys the following equation of motion:%
\begin{equation}
\frac{dS^{\mathbf{i}}}{d\tau }=-R_{~\mathbf{\alpha }}^{\mathbf{i}%
}\varepsilon _{~\mathbf{\beta \gamma }}^{\mathbf{\alpha }}\Omega _{\left(
prec\right) }^{\mathbf{\beta }}S^{\mathbf{\gamma }}.  \label{spineq}
\end{equation}%
Here the precession angular velocity is\footnote{%
Note that that the expression of $\Omega _{\left( prec\right) }^{\mathbf{%
\beta }}$ reduces to -1 times that of Ref. \cite{Bini2017} for $\epsilon =0$.%
}
\begin{equation}
\Omega _{\left( prec\right) }^{\mathbf{\beta }}=\Omega _{\left( p\right) }^{%
\mathbf{\beta }}+\epsilon \omega ^{\mathbf{\beta }},  \label{omegaprec}
\end{equation}%
with%
\begin{equation}
\Omega _{\left( p\right) }^{\mathbf{\beta }}=-\Omega _{\left( orb\right) }^{%
\mathbf{\beta }}+\Omega ^{\mathbf{\beta }},
\end{equation}%
where $\Omega _{\left( orb\right) }^{\mathbf{\beta }}$ defined [see also
Ref. \cite{Bini2017}] as
\begin{equation}
\left( R^{-1}\right) _{~~\mathbf{j}}^{\mathbf{\alpha }}\frac{dR_{~\mathbf{%
\beta }}^{\mathbf{j}}}{d\tau }=\varepsilon _{~\mathbf{\gamma \beta }}^{%
\mathbf{\alpha }}\Omega _{\left( orb\right) }^{\mathbf{\gamma }},
\label{Omegaorb}
\end{equation}%
and $\epsilon =0$ in the zero 3-momentum frame, while $\epsilon =1$ in the
comoving frame. The angular velocity $\Omega _{\left( prec\right) }^{\mathbf{%
\beta }}$ describes the spin precession\ in the Cartesian-like frame. The
Cartesian-like triad components of $\Omega _{\left( prec\right) }^{\mathbf{%
\beta }}$ are obtained from Equation (\ref{Cart}) with notation change $%
S\rightarrow \Omega _{\left( prec\right) }$.

The quantity $\Omega _{\left( p\right) }$ can also be expressed in terms of
the inner product of the Cartesian-like triad vectors $E_{\mathbf{i}}$ and
their derivatives along the considered worldline as
\begin{equation}
\Omega _{\left( p\right) }^{\mathbf{i}}\equiv R_{~\mathbf{\beta }}^{\mathbf{i%
}}\Omega _{\left( p\right) }^{\mathbf{\beta }}=-\frac{1}{2}\varepsilon ^{%
\mathbf{ijk}}E_{\mathbf{j}}\cdot \frac{DE_{\mathbf{k}}}{d\tau }.
\label{OmegaPrecDef}
\end{equation}%
This expression is analogous with Equation (\ref{OmegaDef}). The angular
velocities $\Omega _{\left( p\right) }^{\mathbf{i}}\left( e,U\right) $ and $%
\Omega _{\left( p\right) }^{\mathbf{i}^{\prime }}\left( f,U\right) $ defined
in terms of $E_{\mathbf{i}}\left( e,U\right) $ and $E_{\mathbf{i}^{\prime
}}\left( f,U\right) $, respectively, are related by
\begin{eqnarray}
&&T_{~\ \mathbf{i}}^{\mathbf{k}^{\prime }}\Omega _{\left( p\right) }^{%
\mathbf{i}}\left( e,U\right) +S_{~\mathbf{\alpha }}^{\mathbf{k}^{\prime
}}\Omega _{\left( orb\right) }^{\mathbf{\alpha }}\left( e,U\right)   \notag
\\
&=&\Omega _{\left( p\right) }^{\mathbf{k}^{\prime }}\left( f,U\right)
+R_{\left( f\right) \,\mathbf{\beta }^{\prime }}^{\mathbf{k}^{\prime
}}\Omega _{\left( orb\right) }^{\mathbf{\beta }^{\prime }}\left( f,U\right)
+S_{~\mathbf{\alpha }}^{\mathbf{k}^{\prime }}\Omega _{\left( \mathcal{R}%
\right) }^{\mathbf{\alpha }},  \label{Omptrafo}
\end{eqnarray}%
with%
\begin{equation}
T_{~\ \mathbf{i}}^{\mathbf{k}^{\prime }}\equiv \left( R_{\left( e\right)
}^{-1}\right) _{~\mathbf{i}}^{\mathbf{\beta }}\mathcal{R}_{~\mathbf{\beta }%
}^{\mathbf{\alpha }^{\prime }}R_{\left( f\right) \,\mathbf{\alpha }^{\prime
}}^{\mathbf{k}^{\prime }},~S_{~\ \mathbf{\alpha }}^{\mathbf{k}^{\prime
}}\equiv R_{\left( f\right) \,\mathbf{\beta }^{\prime }}^{\mathbf{k}^{\prime
}}\mathcal{R}_{~\mathbf{\alpha }}^{\mathbf{\beta }^{\prime }}.
\end{equation}%
Noting that $\omega ^{\alpha }$ in Equation (\ref{omegadef}) transforms as a
vector for real rotation $\mathcal{R}_{~\mathbf{\alpha }}^{\mathbf{\beta }%
^{\prime }}$. The transformation rules for $\Omega _{\left( prec\right) }^{%
\mathbf{\beta }}$ and $\Omega _{\left( prec\right) }^{\mathbf{i}}$\ follow
from the definitions (\ref{omegaprec}) and (\ref{Omptrafo}).

\section{Numerical investigations\label{num}}

The orbit of the spinning body will be represented in the coordinate space:%
\begin{equation}
x=r\cos \phi \sin \theta ,~y=r\sin \phi \sin \theta ,~z=r\cos \theta .
\label{Cartesian}
\end{equation}%
We characterize the instantaneous plane of the motion in the ($x$, $y$, $z$%
)-space by the unit vector:%
\begin{equation}
\mathbf{l}=\frac{\mathbf{R}\times \mathbf{V}}{\left\vert \mathbf{R}\times
\mathbf{V}\right\vert },  \label{ldef}
\end{equation}%
where $\times $ is the cross product in Euclidean 3-space, $\mathbf{R}$ is
the position vector with components $R^{\mathbf{x}}=x$,$~R^{\mathbf{y}}=y$,$%
~R^{\mathbf{z}}=z$, and $\mathbf{V}$ is a spatial velocity vector with%
\footnote{%
Noting that we could use any timelike parameter in the definition (\ref{vel}%
) due to the normalisation in Equation (\ref{ldef}).}%
\begin{equation}
V^{\mathbf{x}}=\frac{dx}{d\tau },~V^{\mathbf{y}}=\frac{dy}{d\tau },~V^{%
\mathbf{z}}=\frac{dz}{d\tau }.  \label{vel}
\end{equation}%
The absolute value in the denominator denotes the \textquotedblleft
Euclidean length\textquotedblright\ of the numerator. Since the considered
spacetimes are asymptotically flat, the quantity $l^{\mathbf{i}}$ coincides
with the direction of the orbital angular momentum\footnote{%
We mention that the natural definition of orbital angular momentum around $%
x_{0}$ would be%
\begin{equation*}
L^{ab}=-\sigma ^{\lbrack a}\left( x,x_{0}\right) p^{b]}\left( \tau \right) .
\end{equation*}%
Here $\sigma ^{a}$ is a generalized position vector which can be computed
from the Synge's world function \cite{Synge}. However, there are only few
metrics for which the exact world function is known \cite%
{Ruse1930,Gunther1965,Buchdahl1972,BuchdahlWarner1980,John1984,John1989,Roberts1999}%
.} at spatial infinity.

The initial data for the spin vector will be characterized by its magnitude
and two angles in the boosted SO Cartesian-like frame as%
\begin{equation}
S=S^{\mathbf{i}}E_{\mathbf{i}}\left( e,u\right) ,
\end{equation}%
with%
\begin{equation}
S^{\mathbf{i}}=\left\vert S\right\vert \left( \cos \phi ^{\left( S\right)
}\sin \theta ^{\left( S\right) },\sin \phi ^{\left( S\right) }\sin \theta
^{\left( S\right) },\cos \theta ^{\left( S\right) }\right) .  \label{Sinit}
\end{equation}

Since we use dimensionless quantities during the numerical investigation,
the parameters $\mu $, $a$, $m$ and $M$ only appear through the ratios $%
a/\mu $ and $m/M$. We choose the initial data set in the TD SSC as $%
p_{\left( TD\right) }^{a}/M$ and $S^{a}/\mu M$ (by Equation (\ref{Sinit})),
then the initial spin tensor is derived from the inverse of (\ref{SvecT}),
while $m\left( 0\right) /M\left( 0\right) $ and the four velocity $u_{\left(
TD\right) }^{a}$ of the centroid from (\ref{u}).

The SSC choice determining the representative worldline of a spinning test
body corresponds to a gauge choice in an action approach \cite{Steinhoff2015}%
. In the following we will consider the evolution of the spin precessional
angular velocity and will check its dependence on the SSC choice. For the
numerical comparison, we will use the Frenkel--Mathisson--Pirani (FMP) SSC
which imposes $u_{a}S^{ab}=0$. The definition of the spin vector is $%
s^{a}=-\eta ^{abcd}u_{b}S_{cd}/2$, which is Fermi-Walker transported along
the worldline of the centroid making the FMP SSC preferred from mathematical
point of view \cite{Costa2012,Costa2015,Costa2016}. Its frame components
obey the same precessional equation in the comoving frame like the TD spin
vector $S^{a}$ in the zero 3-momentum frame (\ref{SevoTD}). In the FMP SSC,
there exists also a velocity-momentum relation, Equation (19) of Ref. \cite%
{Costa2018}, like in the TD SSC. Hence the initial data set $\left\{
x^{a},p^{a},S^{ab}\right\} |_{\tau _{in}}$ provides a unique solution of the
MPD equation with FMP SSC. However, we must mention that, this
velocity-momentum relation does not automatically ensure that $u_{a}S^{ab}=0$
for arbitrary $p^{a}$ and $S^{ab}$. In order to ensure this, we have a
constraint between the four momentum and the spin tensor emerging from the
contraction of this equation with $S_{ab}$. In addition, the data set $%
\left\{ x^{a},p^{a},S^{ab}\right\} |_{\tau _{in}}$ cannot be inverted for
the set $\left\{ x^{a},m,u^{a},S^{ab}\right\} |_{\tau _{in}}$ like in the TD
SSC. One needs the data set $\left\{ x^{a},m,u^{a},S^{ab},a^{a}\right\}
|_{\tau _{in}}$ to fix the trajectory. For a set $\left\{
x^{a},m,u^{a},S^{ab}\right\} |_{\tau _{in}}$, we can obtain a non-helical
and infinite number of helical trajectories for 
\begin{widetext}
\newpage
\begin{figure}[H]
\begin{center}
\includegraphics[width=5cm]{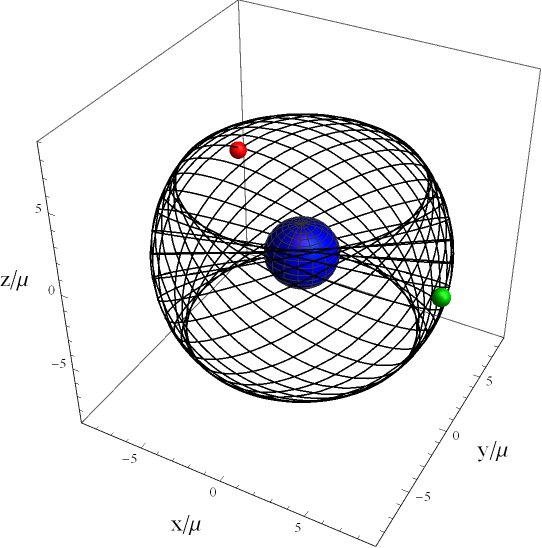} \hfill %
\includegraphics[width=5cm]{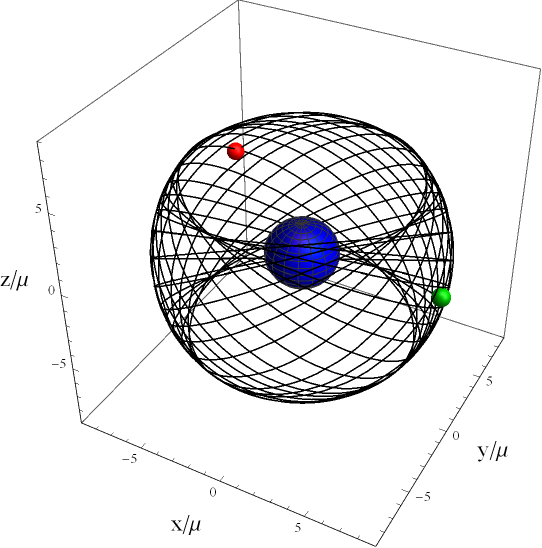} \hfill %
\includegraphics[width=5cm]{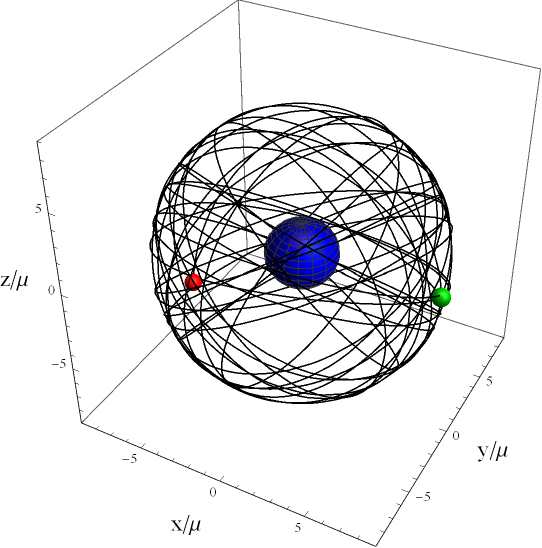} \\[0pt]
\includegraphics[width=5cm]{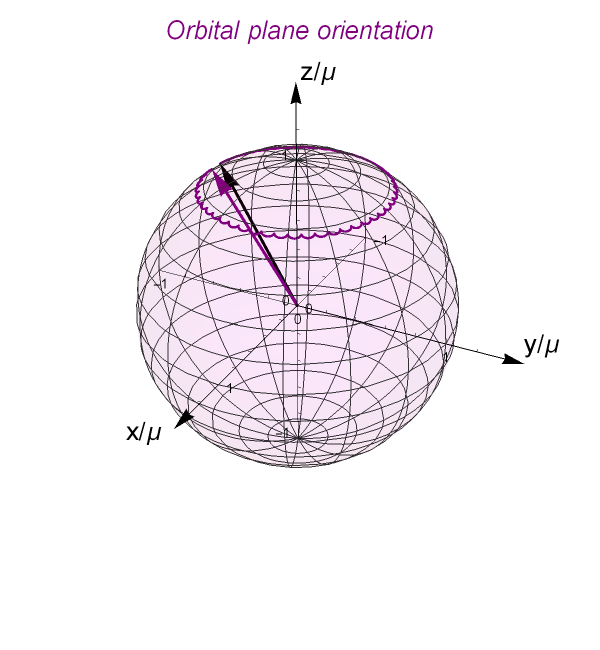} \hfill %
\includegraphics[width=5cm]{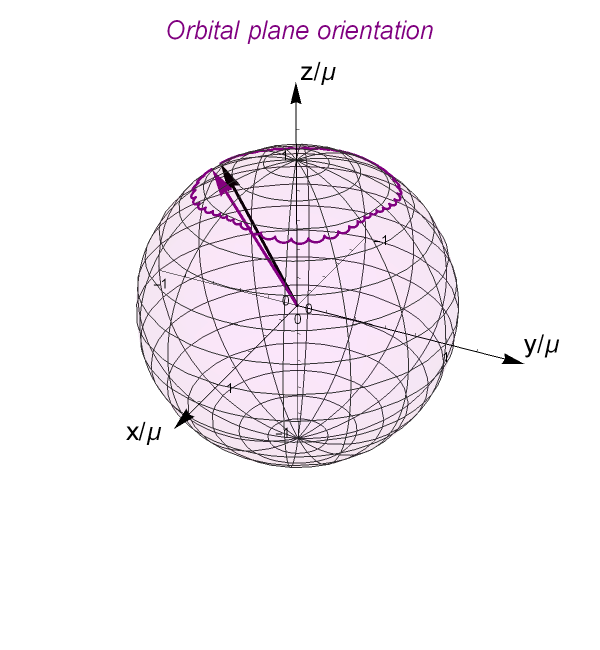} \hfill %
\includegraphics[width=5cm]{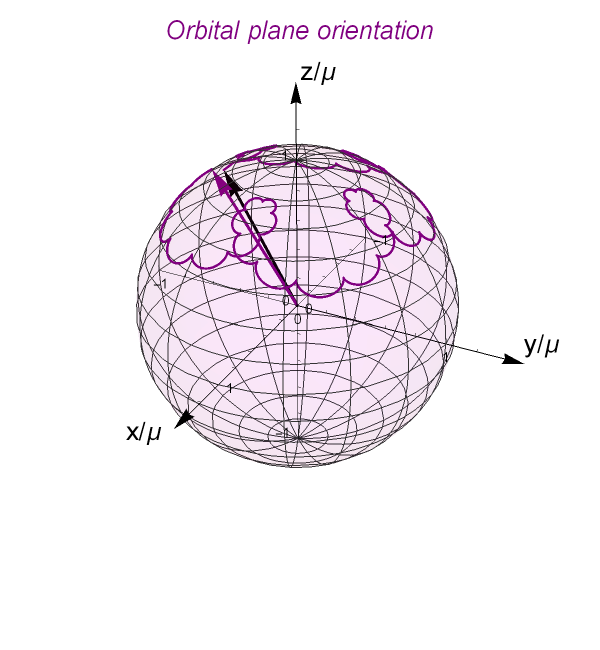} \\[0pt]
\vskip -1cm \includegraphics[width=5cm]{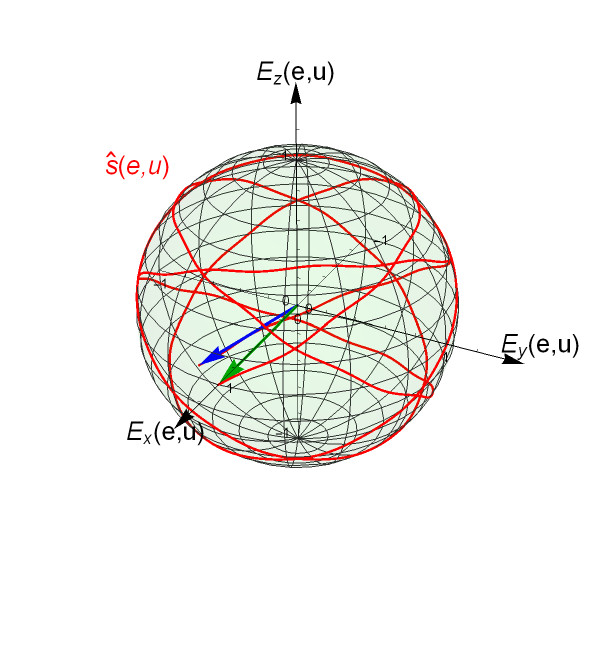}\hfill %
\includegraphics[width=5cm]{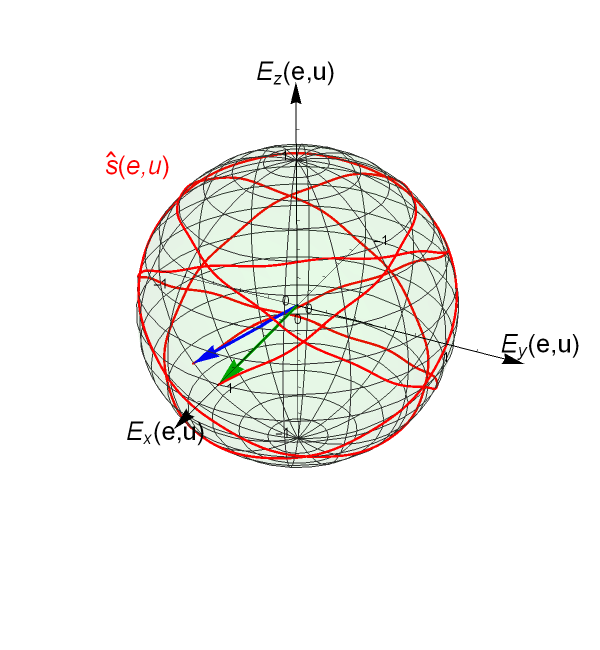} \hfill %
\includegraphics[width=5cm]{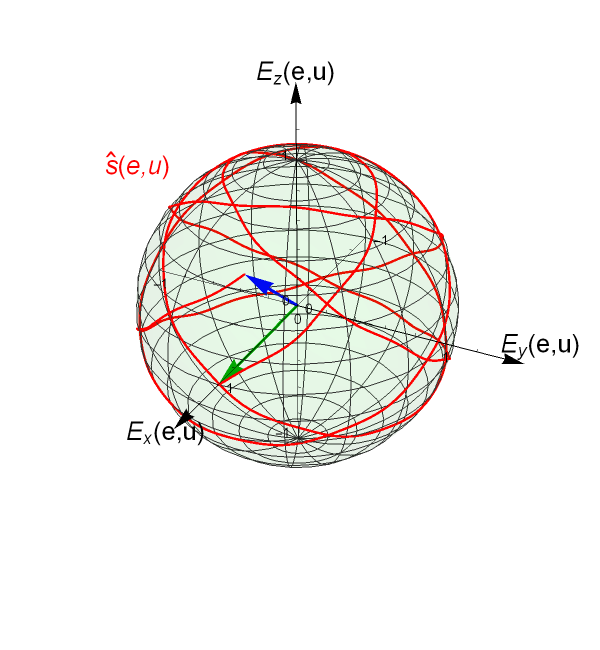} \\[0pt]
\vskip -1cm \includegraphics[width=5cm]{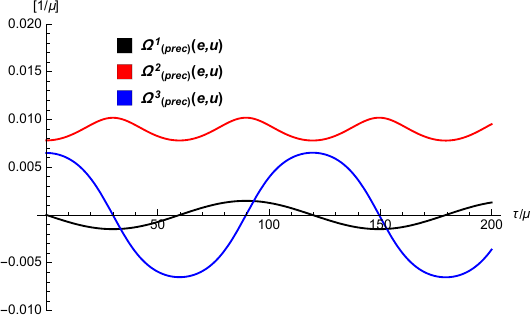} \hfill %
\includegraphics[width=5cm]{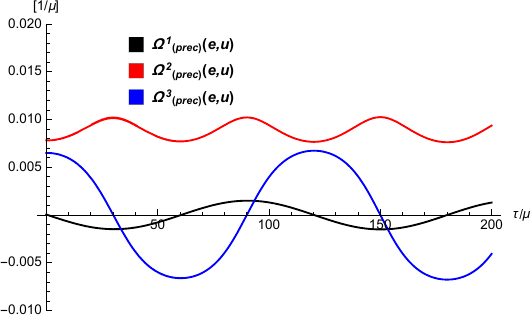} \hfill %
\includegraphics[width=5cm]{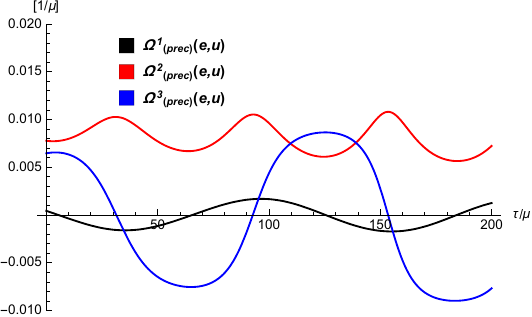} \\[0pt]
\includegraphics[width=5cm]{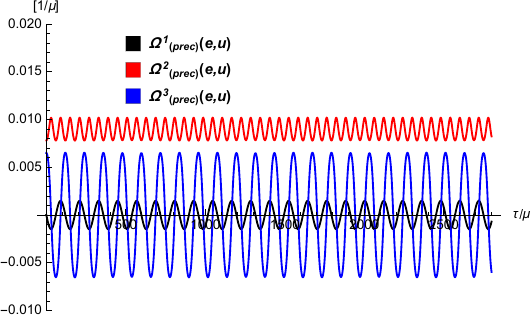} \hfill %
\includegraphics[width=5cm]{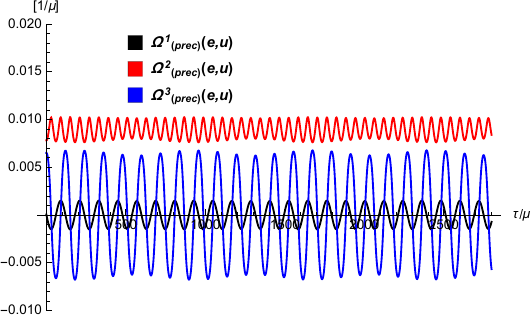} \hfill %
\includegraphics[width=5cm]{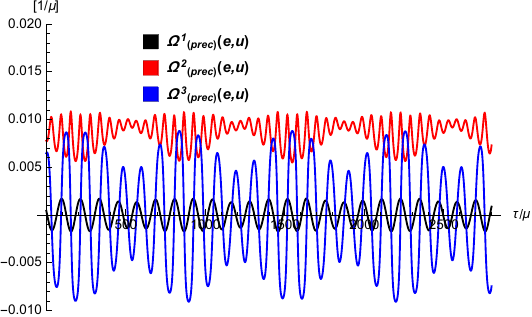} \\[0pt]
\includegraphics[width=5cm]{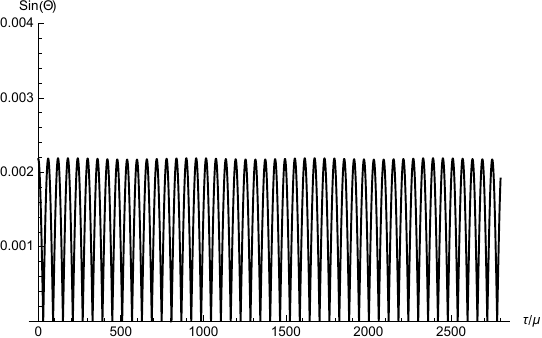} \hfill %
\includegraphics[width=5cm]{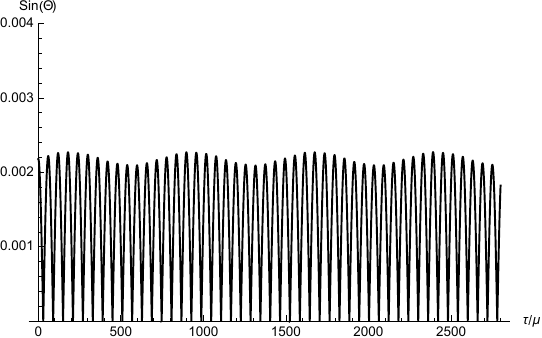} \hfill %
\includegraphics[width=5cm]{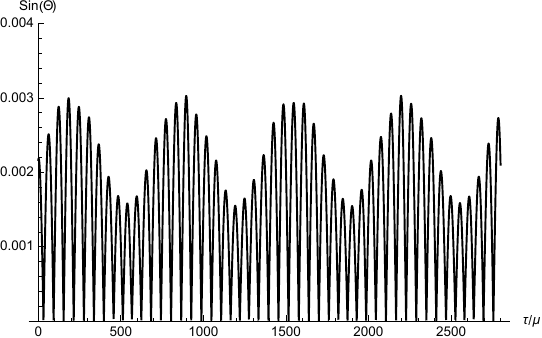}
\end{center}
\par
\vskip -0.5cm \caption{(color online). The evolution of spinning
body moving on spherical-like orbits around the Kerr black hole with
$a=0.5\protect\mu $. From left to right the magnitude of the body's
spin increases as $\left\vert S\right\vert /\protect\mu M=0.01$,
$0.1$ and $0.9$. The rows represent the following: 1. the orbit in
coordinate space ($x/\protect\mu $,$y/\protect\mu $,$z/\protect\mu
$) (the ergosphere of the central black hole is marked by blue and
the initial and the final positions of the spinning body are denoted
by green and red dots, respectively,), 2. the instantaneous orbital
plane orientation $l^{\mathbf{i}}$ (initial and final directions are
marked by purple and black arrows, respectively), 3. unit spin
vector in the boosted SO comoving Cartesian-like frame
$E_{\mathbf{i}}\left( e,u\right) $ (initial and final spin
directions are marked by green and blue arrows,
respectively), 4. and 5. $\Omega _{\left( prec\right) }^{\mathbf{\protect%
\alpha }}\left( e,u\right) $ on shorter and longer timescales,
respectively,
6. $\sin \Theta $. The initial place of the body is $r(0)=8\protect\mu $, $%
\protect\theta (0)=\protect\pi /2$ and $\protect\phi (0)=0$. The
direction of the initial spin vector is given by $\protect\theta
^{\left( S\right) }\left( 0\right) =\protect\pi /2$ and
$\protect\phi ^{\left( S\right) }\left( 0\right)=0$ in the boosted
SO frame (resulting in $S^{r}\left(
0\right) /\left\vert S\right\vert =0.8682$, $\protect\mu S^{\protect\theta %
}\left( 0\right) /\left\vert S\right\vert =0$ and $\protect\mu S^{\protect%
\phi }\left( 0\right) /\left\vert S\right\vert =0$ in
Boyer-Lindquist coordinates). The four momentum $p_{\left( TD\right)
}^{a}/M$ is chosen for
the TD SSC as $p_{\left( TD\right) }^{r}(0)/M\left( 0\right) =0$, $\protect%
\mu p_{\left( TD\right) }^{\protect\theta }\left( 0\right) /M\left(
0\right) =0.0442$ and $\protect\mu p_{\left( TD\right)
}^{\protect\phi }(0)/M\left( 0\right) =-0.0316$. The initial
centroid four velocity $u_{\left( TD\right) }^{a}$ is determined
from Equation (\protect\ref{u}).} \label{quasicircularFMP}
\end{figure}
\end{widetext}
different $a^{a}$. In
principle all worldlines where the conditions $u_{a}u^{a}=-1$, $u_{a}S^{a}=0$
and $p_{a}S^{a}=0$ are satisfied can be used for representation of the
moving body. Since the tangent vector of the centroid orbit occurs in the
spin precessional equation through $E_{\mathbf{A}}$ and their derivatives,
the spin axis may describe very complicated motion in such observer's frame,
which follows a helical trajectory. In order to characterize the self
rotation of the body in the easiest way possible, the helical trajectories
should be avoided. However, there is no generic rule for determination of
the non-helical trajectory. According to the Authors' knowledge, the best
ansatz is suggested in Ref. \cite{Costa2015} as taking%
\begin{equation}
p^{a}=mu^{a}+S^{ab}\frac{F_{b}}{m}.  \label{nonhel}
\end{equation}%
In this case $a^{a}\propto F^{a}/m$ at leading order in spin, which is
plausible for a non-helical trajectory since $a^{a}\propto \mathcal{O}\left(
S^{-1}\right) $ for the helical ones. However, the ansatz (\ref{nonhel})
cannot be imposed as a constraint for the dynamics with significant spin
magnitude in the consideration. We require the ansatz (\ref{nonhel}) for
setting initial conditions in the numerical investigations. This is not
forbidden because (\ref{nonhel}) is consistent with the algebraic
velocity-momentum equation. The corresponding initial data set in the FMP
SSC are chosen by identifying the initial centroid four velocity and spin
vector as $u_{\left( FMP\right) }^{a}=u_{\left( TD\right) }^{a}$ and $%
s^{a}/m=S^{a}/M$. Then the initial spin tensor and $p_{\left( FMP\right)
}^{a}/m$ are computed from $S^{ab}=\eta _{~\ cd}^{ab}u^{c}s^{d}$ and (\ref%
{nonhel}), respectively. Bringing forward the result of the SSC
dependence, we have found that the evolutions of the spin vectors
defined in the TD and the FMP SSCs are barely distinguishable from
each other in all cases. The differences in the evolutions of the
different considered quantities considered in the subsequent
subsections remains below 1\%. This is in agreement with result of
Ref. \cite{TLA}, where the evolution of test bodies moving on
circular equatorial orbits around a Schwarzschild black hole were
investigated.

\subsection{Spinning bodies moving in the Kerr spacetime\label{Kerr}}

In this subsection, we set $\mu _{em}=0$ and $a/\mu <1$, i.e. the background
is a Kerr black hole's spacetime. Figure \ref{quasicircularFMP} shows
spherical-like orbits. The initial values are listed in the caption. The
orbits, the black curves in the upper row, are shown in the coordinate space
($x/\mu $,$y/\mu $,$z/\mu $) defined in Equation (\ref{Cartesian}). The
initial and the final positions of the body are marked by green and red
dots, respectively. The initial position is in the equatorial plane $\theta
\left( 0\right) =\pi /2$ at $r\left( 0\right) =8\mu $ and $\phi \left(
0\right) =0$. The blue surface at the center depicts the outer bound of the
Kerr black hole's ergosphere. In the columns from left to right, the spin
magnitude $\left\vert S\right\vert /\mu M$ variates as $0.01$, $0.1$ and $%
0.9 $, respectively, while the other initial values are fixed. For small
spin, the orbit is spherical ($\dot{r}=0$) and reproduces Figure 3 of \cite%
{Bini2017}. For higher spins (second and third columns) the orbit becomes
less and less spherical, but because of $\dot{r}\ll 1$, it is
spherical-like. On the purplish spheres in the second row, the evolutions of
the kinematical quantity defined in Equation (\ref{ldef}) are shown under
the corresponding orbits. Their initial and final directions are marked by
purple and black arrows, respectively. The evolution of this vector clearly
shows that the increasing spin magnitude due to the nonvanishing
spin-curvature coupling (i.e. the non-vanishing right hand side of Equation (%
\ref{MPD1})) in the spin precession, which was not included in the
investigation of Ref. \cite{Bini2017}) affects the orbit. On the greenish
spheres in the third row, the evolutions of the spin direction are
represented in the boosted SO frame $E_{\mathbf{i}}\left( e,u\right) $. The
initial and final spin directions are marked by green and blue arrows,
respectively. In Boyer-Lindquist coordinates, the initial spin four vector $%
S^{a}$ has only non-vanishing component $S^{r}$. The fourth and fifth rows
image the evolutions of spin precessional angular velocity $\Omega _{\left(
prec\right) }^{\mathbf{\alpha }}\left( e,u\right) $ on shorter and longer
timescales, respectively. For $\left\vert S\right\vert /\mu M=0.01$, the
frame components of this angular velocity oscillates (see also Figure 3 of
Ref. \cite{Bini2017} and remembering for that the definition of $\Omega
_{\left( prec\right) }^{\mathbf{\beta }}$ carries an extra sign). For $%
\left\vert S\right\vert /\mu M=0.1$ and $0.9$, an amplitude modulation
occurs. This is also a clear sign of the spin-curvature effect. We mention
that, the evolution of $\Omega _{\left( prec\right) }^{\mathbf{\alpha }%
}\left( f,u\right) $ differs less than 1\% from that of $\Omega _{\left(
prec\right) }^{\mathbf{\alpha }}\left( e,u\right) $. This is because the
boosted SO and ZAMO frames are almost the same, i.e. the rotation angle $%
\Theta $ between them is small as shown in the last row. We also mention
that, all precessional angular velocities $\Omega _{\left( prec\right) }^{%
\mathbf{\alpha }}\left( e,p/M\right) $, $\Omega _{\left( prec\right) }^{%
\mathbf{\alpha }}\left( e,u\right) $, $\Omega _{\left( prec\right) }^{%
\mathbf{\alpha }}\left( f,p/M\right) $ and $\Omega _{\left( prec\right) }^{%
\mathbf{\alpha }}\left( f,u\right) $ in the frames $E_{\mathbf{\alpha }%
}\left( e,p/M\right) $, $E_{\mathbf{\alpha }}\left( e,u\right) $, $E_{%
\mathbf{\alpha }}\left( f,p/M\right) $ and $E_{\mathbf{\alpha }}\left(
f,u\right) $, respectively, describe the same evolutions within 1\%. The
Boyer-Lindquist components of the spin vector are frame independent
quantities. Their evolutions are presented on Figure \ref%
{quasicircularTDspincomp}. The blue, the green and the red curves belong to
the different spin magnitude cases $\left\vert S\right\vert /\mu M=0.01$, $%
\left\vert S\right\vert /\mu M=0.1$ and $\left\vert S\right\vert /\mu M=0.9$%
, respectively. An amplitude modulation due to the spin-curvature coupling
occurs in the oscillation around a harmonic evolution of the $\theta$%
-component, which can be mostly seen along the red curve.

In the following, we will consider zoom-whirl and unbound orbits passing
over the ergosphere. In all cases we choose such initial conditions that the
body moves in the equatorial plane for negligible spin magnitude. Hence the
deviation of the trajectory from this plane is a clear sign of the
spin-curvature effect. Zoom-whirl orbits of a nonspinning test body around a
spinning black hole were already investigated in Refs. \cite%
{Glampedakis,Glampedakis2002,Glampedakis2005,Levin1,Levin2}. Those orbits
did not passed through the ergosphere for which we will focus. In addition,
when the test body is spinning, some effects from the spin-curvature
coupling are waited which we will consider. In addition, zoom-whirl orbits
of comparable mass black holes, when only one of them is spinning, were
analyzed in Ref. \cite{Levin3} within the framework of PN approximation.
Hyperbolic orbits of spinning 
\begin{widetext}
\newpage
\begin{figure}[H]
\begin{center}
\includegraphics[width=4cm]{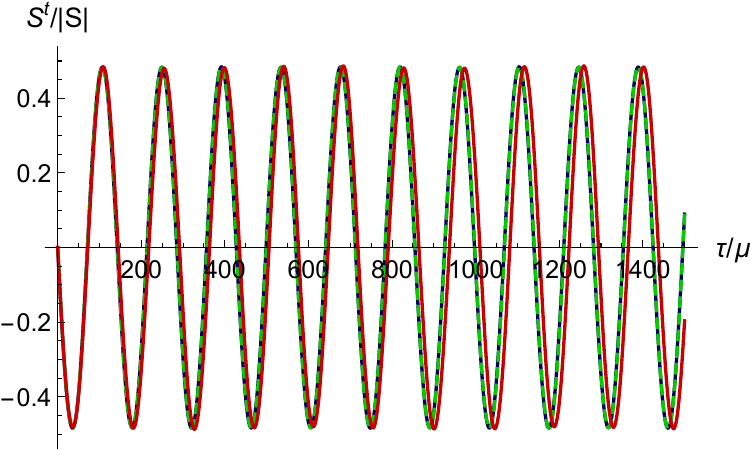} \hfill % %
\includegraphics[width=4cm]{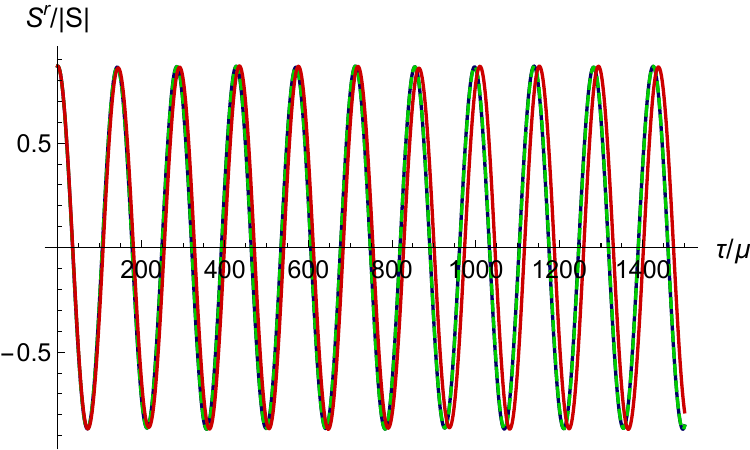} \hfill %
\includegraphics[width=4cm]{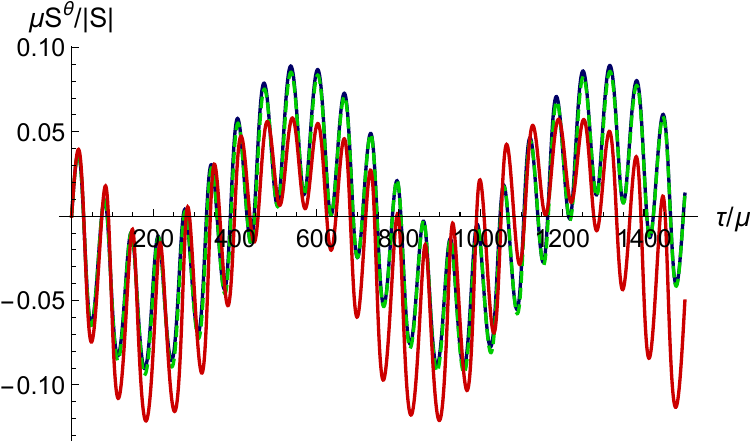} \hfill %
\includegraphics[width=4cm]{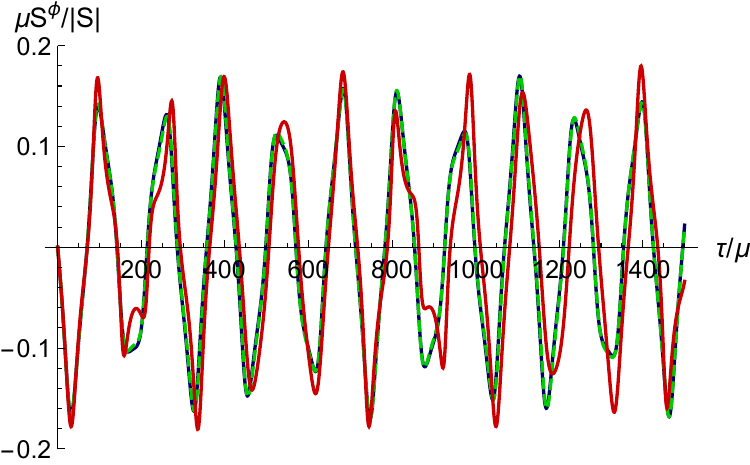}
\end{center}
\par
\vskip -0.5cm \caption{(color online). The evolutions of the
Boyer-Lindquist coordinate components of the unit spin vector are
shown. The blue and the green curves belonging to the spin magnitude
$\left\vert S\right\vert/ \protect\mu M=0.01$ and $\left\vert
S\right\vert/\protect\mu M=0.1$, respectively, almost cover each
other. The red curve represents the high spin magnitude case
$\left\vert S\right\vert /\protect\mu M=0.9$. An amplitude
modulation occurs in the oscillation around a harmonic evolution of
the $\protect\theta$-component which can be mostly seen along the
red curve.} \label{quasicircularTDspincomp}
\end{figure}
\begin{figure}[H]
\begin{center}
\includegraphics[width=4.9cm]{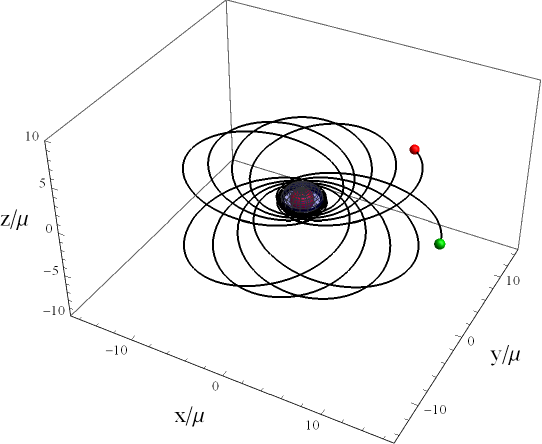}\hskip 1.5cm %
\includegraphics[width=4.9cm]{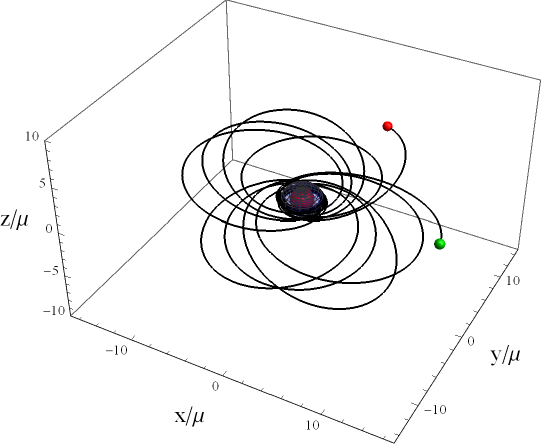} \\[0pt]
\includegraphics[width=4.9cm]{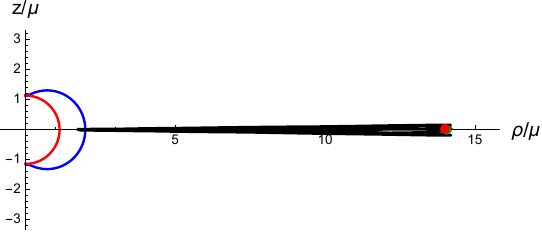}\hskip 1.5cm %
\includegraphics[width=4.9cm]{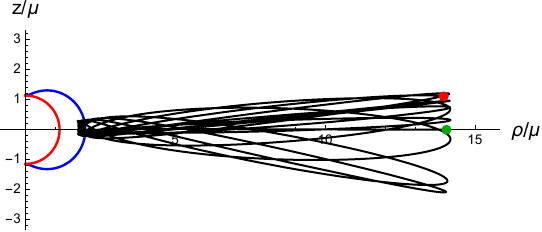} \\[0pt]
\includegraphics[width=4.9cm]{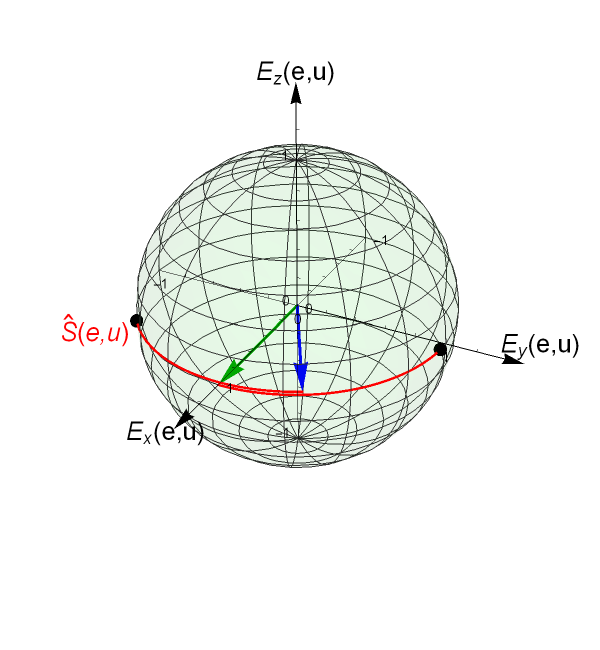}\hskip 1.5cm %
\includegraphics[width=4.9cm]{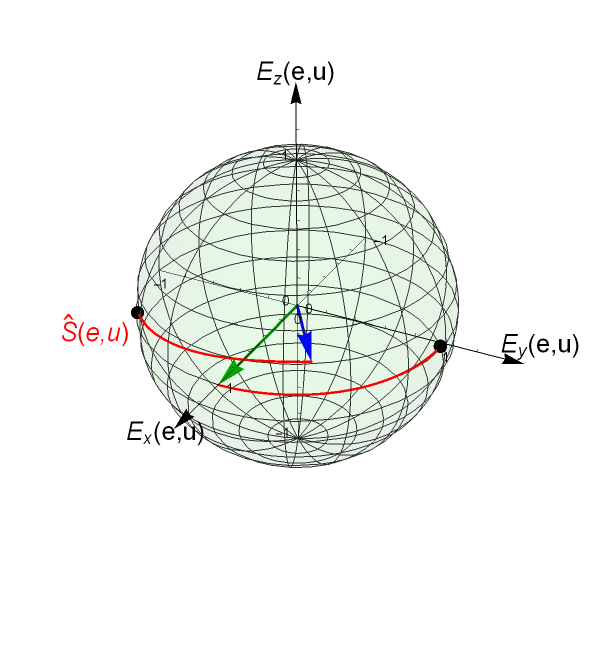}\\[0pt]
\vskip -1cm \includegraphics[width=4.9cm]{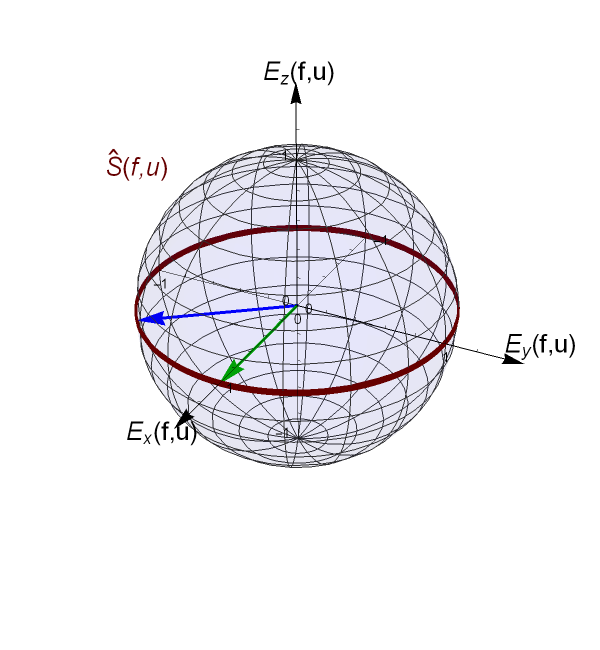}\hskip 1cm %
\includegraphics[width=4.9cm]{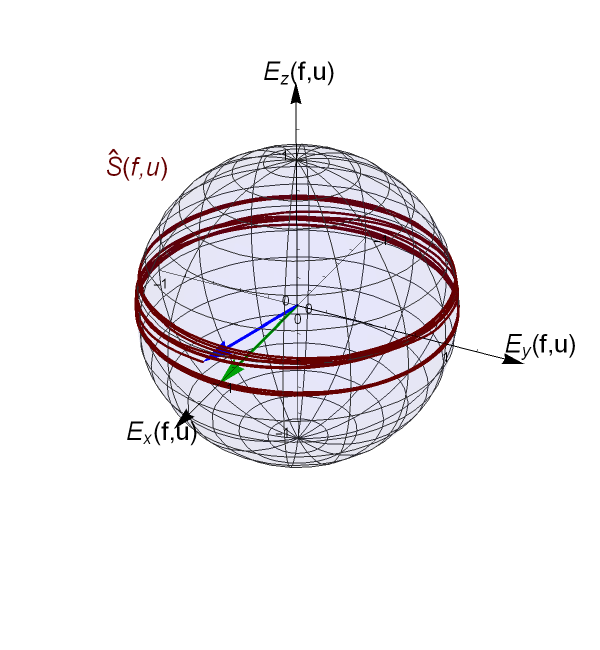}
\end{center}
\par
\vskip -1cm \caption{(color online). The evolution of spinning body
moving on zoom-whirl orbits around the Kerr black hole with
$a=0.99\protect\mu $. The magnitudes of the body's spin are
$\left\vert S\right\vert /\protect\mu M=0.01$ (left panel) and $0.1$
(right panel). The rows represent the following: 1. the
orbit in coordinate space ($x/\protect\mu $,$y/\protect\mu $,$z/\protect\mu $%
) (outer and inner bounds of the ergosphere of the central black
hole is marked by blue and red surfaces, respectively, and initial
and final positions of the spinning body are denoted by green and
red dots,
respectively), 2. the orbit in the coordinate space $\protect\rho /\protect%
\mu =r\sin \protect\theta /\protect\mu $ and $z/\protect\mu =r\cos \protect%
\theta /\protect\mu $\textbf{\ }with marked initial and final
positions and bounds of the ergosphere, 3. the unit spin vector in
the boosted SO Cartesian-like comoving frame $E_{\mathbf{i}}\left(
e,u\right) $ on a shorter timescale including the first whirling
period, and 4. the unit spin
vector in the boosted ZAMO Cartesian-like comoving frame $E_{\mathbf{i}%
}\left( f,u\right) $ on the total timescale (initial and final spin
directions are marked by green and blue arrows, respectively). The
initial
data set: $t\left( 0\right) =0$, $r(0)=14.05\protect\mu $, $\protect\theta %
(0)=\protect\pi /2$, $\protect\phi (0)=0$, $p^{r}(0)/M=-0.03$, $\protect\mu %
p^{\protect\theta }(0)/M=0$, $\protect\mu p^{\protect\phi }\left(
0\right)
/M=0.012$, $\protect\theta ^{\left( S\right) }\left( 0\right) =\protect\pi %
/2 $ and $\protect\phi ^{\left( S\right) }\left( 0\right) =0$ (the
spatial Boyer-Lindquist coordinate components are $S^{r}\left(
0\right) /\left\vert S\right\vert =0.9293$, $\protect\mu
S^{\protect\theta }\left( 0\right) /\left\vert S\right\vert =0$ and
$\protect\mu S^{\protect\phi }\left( 0\right) /\left\vert
S\right\vert =-0.0002$).} \label{ZW1TD}
\end{figure}
\begin{figure}[H]
\begin{center}
\includegraphics[width=6cm]{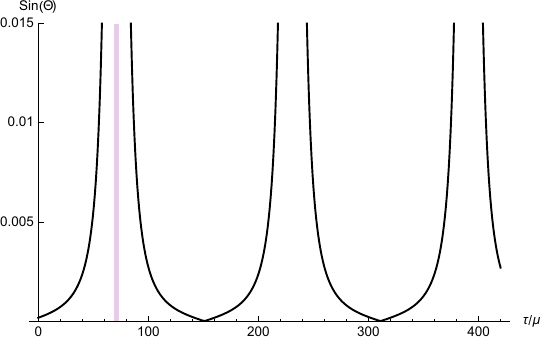}\hskip 1.5cm %
\includegraphics[width=6cm]{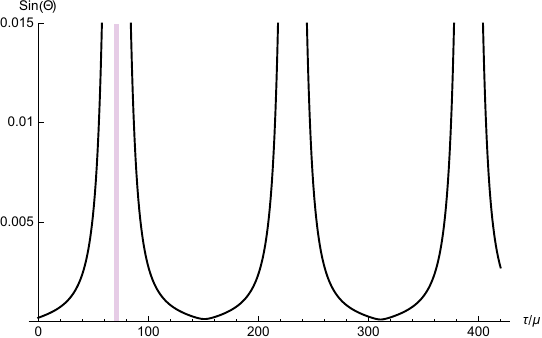} \\[0pt]
\includegraphics[width=6cm]{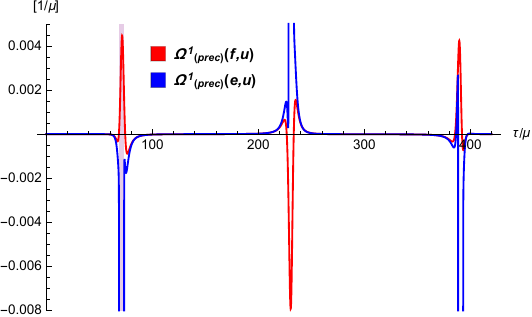}\hskip 1.5cm %
\includegraphics[width=6cm]{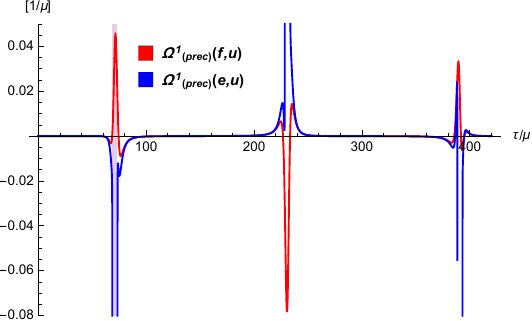} \\[0pt]
\includegraphics[width=6cm]{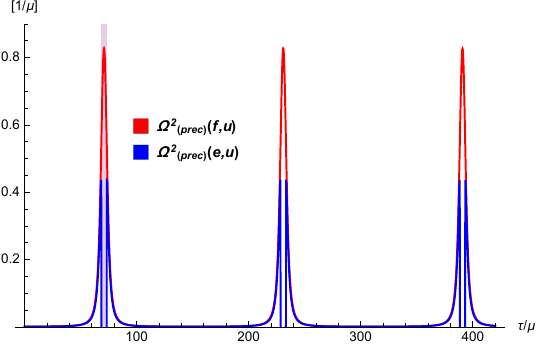}\hskip 1.5cm %
\includegraphics[width=6cm]{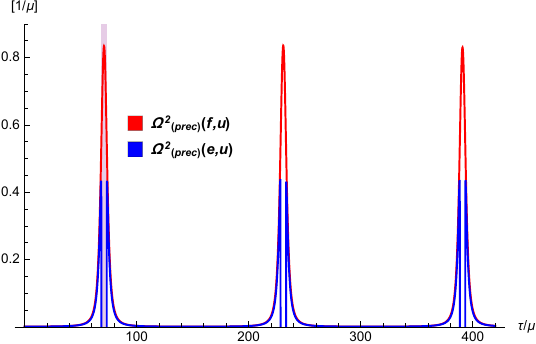} \\[0pt]
\includegraphics[width=6cm]{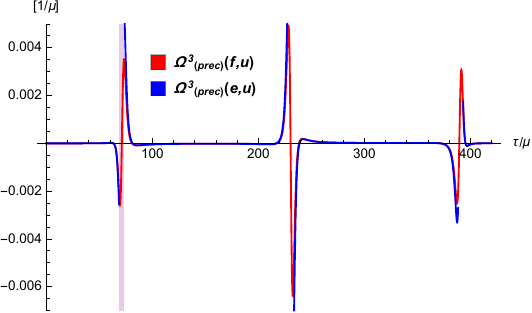}\hskip 1.5cm %
\includegraphics[width=6cm]{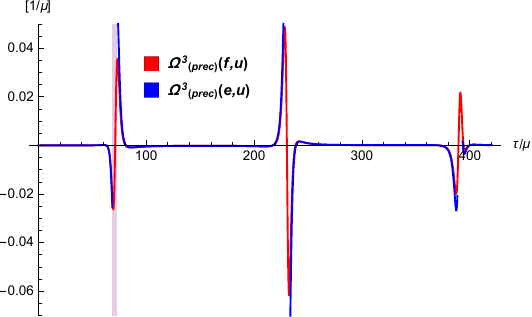}
\end{center}
\caption{(color online). The left and right columns belong to the
same evolution which are shown on Figure \protect\ref{ZW1TD}. The
first row shows $\sin \Theta $. The next three rows present the
evolutions of the spherical triad components of the spin
precessional angular velocities $\Omega
_{\left( prec\right) }^{\mathbf{\protect\alpha }}\left( e,u\right) $ and $%
\Omega _{\left( prec\right) }^{\mathbf{\protect\alpha }}\left(
f,u\right) $.} \label{ZW1TDb}
\end{figure}
\end{widetext}
test bodies based on the MPD equations were
analytically studied in Ref. \cite{BiniHyp}. The perturbations caused by the
spin-curvature coupling in the equatorial orbits were considered in the case
when the spin is parallel to the 
central black hole rotation axis. In this
configuration the spin vector is conserved. Here, in order to discuss
nontrivial spin evolution and spin-curvature coupling effects, we choose the
initial spin direction to be perpendicular to the central black hole
rotation axis.

The first row of Figure \ref{ZW1TD} shows the orbits in the ($x$,$y$,$z$%
)-space for increasing spin magnitude $\left\vert S\right\vert /\mu M=0.01$
(left panel) and $0.1$ (right panel). The other initial values listed in the
caption are the same. The blue and red surfaces at the center depict the
outer and interior bounds of the ergosphere, respectively (i.e. the outer
stationary limit surface and the outer event horizon). The initial and final
positions of the body are marked by green and red dots, respectively. The
initial position is in the equatorial plane $\theta \left( 0\right) =\pi /2$
at $r\left( 0\right) =14.05\mu $ and $\phi \left( 0\right) =0$, and both the
initial four momentum and centroid four velocity have vanishing $\theta $%
-component. With this initial location and four velocity a non-spinning
particle moves in the equatorial plane. However, since the spin direction is
not parallel with the rotation axis of the central black hole, the body's
centroid leaves the equatorial plane due to the effect of spin-curvature
coupling. This is highlighted in the second row representing the orbits in
coordinates $\rho /\mu =r\sin \theta /\mu $ and $z/\mu =r\cos \theta /\mu $.
The bounds of the ergosphere are drawn by blue and red curves. The body is
inside the ergosphere when it whirls around the 
\begin{widetext}
\newpage
\begin{figure}[H]
\begin{center}
\includegraphics[width=4cm]{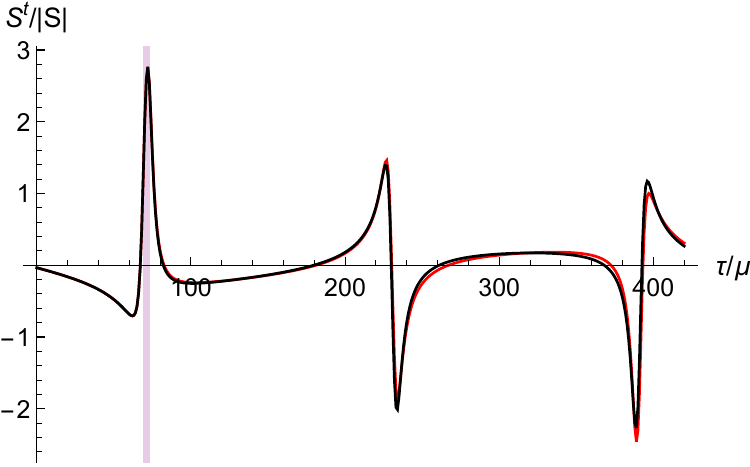}\hfill %
\includegraphics[width=4cm]{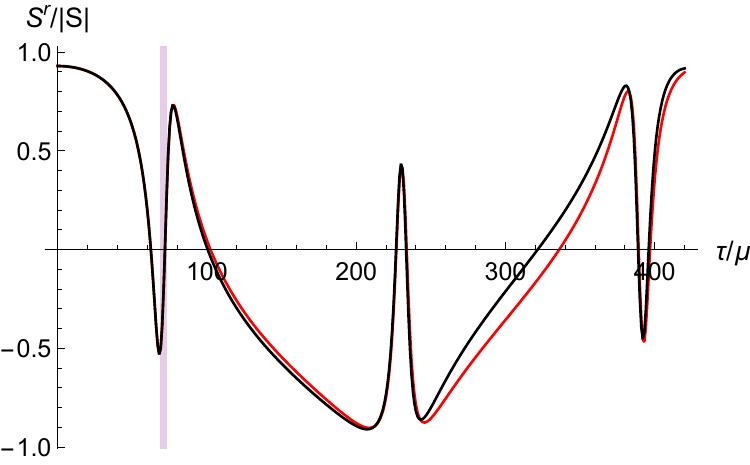}\hfill %
\includegraphics[width=4cm]{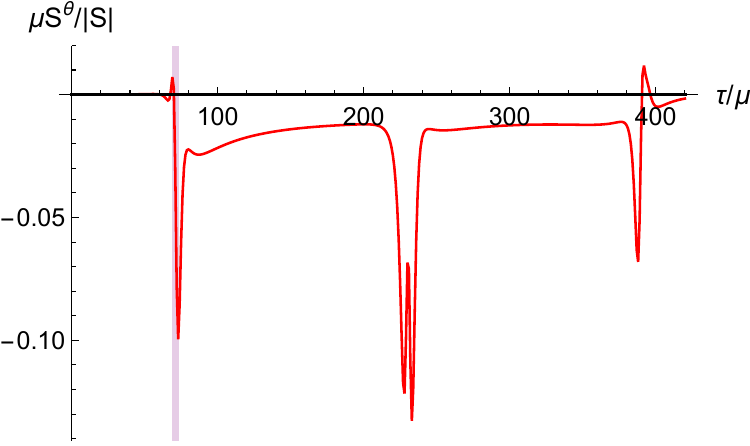}\hfill %
\includegraphics[width=4cm]{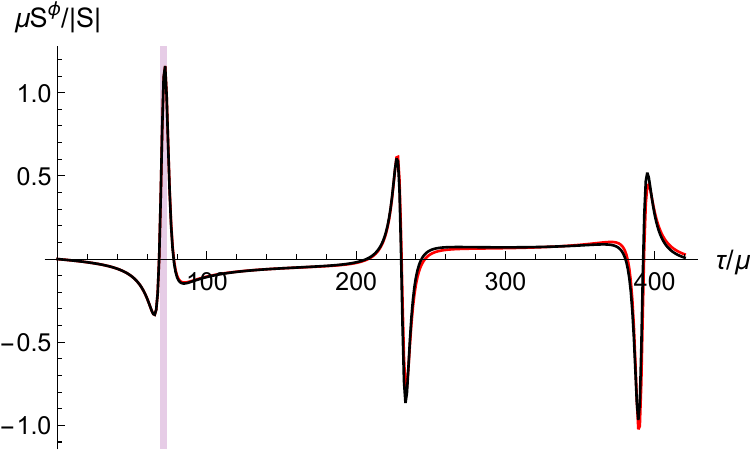}\\[0pt]
\includegraphics[width=4cm]{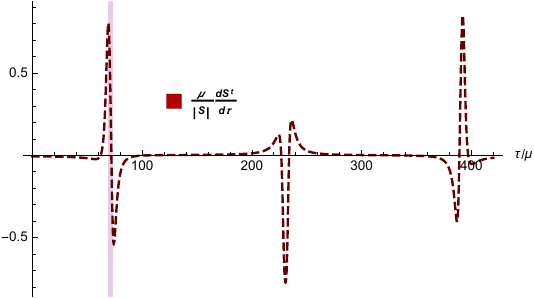}\hfill %
\includegraphics[width=4cm]{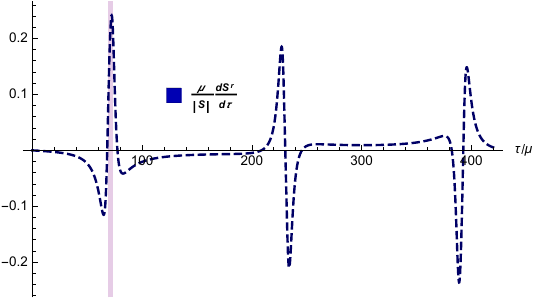}\hfill %
\includegraphics[width=4cm]{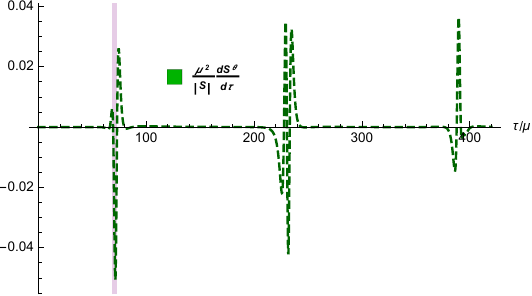}\hfill %
\includegraphics[width=4cm]{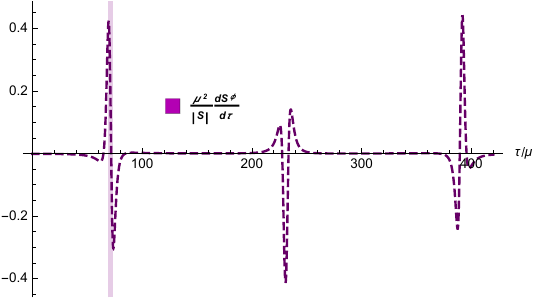}\\[0pt]
\includegraphics[width=4cm]{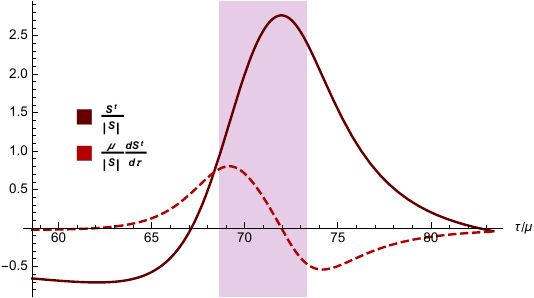}\hfill %
\includegraphics[width=4cm]{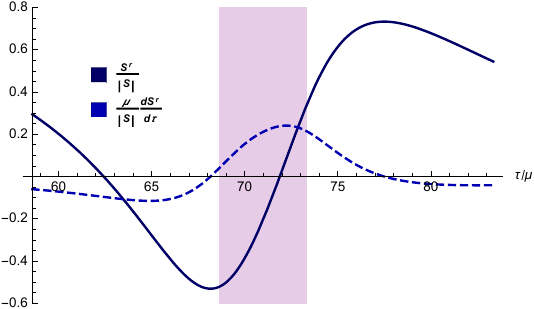}\hfill %
\includegraphics[width=4cm]{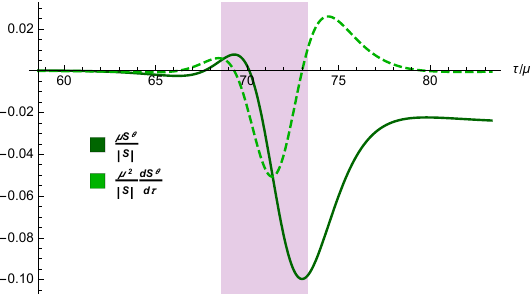}\hfill %
\includegraphics[width=4cm]{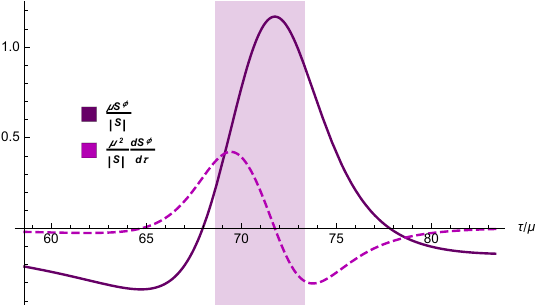}
\end{center}
\caption{(color online). The evolution of the Boyer-Lindquist
coordinate components of the unit spin vector and their derivatives
rescaled to dimensionless variables is presented for that case which
is shown on the right hand sides of Figures \protect\ref{ZW1TD} and
\protect\ref{ZW1TDb}. The first and the second rows show the
evolution on a timescale which includes the first three whirling
period when the body is inside the ergosphere. The third row zooms
in on that evolution period where the body is first in the
ergosphere which is indicated by the purplish shadow on all panels.
In the first row, the black and the red curves represent the
evolutions without and with spin-curvature coupling, respectively.}
\label{ZW1TDeff}
\end{figure}
\end{widetext}
central Kerr black hole.
This happens during all whirling period. The unit spin vector evolutions in
the boosted SO Cartesian-like comoving frame ($E_{\mathbf{i}}\left(
e,u\right) $) during a timescale including the first 
whirling period are
shown in the third row. The initial and final directions are marked by green
and blue arrows, respectively. The rotation of the projection of\ spin
vector in the plane ($E_{\mathbf{x}}\left( e,u\right) $, $E_{\mathbf{y}%
}\left( e,u\right) $) is counterclockwise in both cases. In the boosted SO
frame the evolution is not continuous due to the motion through the
ergosphere. The black dots denote the spin directions when the body first
enters and leaves the ergosphere. The magnitude of the jump shows that the
spin direction changed significantly inside the ergosphere. The fourth row
shows the unit spin vector evolution on the total timescale in the boosted
ZAMO Cartesian-like comoving frame ($E_{\mathbf{i}}\left( f,u\right) $).
This frame can be used for the spin representation inside the ergosphere,
hence the evolution is continuous. For higher spin, when the spin-curvature
coupling is stronger the deflection of the spin direction moves more out of
the equatorial plane of ZAMO frame. The first row in Figure \ref{ZW1TDb}
shows the rotation angle $\Theta $ between the boosted SO and ZAMO frames.
Here and in the following pictures the purplish shadow indicates the time
interval where the body moves inside the ergosphere during the first
whirling period. The next three rows in Figure \ref{ZW1TDb} depict the
evolutions of $\Omega _{\left( prec\right) }^{\mathbf{\alpha }}\left(
e,u\right) $ and $\Omega _{\left( prec\right) }^{\mathbf{\alpha }}\left(
f,u\right) $. Each row shows one component of these angular velocities. The
red and blue curves represent the precessional angular velocities in the
boosted ZAMO and SO frames, respectively. The blue curves diverge at the
ergosphere where the description in the boosted SO frame fails. The
magnitude of the precessional angular velocities rapidly increases near and
inside the ergosphere and becomes higher for higher spin magnitude. Finally,
we note that the precessional velocities $\Omega _{\left( prec\right) }^{%
\mathbf{\alpha }}\left( e,p/M\right) $ ($\Omega _{\left( prec\right) }^{%
\mathbf{\alpha }}\left( f,p/M\right) $) and $\Omega _{\left( prec\right) }^{%
\mathbf{\alpha }}\left( e,u\right) $ ($\Omega _{\left( prec\right) }^{%
\mathbf{\alpha }}\left( f,u\right) $) describe the same evolutions within
1\%.

From the consideration of the moving body near and inside the ergosphere, we
have found that the spin precession was highly increased. Since the
presented investigation was based on the introduction of ZAMO, and the
precessional angular velocity $\Omega _{\left( prec\right) }^{\mathbf{\alpha
}}\left( f,u\right) $ described the spin evolution with respect to the
boosted ZAMO frame, the highly increased precession effect could be an
observer dependent statement. However, this effect is supported in another
way. The static observers play fundamental role in comparing the variation
of spin direction with respect to the distant stars. The third row of Figure %
\ref{ZW1TD} shows in the boosted SO frame that the jump of the spin
direction (between the black dots) happening during that period when the
body is staying first in the ergosphere. This jump happens during relatively
short period indicated by the purplish shadow on Figure \ref{ZW1TDb}. More
exactly, the evolution period presented in the third row of Figure \ref%
{ZW1TD} is given by $\tau =\left[ 0,94.7\mu \right] $ from which the body is
inside the ergosphere in the interval $\tau =\left[ 68.6\mu ,73.4\mu \right]
$. The evolution of the spin together with these timescales result in the
same conclusion that the precession angular velocity is highly increased in
the ergosphere. In addition, this effect can also be supported without using
any particular reference frame. For the case presented on\ the right hand
side of Figure \ref{ZW1TDb}, we show the evolution of the Boyer-Lindquist
coordinate components of the unit 
\begin{figure}[H]
\begin{center}
\includegraphics[width=3.8cm]{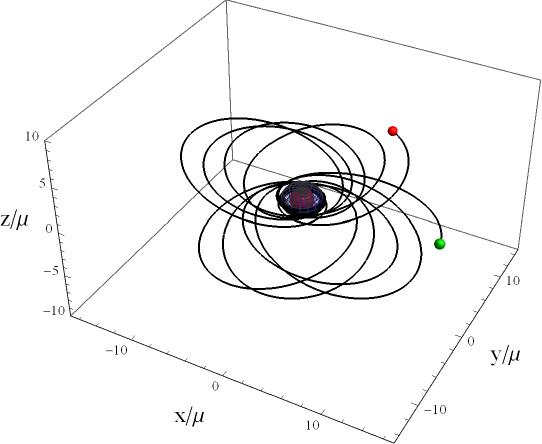} \hskip 0.5cm %
\includegraphics[width=3.8cm]{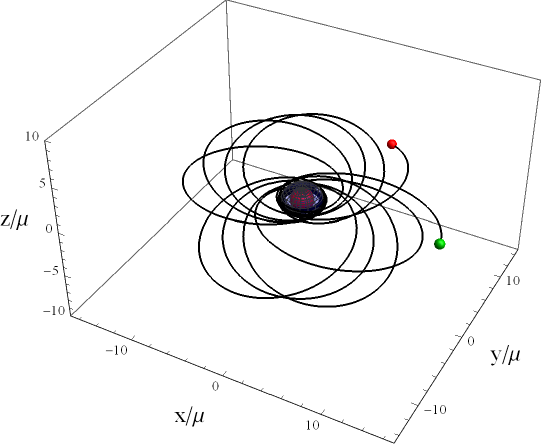} \\[0pt]
\includegraphics[width=3.8cm]{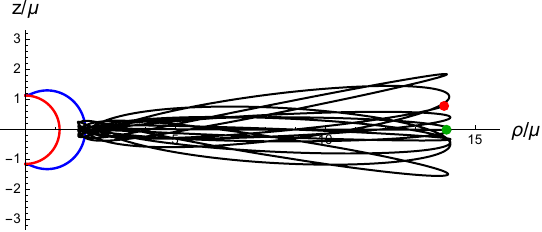} \hskip 0.5cm %
\includegraphics[width=3.8cm]{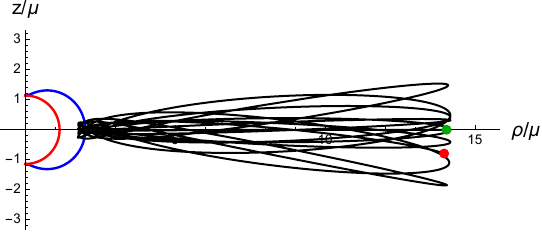} \\[0pt]
\includegraphics[width=3.8cm]{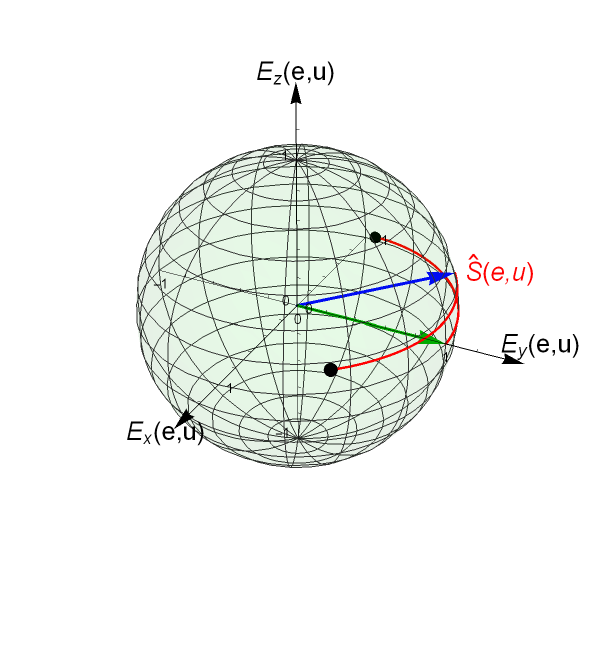} \hskip 0.5cm %
\includegraphics[width=3.8cm]{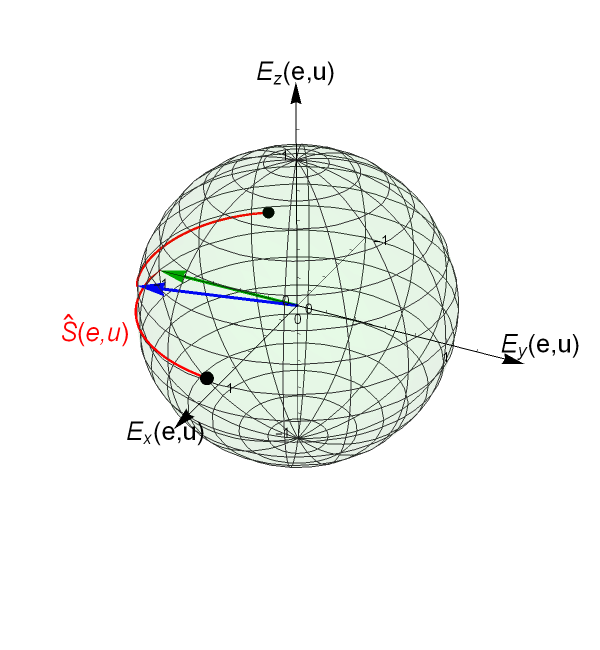} \\[0pt]
\vskip -1cm \includegraphics[width=3.8cm]{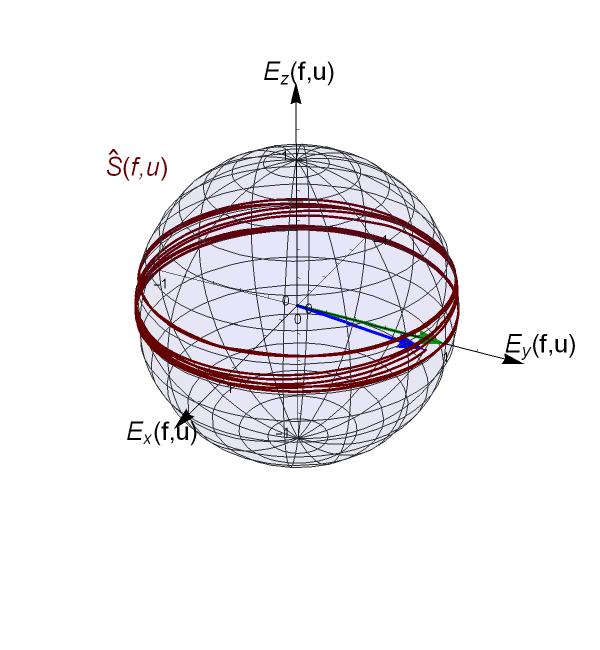} \hskip 0.5cm %
\includegraphics[width=3.8cm]{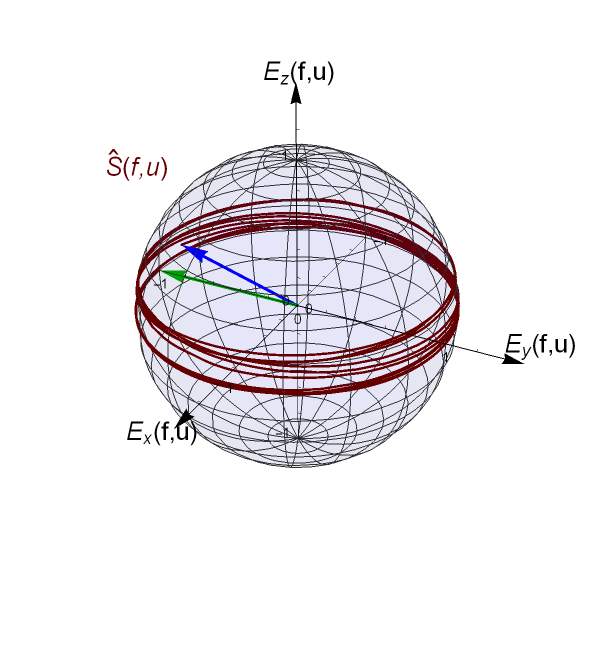} \\[0pt]
\includegraphics[width=3.8cm]{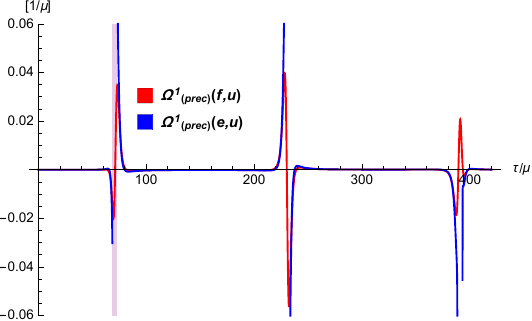} \hskip 0.5cm %
\includegraphics[width=3.8cm]{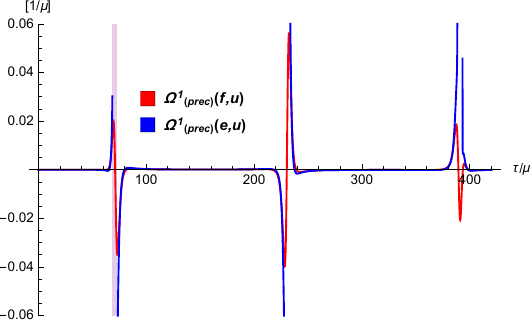} \\[0pt]
\includegraphics[width=3.8cm]{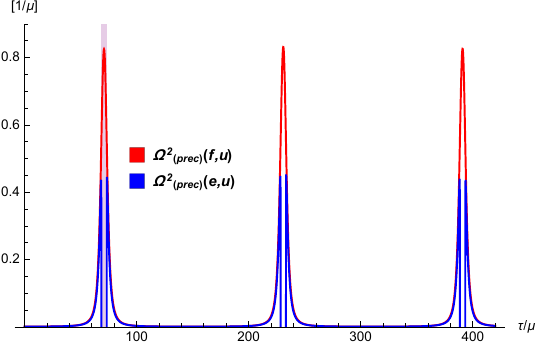} \hskip 0.5cm %
\includegraphics[width=3.8cm]{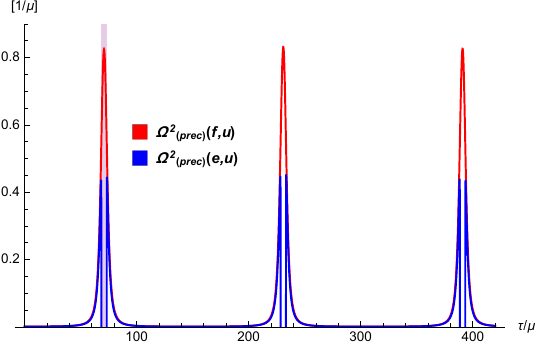} \\[0pt]
\includegraphics[width=3.8cm]{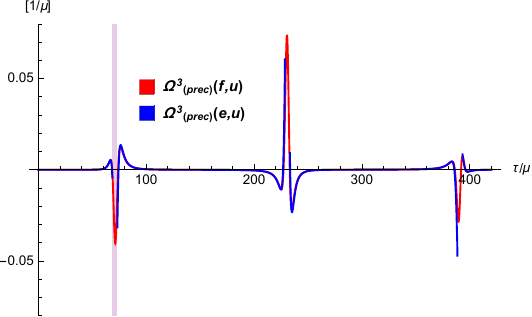} \hskip 0.5cm %
\includegraphics[width=3.8cm]{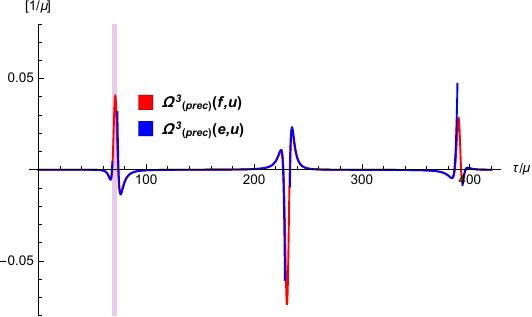}
\end{center}
\par
\vskip -0.5cm \caption{(color online). The same as on the right
column of Figures \protect \ref{ZW1TD} and \protect\ref{ZW1TDb}
(apart from $\sin \Theta $), but the initial direction of the spin
vector is rotated by $\protect\pi /2$ (left column) and by
$-\protect\pi /2$ (right column). (The spatial Boyer-Lindquist
coordinate components of the spin vector are $S^{r}\left( 0\right)
/\left\vert S\right\vert =-0.0025$ (left column), $0.0025$ (right
column), $\protect\mu S^{\protect\theta }\left( 0\right) /\left\vert
S\right\vert =0$ (both left and right columns) and $\protect\mu S^{\protect%
\phi }\left( 0\right) /\left\vert S\right\vert =0.0720$ (left column), $%
-0.0720$ (right column).} \label{ZW3TD}
\end{figure}
spin vector and their derivatives rescaled
to dimensionless variables on Figure \ref{ZW1TDeff}. The first and the
second rows represent the evolution on a timescale which includes the first
three whirling period when the body is inside the ergosphere. 
The third row
zooms in on that evolution period where the body is first inside the
ergosphere. As mentioned, this period is indicated by the purplish shadow.
All panels of Figure \ref{ZW1TDeff} confirm that the rate of change in the
direction of the spin vector is highly increased near and inside the
ergosphere. As a reference, the black curves in the first row represent the
evolution of the unit spin coordinate components when the spin-curvature
coupling is turned off. While the red curves show the evolutions when the
spin-curvature coupling is taken into account. Significant differences in
the evolutions can be seen in the case of the $r$ and $\theta $ coordinate
components. The $\theta $ coordinate component identically vanishes when the
spin-curvature coupling is neglected. Finally, we mention that since $p^{a}$
and $u^{a}$ are not parallel with each other it may happen that $%
u^{a}u_{a}=0 $ \cite{Semerak1999,Toshmatov2019,Toshmatov2020,Benavides2021}
or $p_{a}u^{a}=0$ \cite{DR2,DR3,DR4}. The first case was discussed
previously in Section IIA. The second case is equivalent with becoming the
momentum light-like $p_{a}p^{a}=0$, which can be seen from the contraction
Equation (\ref{u}) with $p_{a}$. We have checked in the Appendix \ref{para}
that the MPD equations are applied only in that domain where such
pathological\ behaviours do not occur.

When the initial direction of the spin vector is opposite with respect to
the case presented in the Figures \ref{ZW1TD} and \ref{ZW1TDb}, while all
other initial conditions are the same, we have found the following. The
centroid trajectory becomes the reflection of the orbit presented on Figure %
\ref{ZW1TD} through the equatorial plane. The instantaneous directions of
the spin vector in the boosted SO (ZAMO) frame can be obtained from the
corresponding picture of Figure \ref{ZW1TD} by a rotation with an angle $\pi
$ about the axis $z$ and $E_{\mathbf{z}}\left( e,u\right) $ ($E_{\mathbf{z}%
}\left( f,u\right) $), respectively. The angle $\Theta $ describes the same
evolution. Finally, the components $\Omega _{\left( prec\right) }^{\mathbf{2}%
}\left( e,u\right) $ and $\Omega _{\left( prec\right) }^{\mathbf{2}}\left(
f,u\right) $ remain unchanged, while $\Omega _{\left( prec\right) }^{\mathbf{%
1}}\left( e,u\right) $, $\Omega _{\left( prec\right) }^{\mathbf{1}}\left(
f,u\right) $, $\Omega _{\left( prec\right) }^{\mathbf{3}}\left( e,u\right) $
and $\Omega _{\left( prec\right) }^{\mathbf{3}}\left( f,u\right) $ get an
extra sign.

On Figure \ref{ZW3TD}, the initial spin direction is rotated by $\pi /2$
(left column) and $-\pi /2$\ (right column) in the plane ($E_{\mathbf{x}%
}\left( e,u\right) $, $E_{\mathbf{y}}\left( e,u\right) $) with respect to
the case presented on Figure \ref{ZW1TD}. These two cases have opposite
initial spin directions leading to the following differences in the orbit
and spin evolutions. The zoom-whirl orbit on the right hand side is the
reflection of the trajectory on the left hand side through the equatorial
plane, which are shown in the first two rows. The spin the evolutions
presented on the left and the right hand sides in the third and fourth rows
are related to each other by a rotation with an angle $\pi $ about the axis
connecting the south and north poles. The evolution of $\Omega _{\left(
prec\right) }^{\mathbf{2}}$ are the same on the left and right hand sides,
while $\Omega _{\left( prec\right) }^{\mathbf{1}}$ and $\Omega _{\left(
prec\right) }^{\mathbf{3}}$ have a sign difference, as it can be seen in the
last three rows.

For the consideration of evolutions of spinning bodies which follow unbound
orbits crossing through the ergosphere, the spin magnitude is chosen as $%
\left\vert S\right\vert /\mu M=0.1$. The initial spin directions on the left
(right) hand side of Figure \ref{HB1TD} are the same as on Figure \ref{ZW1TD}
(on the left hand 
\begin{widetext}
\newpage
\begin{figure}[H]
\begin{center}
\includegraphics[width=4.7cm]{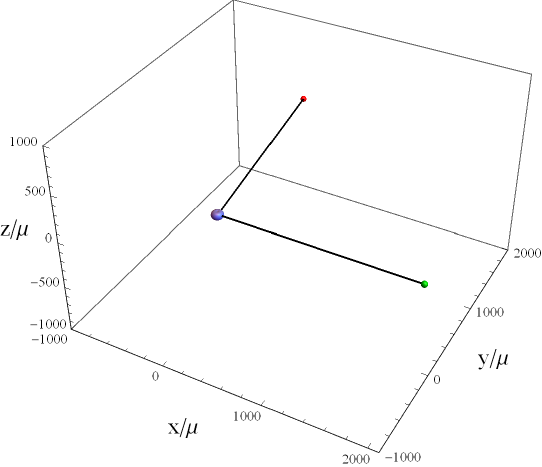} \hskip 1.5cm %
\includegraphics[width=4.7cm]{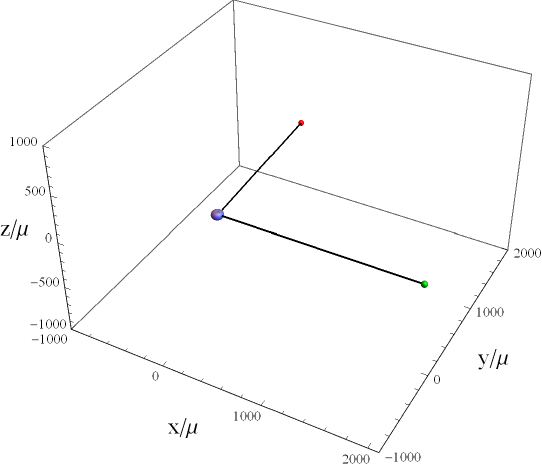} \\[0pt]
\includegraphics[width=4.7cm]{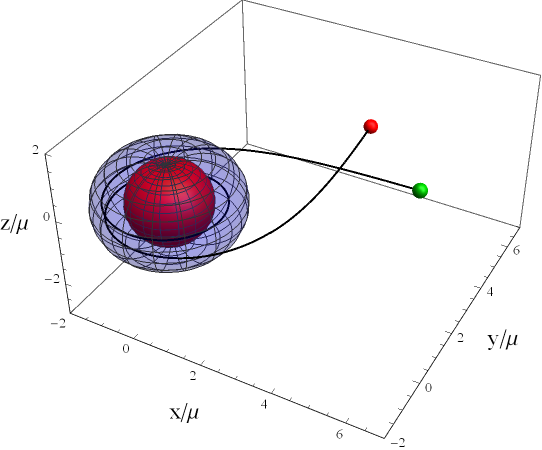} \hskip 1.5cm %
\includegraphics[width=4.7cm]{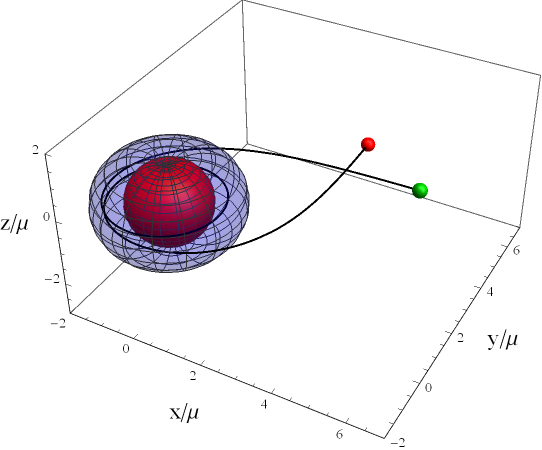} \\[0pt]
\includegraphics[width=4.7cm]{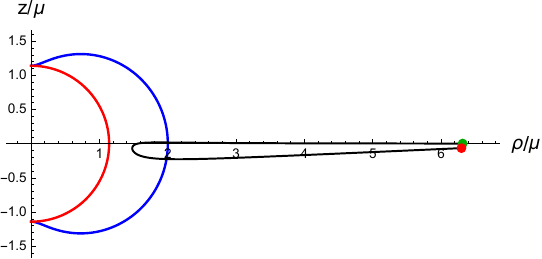} \hskip 1.5cm %
\includegraphics[width=4.7cm]{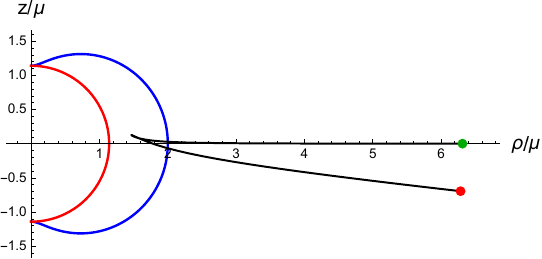} \\[0pt]
\includegraphics[width=4.7cm]{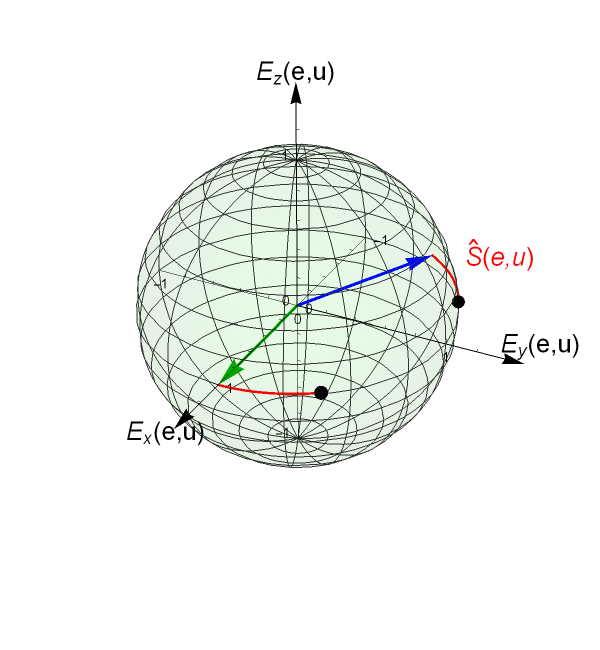} \hskip 1cm %
\includegraphics[width=4.7cm]{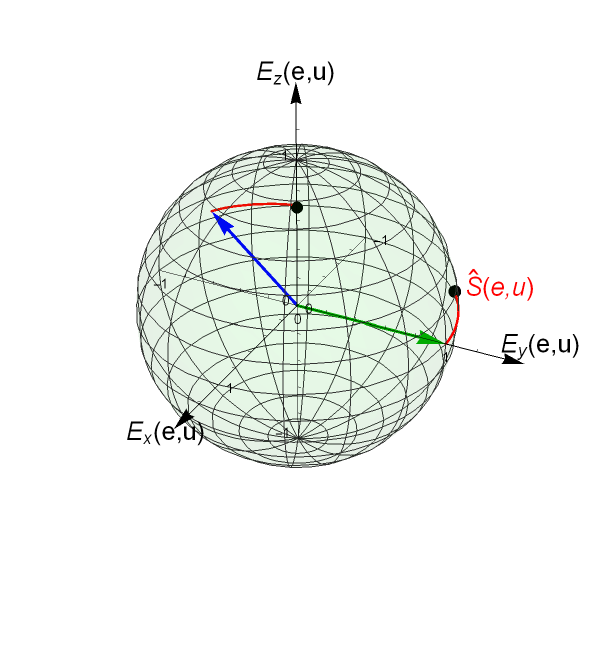} \\[0pt]
\vskip -1cm \includegraphics[width=4.7cm]{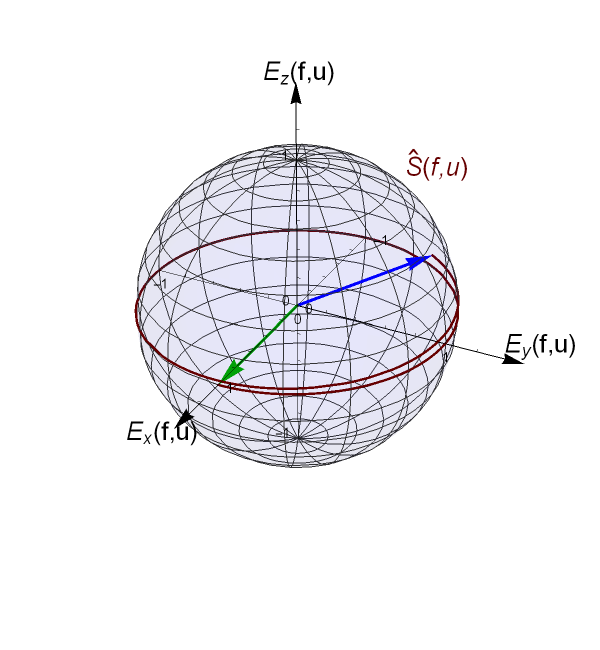} \hskip 1cm %
\includegraphics[width=4.7cm]{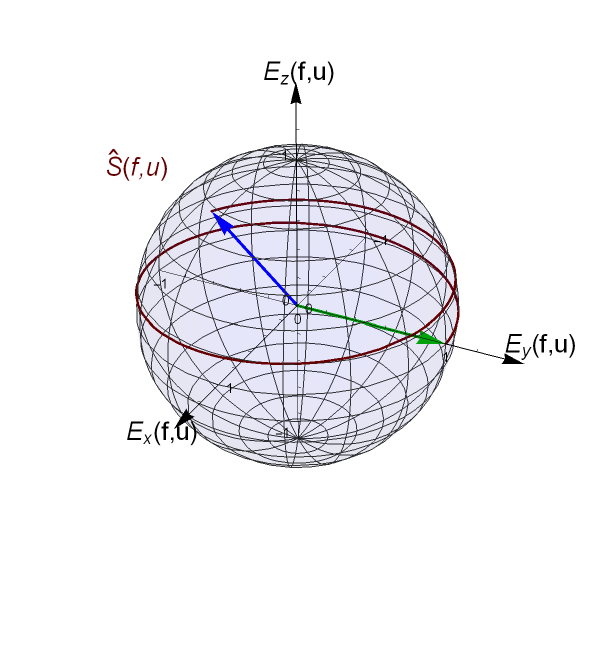}
\end{center}
\par
\vskip -1cm \caption{(color online). The evolutions of spinning body
moving on unbound orbits around Kerr black hole with
$a=0.99\protect\mu $. The spin magnitude chosen as $\left\vert
S\right\vert /\protect\mu M=0.1$. The considered unbound orbits are
shown in the first row. The near black hole parts of
these orbits are represented in the second and third rows\ in ($x/\protect%
\mu $,$y/\protect\mu $,$z/\protect\mu $) and ($\protect\rho /\protect\mu $,$%
z/\protect\mu $) coordinates, respectively. The fourth and fifth
rows present the evolutions of the spin vector in the boosted SO and
ZAMO frames,
respectively. The initial spin direction is determined by $\protect\phi %
^{\left( S\right) }\left( 0\right) =0$ (left col.), $\protect\pi /2$
(right col.) and $\protect\theta ^{\left( S\right) }\left( 0\right)
=\protect\pi /2$
(both cols.). The spatial Boyer-Lindquist coordinate components [$%
S^{r}\left( 0\right) $,$\protect\mu S^{\protect\theta }\left( 0\right) $,$%
\protect\mu S^{\protect\phi }\left( 0\right) $]$/\left\vert
S\right\vert $
of the spin vector are [$-0.0134$,$0$,$-3.1\times 10^{-9}$] and [$-0.000006$,%
$0$,$0.000005$] in the left and right columns, respectively.
Additional
initial data set is $t(0)=0$, $r(0)=2000\protect\mu $, $\protect\theta (0)=%
\protect\pi /2$, $\protect\phi (0)=0$, $p^{r}(0)/M=-0.9$, $\protect\mu p^{%
\protect\phi }\left( 0\right) /M=8\times 10^{-7}$and $\protect\mu p^{\protect%
\theta }(0)/M=0$. The final locations [$t\left( \protect\tau ^{\ast
}\right)
/\protect\mu $,$r(\protect\tau ^{\ast })/\protect\mu $,$\protect\theta (%
\protect\tau ^{\ast })$,$\protect\phi (\protect\tau ^{\ast })$] at $\protect%
\tau ^{\ast }=4433\protect\mu $ are
[$6033.2$,$1999.7$,$1.527$,$14.24$] (left col.) and
[$6033.3$,$1999.7$,$1.671$,$14.25$] (right col.). The final
values of the spatial Boyer-Lindquist coordinate components [$p^{r}(\protect%
\tau ^{\ast })$,$\protect\mu p^{\protect\theta }\left( \protect\tau
^{\ast }\right) $,$\protect\mu p^{\protect\phi }(\protect\tau ^{\ast
})$]$/M$ of
the four momentum are [$0.900002$,$-7.65\times 10^{-8}$,$8.00\times 10^{-7}$%
] (left col.) and [$0.900003$,$-3.48\times 10^{-8}$,$8.01\times
10^{-7}$]
(right col.). The final values of the spatial Boyer-Lindquist components [$%
S^{r}\left( \protect\tau ^{\ast }\right) $,$\protect\mu S^{\protect\theta %
}\left( \protect\tau ^{\ast }\right) $,$\protect\mu S^{\protect\phi
}\left( \protect\tau ^{\ast }\right) $]$/\left\vert S\right\vert $
of the spinvector
are [$0.86$,$-2.9\times 10^{-5}$,$3.8\times 10^{-4}$] (left col.) and [$%
-1.11 $,$-8.6\times 10^{-5}$,$2.7\times 10^{-4}$] (right col.). The
final
spin directions [$\protect\theta ^{\left( S\right) }\left( \protect\tau %
^{\ast }\right) $,$\protect\phi ^{\left( S\right) }\left( \protect\tau %
^{\ast }\right) $] in the boosted SO frame are determined by [$1.48$,$-0.59$%
] (left col.) and [$1.31$,$1.09$] (right col.). The angles [$\protect\theta %
^{\left( l\right) }\left( \protect\tau ^{\ast }\right) $,$\protect\phi %
^{\left( l\right) }\left( \protect\tau ^{\ast }\right) $]
characterizing the
final orbital plane orientations in coordinate space ($x/\protect\mu $,$y/%
\protect\mu $,$z/\protect\mu $) are [$0.11$,$-0.33$] (left col.) and [$0.11$,%
$1.27$] (right col.).} \label{HB1TD}
\end{figure}
\end{widetext}
side of Figure \ref{ZW3TD}). The first row depicts the
unbound orbits in the ($x$,$y$,$z$)-space. The initial data set is chosen 
at
$r\left( \tau =0\right) =2000\mu $ where the body is in the equatorial plane
($\theta \left( \tau =0\right) =\pi /2$ and $\phi \left( \tau =0\right) =0$)
and the centroid four velocity has vanishing $\theta $-component. We
numerically checked that $r\rightarrow \infty $ as $\tau \rightarrow \pm
\infty $. Second and third rows represent the orbits near the black hole in
the ($x$,$y$,$z$) and the ($\rho $,$z$) spaces, respectively. The interval
for $\tau $ is determined by $-5\mu $ before and $+5\mu $ after the body
crossed the outer stationary limit surface. As the body penetrates the
ergosphere, it makes two turns around the black hole, then it leaves the
ergosphere going to the spatial infinity. These evolutions describe such
scattering processes where the center is extremely approached. The deviation
of the trajectory from the equatorial plane is an effect of the
spin-curvature coupling. The fourth and fifth rows image the evolutions of
the unit spin vector represented in the boosted SO and ZAMO frames,
respectively. The deviation of the spin vector direction from the equatorial
plane also occurs because of the spin-curvature coupling. The jump in the
evolution of the spin vector in the boosted SO frame (marked by black dots)
shows that the variation of spin direction takes place mainly inside the
ergosphere. The large part of the variation of spin direction happens during
that period when the body is inside the ergosphere. This time interval is $%
\tau \in \left[ 2214.8\mu ,2218.6\mu \right] $ which is short with respect
to the considered total evolution period $\tau =\left[ 0,4433\mu \right] $.
The final value of the proper time $\tau ^{\ast }=4433\mu $ was chosen in
such a way, that for $\tau >\tau ^{\ast }$\ the spin angles undergo only
unsignificant changes. Figure \ref{HB1TDb} presents the evolutions of $%
\Omega _{\left( prec\right) }^{\mathbf{\alpha }}\left( e,u\right) $ and $%
\Omega _{\left( prec\right) }^{\mathbf{\alpha }}\left( f,u\right) $ for that
time interval which is determined by $-25\mu $ before and $+25\mu $ after
the body crossed the outer stationary limit surface. The purplish shadow
denotes that period when the body is inside the ergosphere where the spin
precessional angular velocity components increases.

The spin-curvature coupling mainly influences the smaller components of the
precessional angular velocity $\Omega _{\left( prec\right) }^{\mathbf{1}%
}\left( f,u\right) $ and $\Omega _{\left( prec\right) }^{\mathbf{3}}\left(
f,u\right) $, as it can be seen in the first row of Figure \ref{HB1TDc}. The
black curve represents the evolutions without the spin-curvature coupling.
In the case of the red curves, the spin-curvature coupling is taken into
account, and they are the same as in the second column of Figure \ref{HB1TDb}%
. The spin-curvature coupling increases the amplitude of the precessional
angular velocity components.

The reparametrization invariance of the representative worldline also
implies a gauge freedom \cite{HansonRegge1974}. Usually, the following
choices for this timelike parameter are applied in the literature: $i)$ the
proper time ($u^{a}u_{a}=-1$) \cite{Semerak1999,Bini2017} also used in this
paper; $ii)$ the parameter determined by the normalization $u_{a}p^{a}/M=-1$
\cite{EhlersRudolf1977,Tanaka96}; $iii)$ the coordinate time $t$ \cite%
{Hojman1977,Hojman2012}. Employing the TD SSC, considerable differences were
not found in both the orbit and the spin dynamics when using the parameters
either $i)$ or $ii)$ \cite{LukesGerakopoulos}. The orbit and the spin
evolutions are unaffected when using the coordinate time $t$ instead of the
proper time $\tau $. However, the precessional angular velocity is changed
for $\Omega _{\left( prec\right) }^{\mathbf{\alpha }}/u^{0}$ which is shown
in the second row of Figure \ref{HB1TDc} as a function of $t$. The black and
the red curves represent the evolution without and with the spin-curvature
coupling. We can conclude the same effects when we have considered the spin
evolution with respect to the proper time.

The relatively rapid change in the direction of the spin vector can also be
confirmed without using any particular reference frame. In the first row of
Figure \ref{HB1TDeff}, we present the evolutions of the Boyer-Lindquist
coordinate components of the unit spin vector for the case imaged on the
right hand sides of Figures \ref{HB1TD} and \ref{HB1TDb}. The spin-curvature
coupling is included in the evolutions depicted by the red curves. The black
curves represent the corresponding evolutions when this coupling is turned
off. The effect of the spin-curvature coupling can be seen in the evolution
of $S^{t}$, $S^{r}$ and $S^{\theta }$ components. The latter vanishes
identically in the absence of the spin-curvature coupling. However, if the
spin-curvature coupling is included in the analysis, the $S^{\theta}$
component deviates significantly from zero when the body is close to the
central black hole. In addition, the effect of the spin-curvature coupling
remains in the $S^{t}$ and $S^{r}$ components far from the central black
hole. They approach another constant values when the spin-curvature coupling
is taken into account. The evolutions of the components of the unit spin
vector and their derivatives rescaled to dimensionless variables on a
smaller timescale, when the spinning test body is close the central black
hole are represented in the second row. All panels supports a relatively
rapid change of the spin vector near and inside the ergosphere. Finally, we
mention that the MPD equations were applied only in its validity domain,
this check is given in the Appendix \ref{para}.

In a wider range of initial conditions, the final values of the polar $%
\theta ^{\left( S\right) }$ and azimuthal $\phi ^{\left( S\right) }$
spin angles (the scattering angles) are represented on Figure
\ref{3dEL} as functions of gauge
invariant, dimensionless energy $\hat{E}=E/M$ and angular momentum $\hat{J}%
_{z}=J_{z}/\mu M$. The small black dots in the plane of the initial spin
angles ($\theta ^{\left( S\right) }\left( 0\right) =\pi /2$ and $\phi
^{\left( S\right) }\left( 0\right) =0$) indicate the region, where the body
crosses the event horizon of the Kerr black hole. Then, instead of a
scattering process, the body falls into the black hole. Close to the left
corner, i.e. at smaller $\hat{E}$ and higher $\hat{J}_{z}$ values, the body
approaches the central black hole less than for higher $\hat{E}$ and/or for
smaller $\hat{J}_{z}$ values. As a consequence, the precession and hence the
variation of the spin angles are both small. However, close to the diagonal
in the $\hat{E}$, $\hat{J}_{z}$ plane indicated by the edge of the black
dots region, the body enters into the ergosphere, and due to the high
precession there, the spin angles undergo a relatively large change. In all
case, the initial values are chosen such that, if the spin-curvature
coupling is neglected, the polar angle $\theta ^{\left( S\right) }$ remains $%
\pi /2$ during the whole evolution, and the spin precession influences only $%
\phi ^{\left( S\right) }$. Hence, the variation of $\theta ^{\left( S\right)
}$ shown on the left panel is a clear effect of the spin-curvature coupling.
We 
\begin{widetext}
\newpage
\begin{figure}[H]
\begin{center}
\includegraphics[width=5cm]{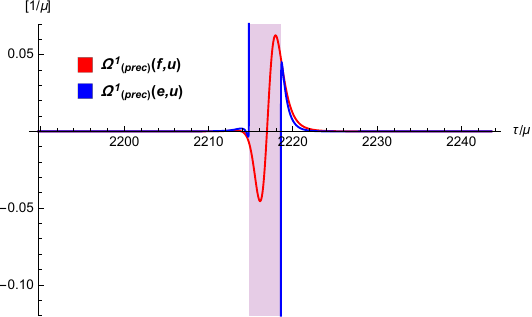} \hskip 1.5cm %
\includegraphics[width=5cm]{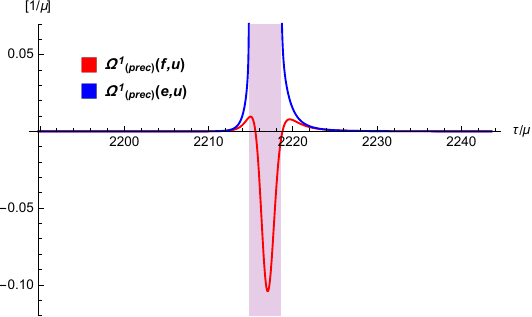} \\[0pt]
\includegraphics[width=5cm]{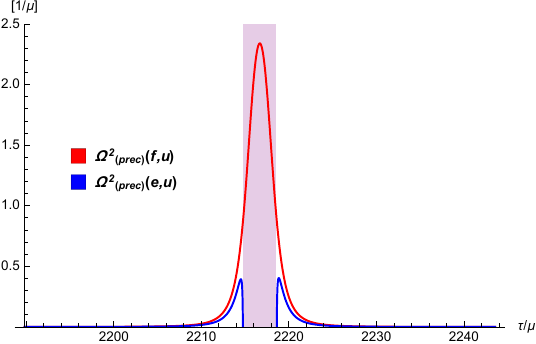} \hskip 1.5cm %
\includegraphics[width=5cm]{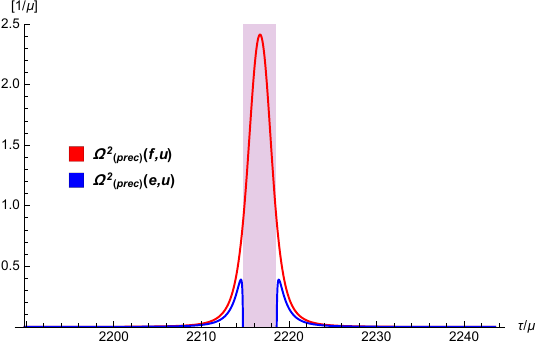} \\[0pt]
\includegraphics[width=5cm]{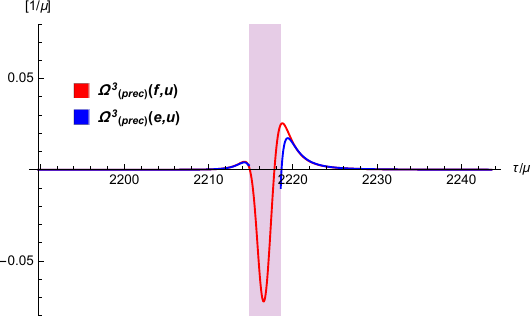} \hskip 1.5cm %
\includegraphics[width=5cm]{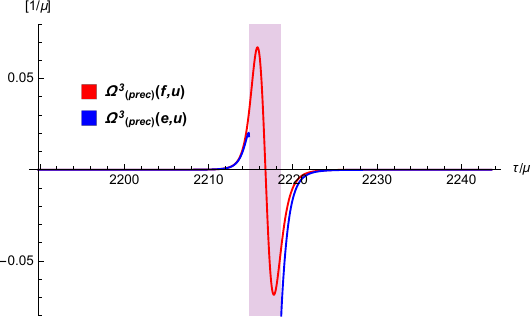}
\end{center}
\par
\vskip -0.5cm \caption{(color online). On the left and right columns
the evolutions of the spherical triad components of the spin
precessional angular velocities are presented along those orbits
which are shown in the left and right columns of Figure
\protect\ref{HB1TD}, respectively.} \label{HB1TDb}
\end{figure}

\begin{figure}[H]
\begin{center}
\includegraphics[width=4.7cm]{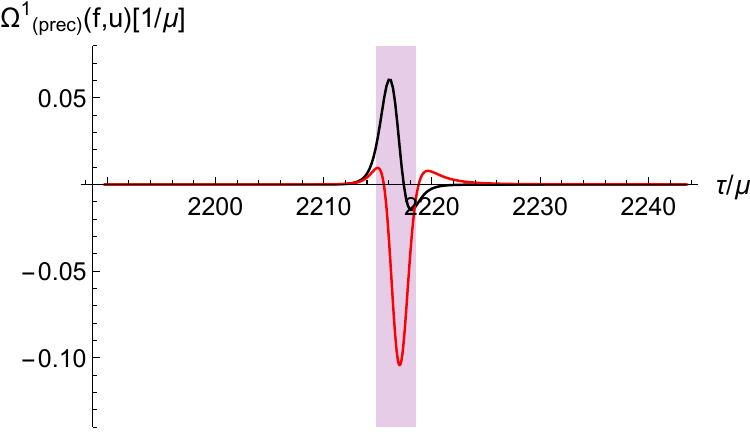} \hskip 1.5cm %
\includegraphics[width=4.7cm]{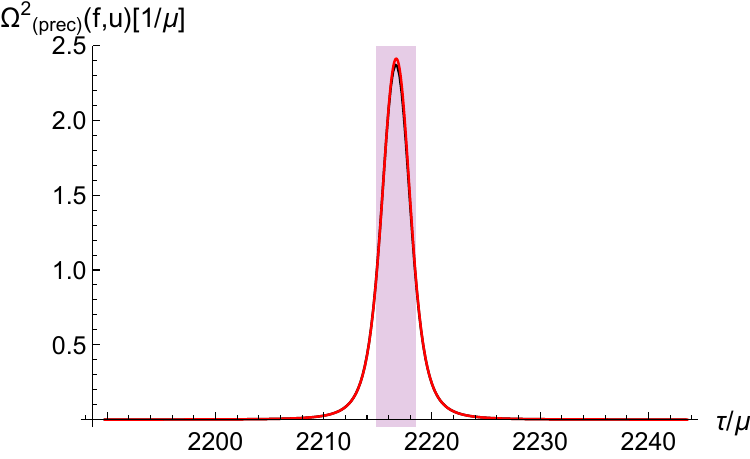} \hskip 1.5cm %
\includegraphics[width=4.7cm]{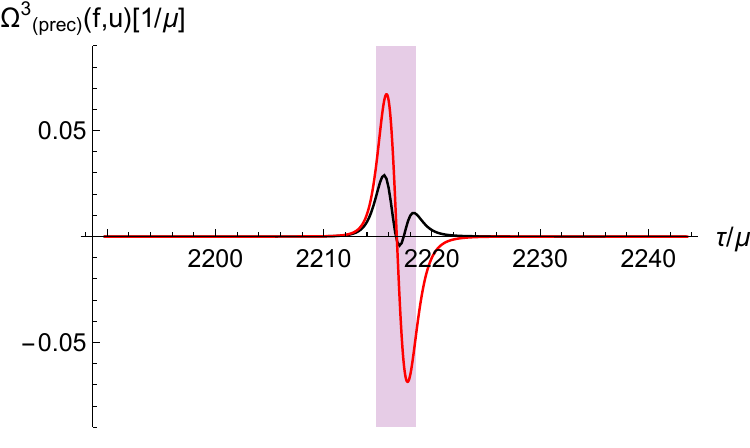} \\[0pt]
\includegraphics[width=4.7cm]{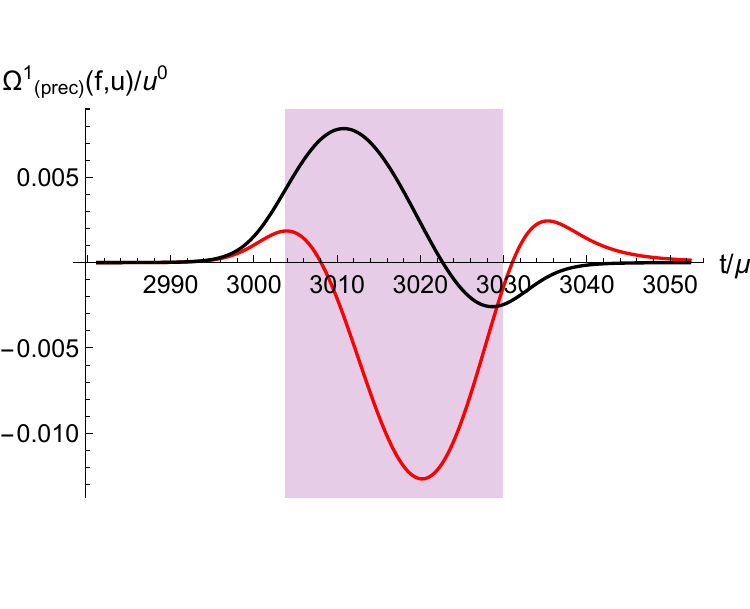} \hskip 1.5cm %
\includegraphics[width=4.7cm]{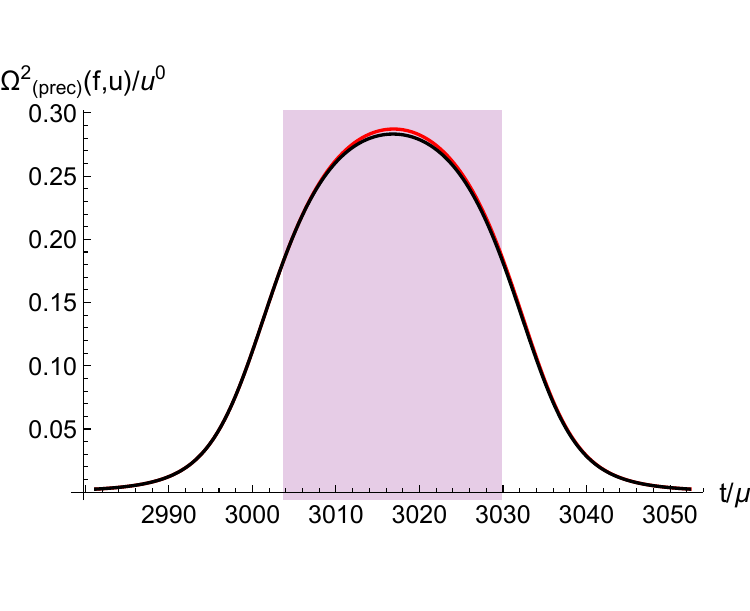} \hskip 1.5cm %
\includegraphics[width=4.7cm]{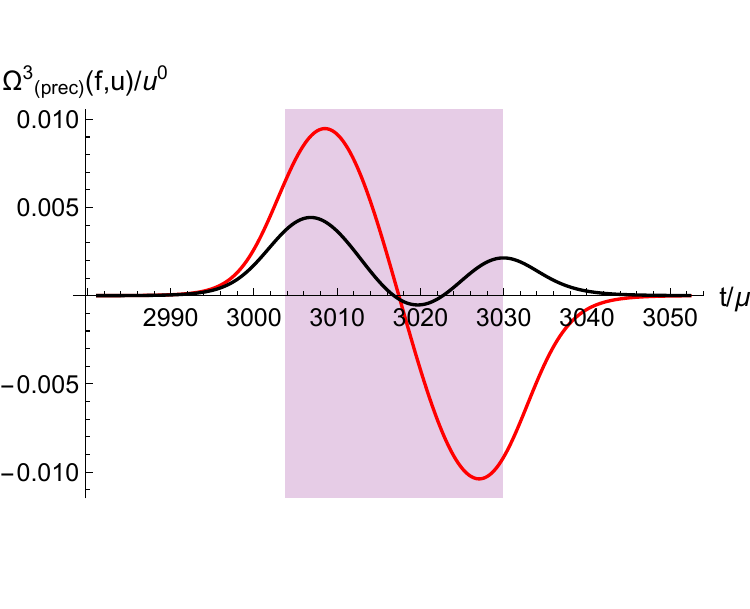}
\end{center}
\par
\vskip -0.5cm \caption{(color online). In the first line, the black
and the red curves show the precessional angular velocity spherical
frame components without and with spin-curvature coupling,
respectively. The second line shows them when the spin evolution is
considered as a function of the coordinate time $t$. These
evolutions belong to the case which is presented in the right hand
sides of Figures \protect\ref{HB1TD} and \protect\ref{HB1TDb}.}
\label{HB1TDc}
\end{figure}
\end{widetext}
mention that, both functions $\theta ^{\left( S\right) }$ and $\phi
^{\left( S\right) }$ steeply increase as approaching the edge of the black
dots region. Those 
maxima, which can be seen on the panels, belong to the
chosen grid in the $\hat{E}$, $\hat{J}_{z}$ plane.

\hskip 1cm

\subsection{\textbf{Spinning bodies moving on zoom-whirl orbits in rotating
regular black hole spacetimes}\label{reg}}

In this subsection, we set $\mu =0$, $\gamma =3$ and $a=0.99\mu _{em}$. The
background is either a regular, rotating Bardeen-like ($\nu =2$) or
Hayward-like ($\nu =3$) black hole spacetime. For $\nu =2$ and $\nu =3$, the
spacetime contains a black hole for $q\leq 0.081$ and $q\leq 0.216$,
respectively. 
\begin{widetext}
\newpage
\begin{figure}[H]
\begin{center}
\includegraphics[width=4cm]{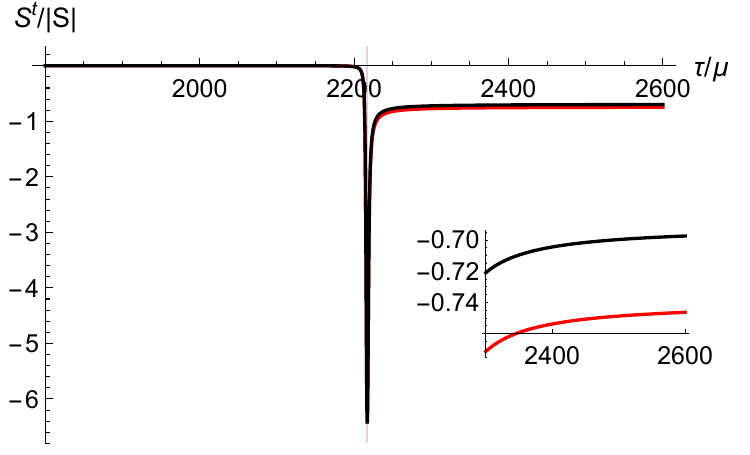}\hfill %
\includegraphics[width=4cm]{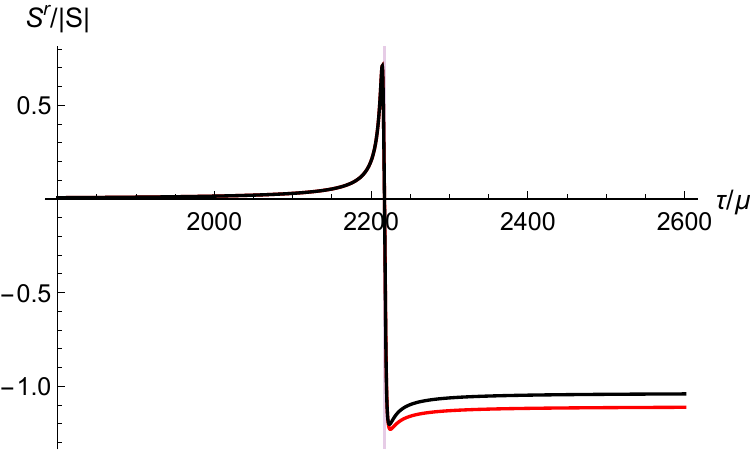}\hfill %
\includegraphics[width=4cm]{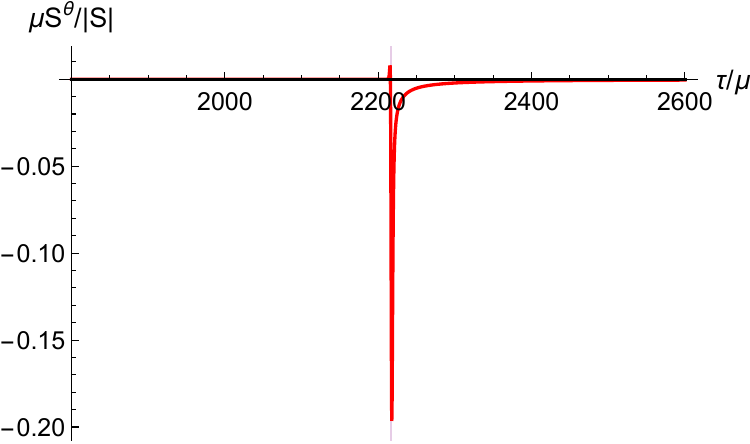}\hfill %
\includegraphics[width=4cm]{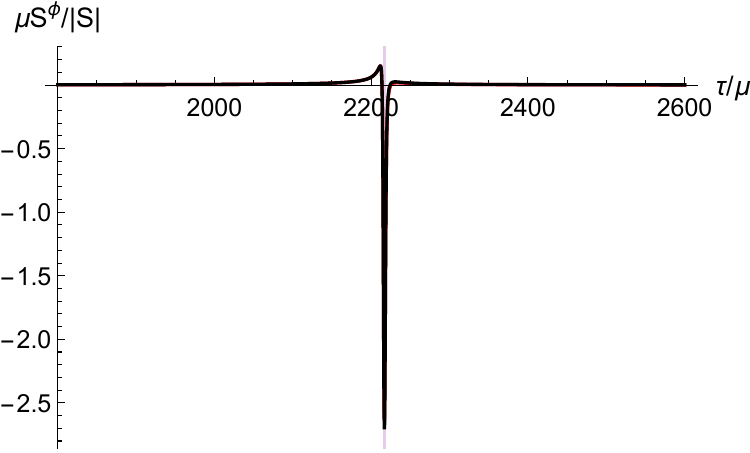}\\[0pt]
\includegraphics[width=4cm]{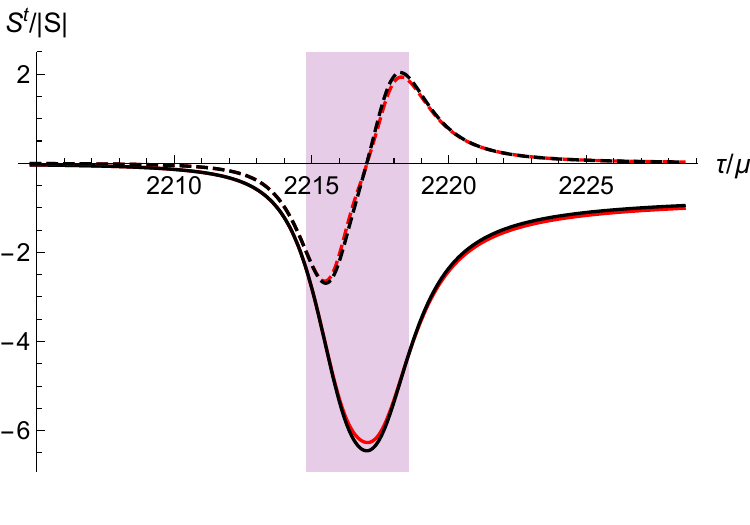}\hfill %
\includegraphics[width=4cm]{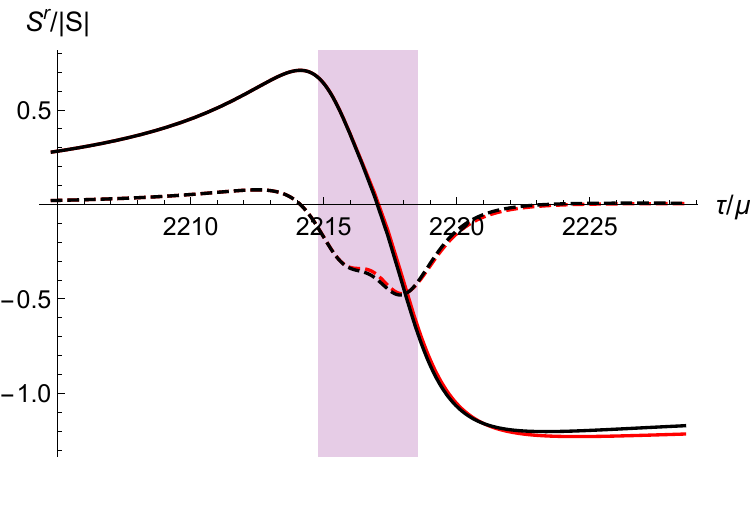}\hfill %
\includegraphics[width=4cm]{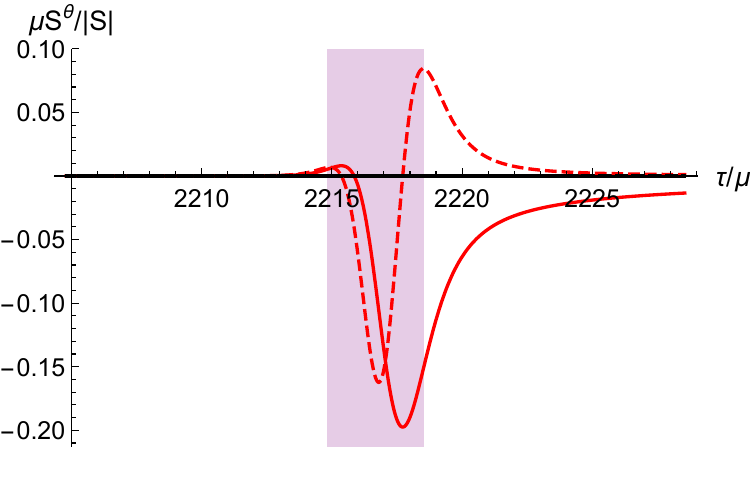}\hfill %
\includegraphics[width=4cm]{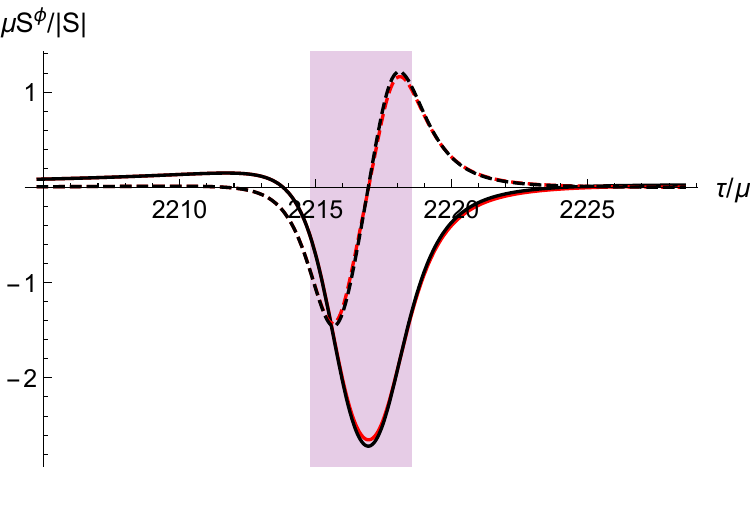}
\end{center}
\par
\vskip -0.5cm \caption{(color online). In the first line, the black
and the red curves present the Boyer-Lindquist coordinate components
of the unit spin vector without and with spin-curvature coupling. In
case of $S^{t}$, the relatively small deviation of the curves when
the test body is moving away from the central black hole is shown in
a small box. The second line presents the evolutions of the unit
spin vector and their derivatives rescaled to dimensionless
variables when the spinning test body is close to the central black
hole. These evolutions belong to the case, which is presented in the
right hand sides of Figures \protect\ref{HB1TD} and
\protect\ref{HB1TDb} and also in Figure \protect\ref{HB1TDc}. The
time interval, when the body is inside the ergosphere, is indicated
by a purplish shadow on all panels.} \label{HB1TDeff}
\end{figure}

\begin{figure}[H]
\begin{center}
\hskip -0.0cm
\includegraphics[width=8.80cm]{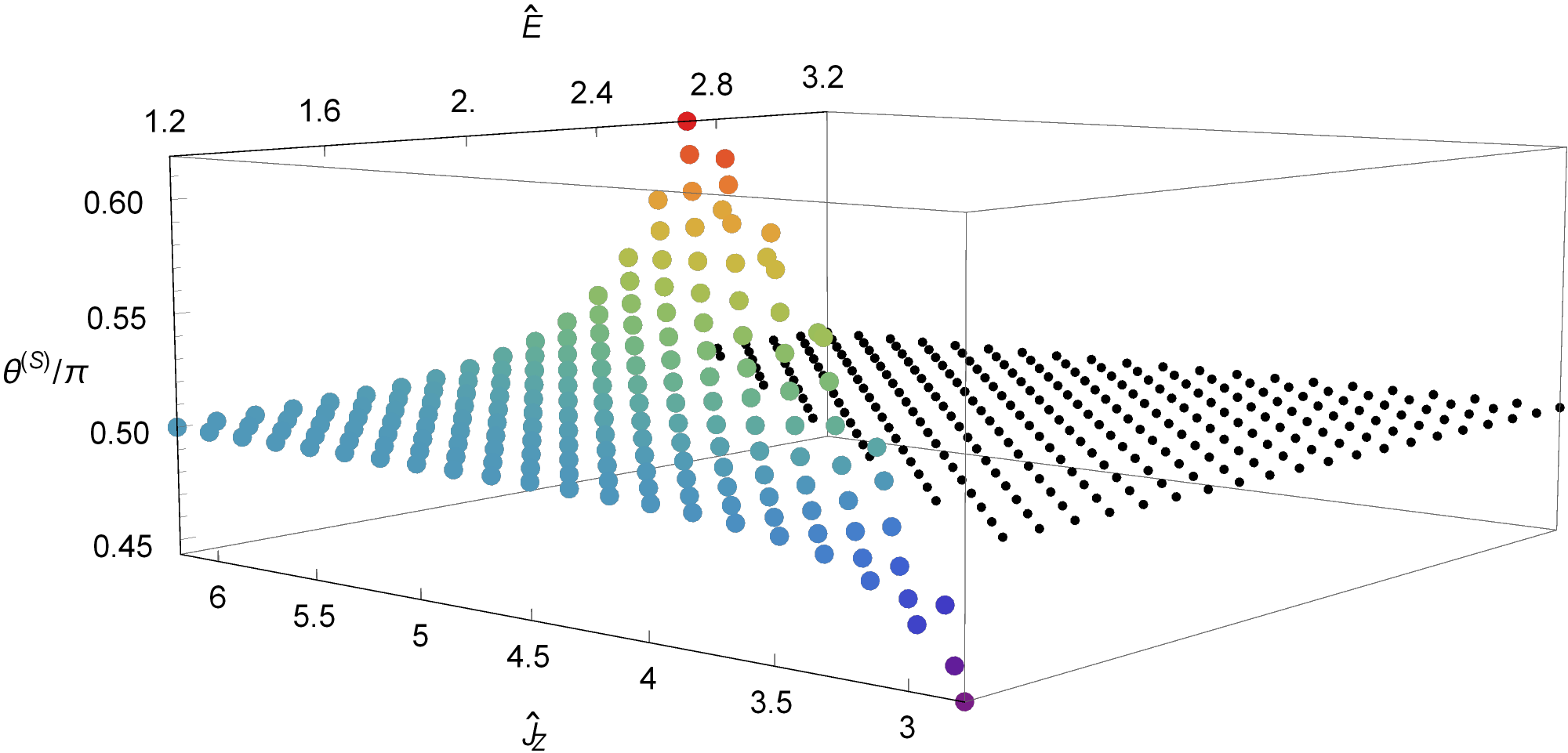}\hfill \hskip %
-1.0cm
\includegraphics[width=8.80cm]{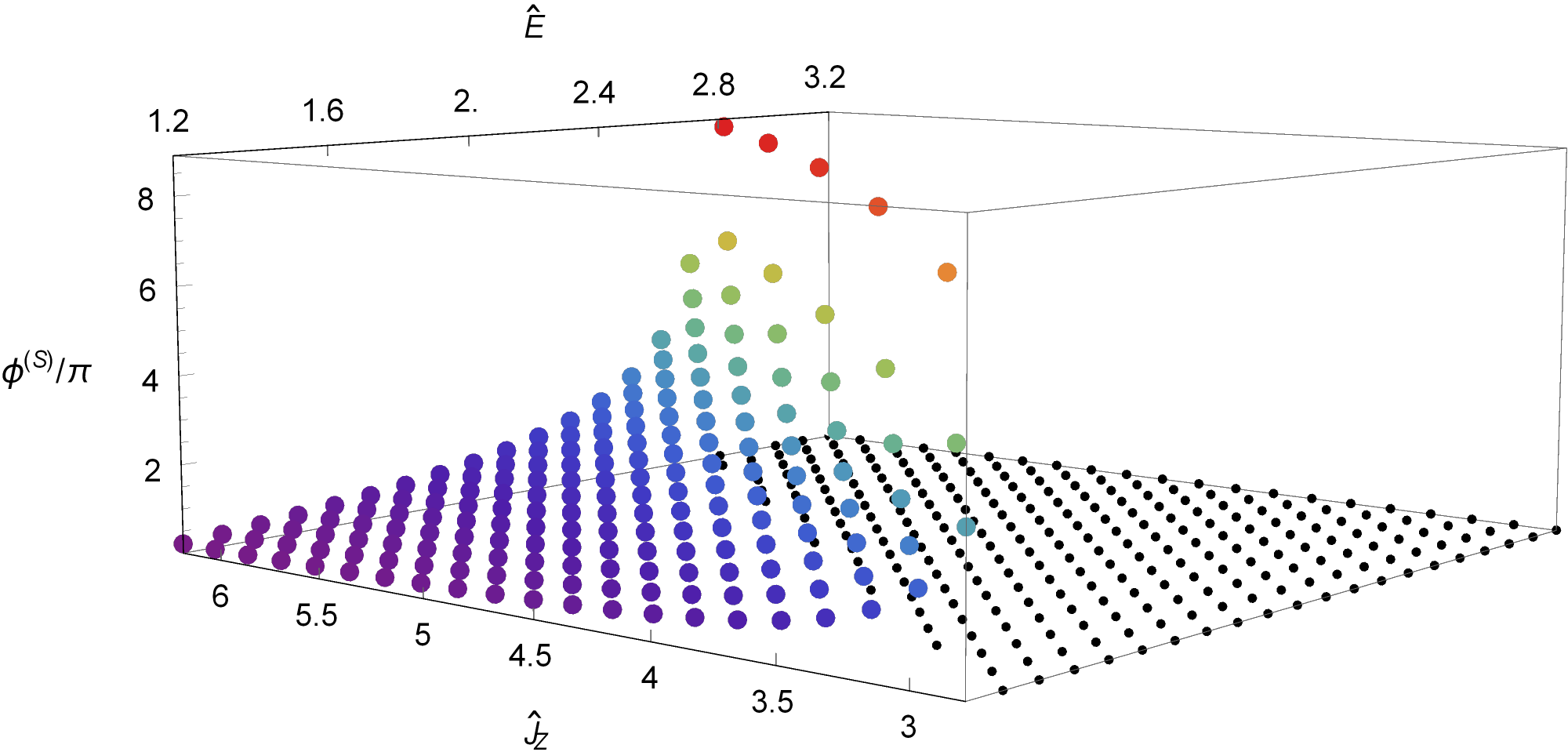}\hfill
\end{center}
\par
\vskip -0.5cm \caption{(color online). The left and right panels
present the final value of the spin angles $\protect\theta ^{\left(
S\right) }\left( \protect\tau ^{\ast }\right) $ and $\protect\phi
^{\left( S\right) }\left( \protect\tau ^{\ast }\right) $,
respectively, as functions of the dimensionless energy $\hat{E}=E/M$
and angular momentum $\hat{J}_{z}=J_{z}/\protect\mu M$. The final
values were computed at $\protect\tau ^{\ast }=4433\protect\mu $. We
have checked that the spin angles undergo only unsignificant changes
for $\protect\tau
>\protect\tau ^{\ast }$. The initial spin is given by $\left\vert
S\right\vert /\protect\mu M=0.1$, $\protect\theta ^{\left( S\right)
}\left( 0\right) =\protect\pi /2$ and $\protect\phi ^{\left(
S\right) }\left( 0\right) =0$. The initial momentum has vanishing
$\protect\theta $ component: $p^{\protect\theta }\left( 0\right)
/M=0$, and its additional components were determined from $\hat{E}$,
$\hat{J}_{z}$ and $p^{a}p_{a}/M^{2}=-1$. The small black dots in the
plane of the initial spin angles represent the region, where the
body crosses the event horizon of the Kerr black hole, hence,
unbound orbits do not develop.} \label{3dEL}
\end{figure}
\end{widetext}
We consider three cases: ($\nu =2$,$q=0.081$), 
($\nu =3$,$%
q=0.081$) and ($\nu =3$,$q=0.216$). For these parameters the regular black
holes have two stationary limit surfaces and event horizons. In addition,
the spin magnitude for the moving body is chosen as $\left\vert S\right\vert
/\mu M=0.1$.

\vskip -0.03cm
On Figure \ref{ZW2regular}, zoom-whirl orbits in different regular rotating
black hole spacetimes are presented. The columns from left to right
correspond to ($\nu =2$,$q=0.081$), ($\nu =3$,$q=0.081$) and ($\nu =3$,$%
q=0.216$). With the notation change $\mu \rightarrow \mu _{em}$, the initial
values are chosen the same as in the second column of Figure \ref{ZW1TD}.
Each row represents the same quantity which was shown on Figure \ref{ZW3TD}.
The first two columns show that both the orbit and the spin evolutions are
significantly different in the cases of the Bardeen-like and Hayward-like
black holes for the same $\mu _{em}$ and $q$ values. In addition, the second
and the third columns show in the case of Hayward-like background that these
evolutions are also sensitive for the value of $q$. The way of deviation of
the orbit from the equatorial plane, which is the effect of the
spin-curvature coupling, is also very sensitive for the parameters of the
regular black holes. The spin vector evolutions including the first whirling
period in the boosted SO frame is presented in the third row. The black dots
represent a jump in the evolution. The part of the evolution which is not
shown takes place inside the ergosphere. The amount of the jumps is somewhat
different for each cases. The fourth row shows the total evolution of the
spin vector in the boosted ZAMO frame. The final directions (blue arrows) of
the spin direction are significantly different. The evolutions of the
spherical frame components of the precessional angular velocity including
the first three whirling period are shown in the last three rows. These are
perturbatively different for the different regular black holes. However, the
effects of these small differences add up over the evolution.

Finally, we mention that a consideration of unbound orbits about regular
black holes can be found in Ref. \cite{KMhyp}.

\begin{widetext}
\newpage
\begin{figure}[H]
\begin{center}
\includegraphics[width=5cm]{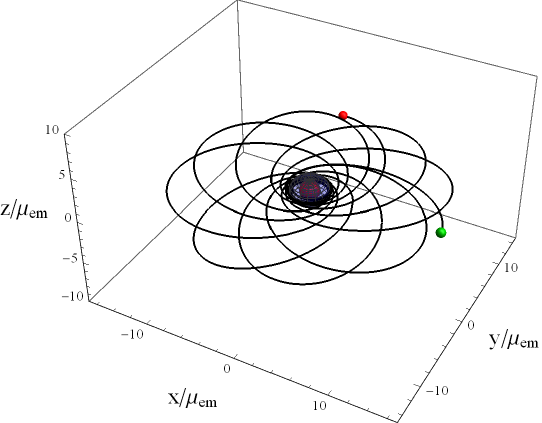}\hfill %
\includegraphics[width=5cm]{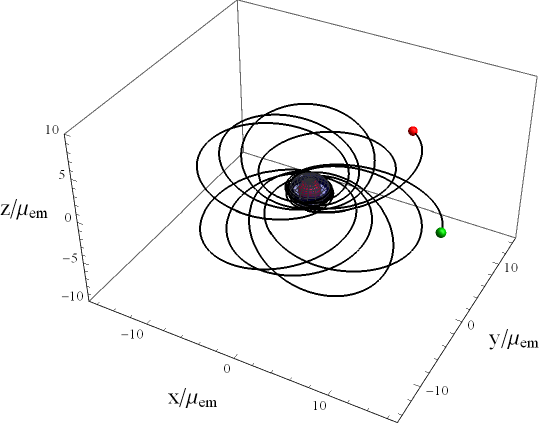}\hfill %
\includegraphics[width=5cm]{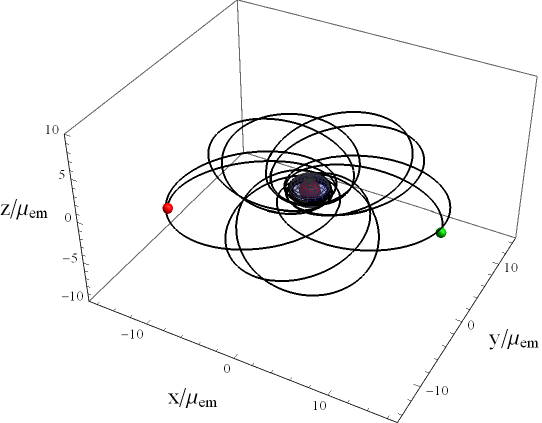}\hfill \\[0pt]
\includegraphics[width=5cm]{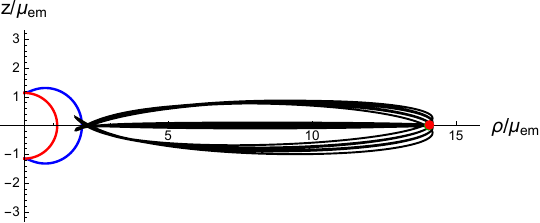}\hfill %
\includegraphics[width=5cm]{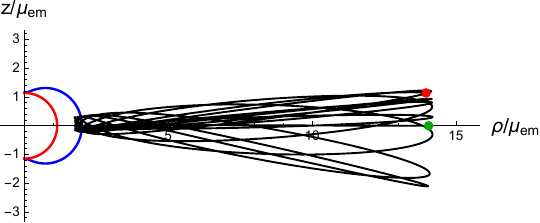}\hfill %
\includegraphics[width=5cm]{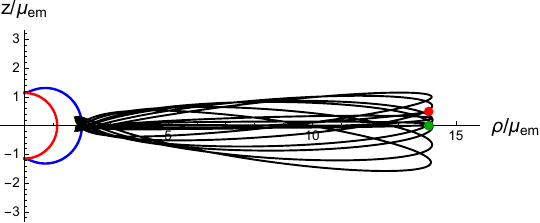}\hfill \\[0pt]
\includegraphics[width=5cm]{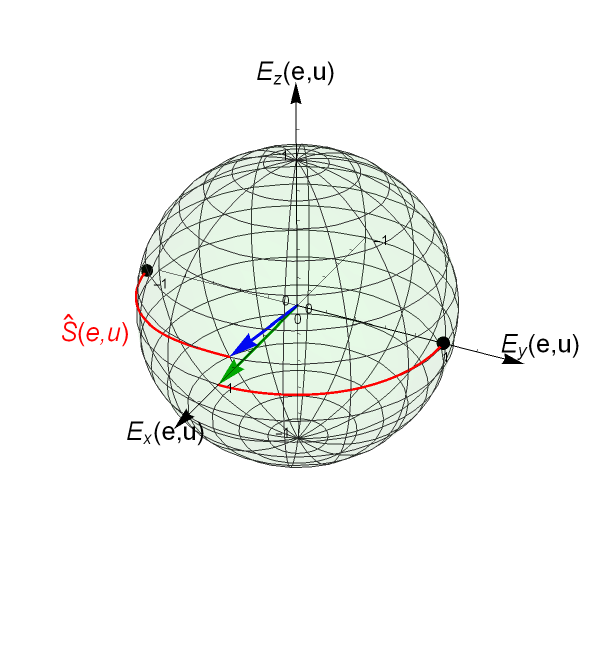}\hfill %
\includegraphics[width=5cm]{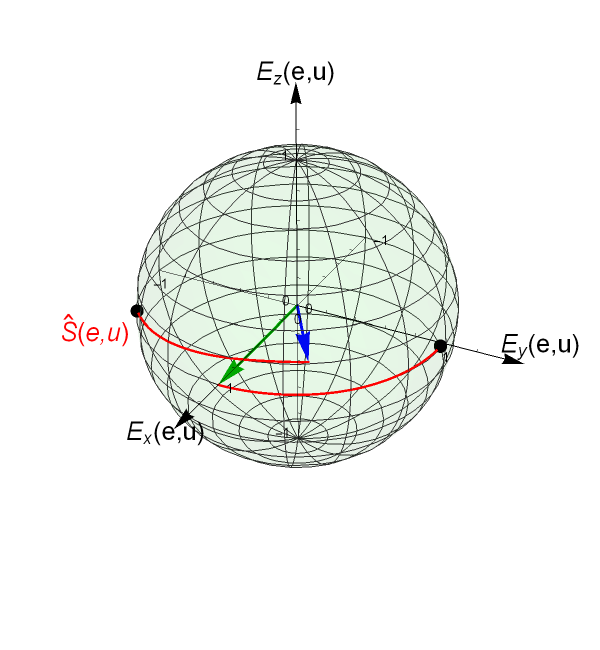}\hfill %
\includegraphics[width=5cm]{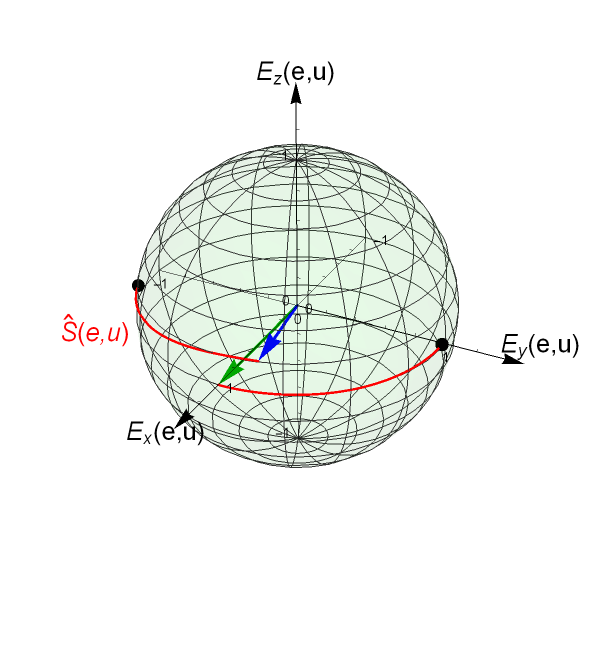}\hfill \\[0pt%
]
\vskip -1.0cm \includegraphics[width=5cm]{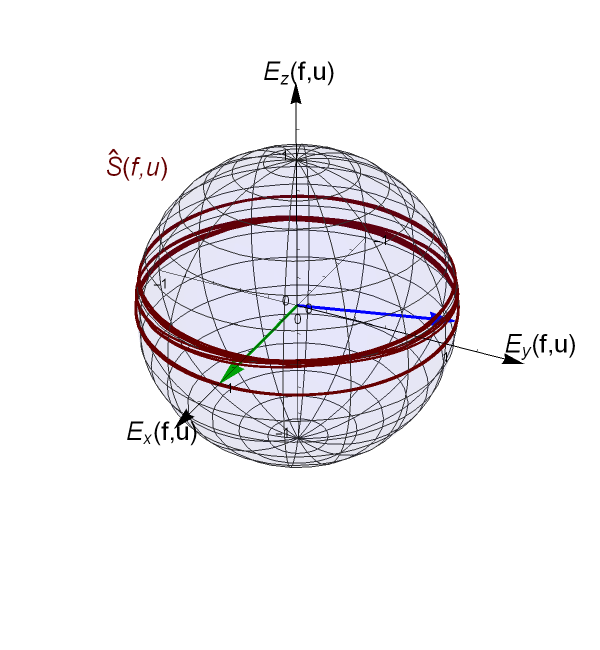}%
\hfill \includegraphics[width=5cm]{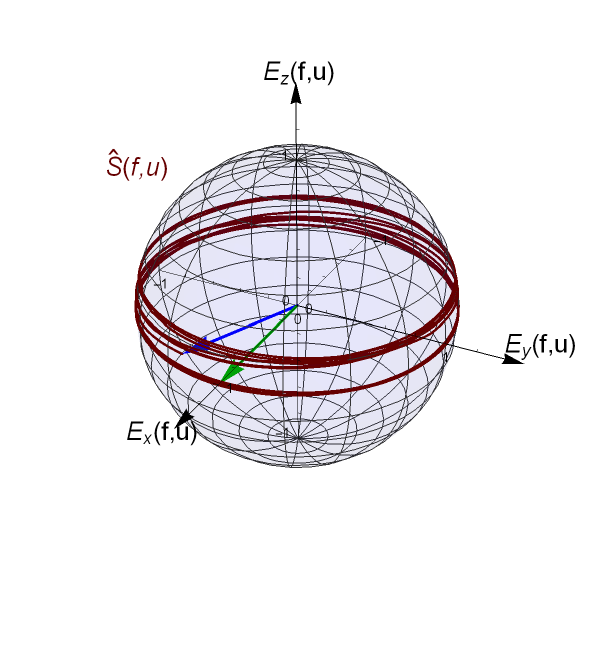}\hfill %
\includegraphics[width=5cm]{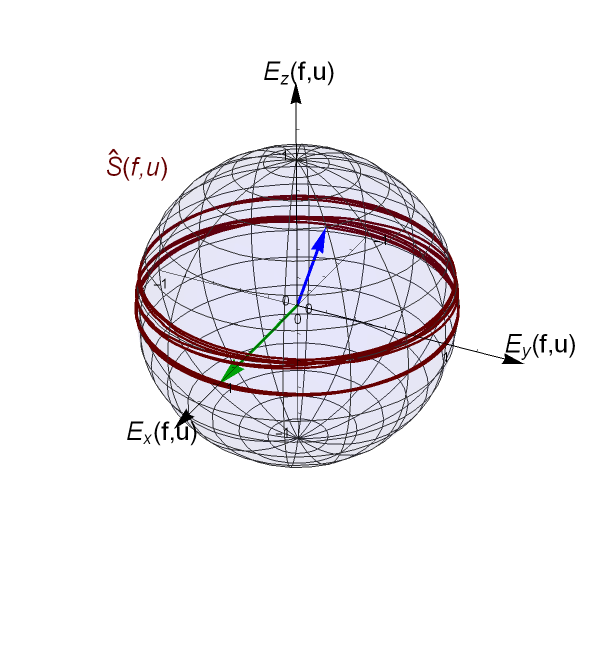}\hfill \\[0pt]
\vskip -1.0cm \includegraphics[width=5cm]{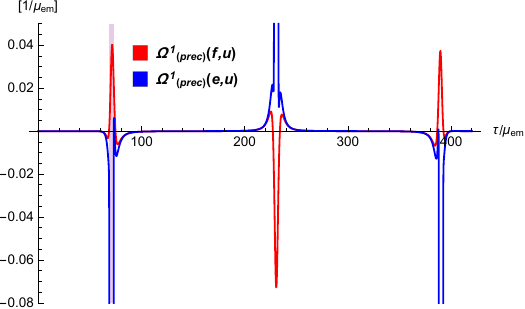}%
\hfill \includegraphics[width=5cm]{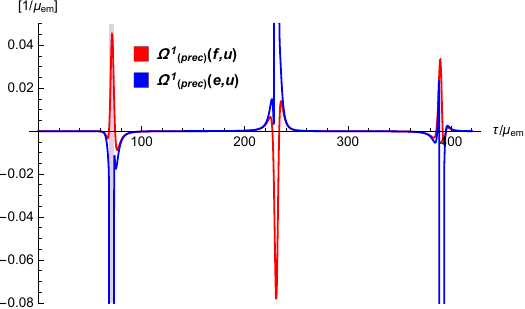}\hfill %
\includegraphics[width=5cm]{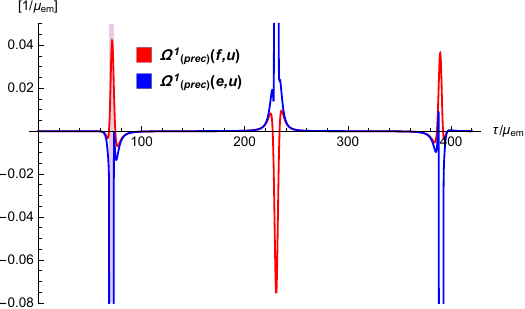}\hfill \\[0pt]
\includegraphics[width=5cm]{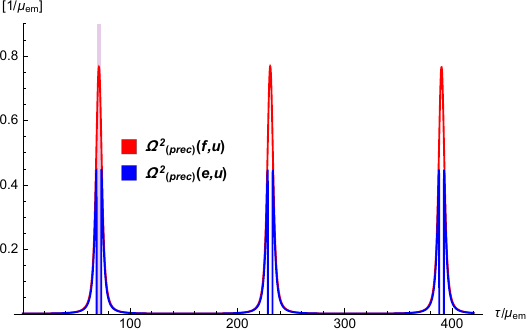}\hfill %
\includegraphics[width=5cm]{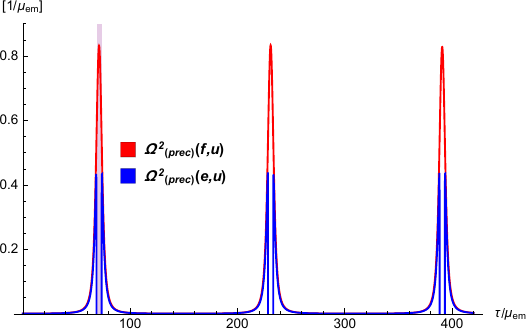}\hfill %
\includegraphics[width=5cm]{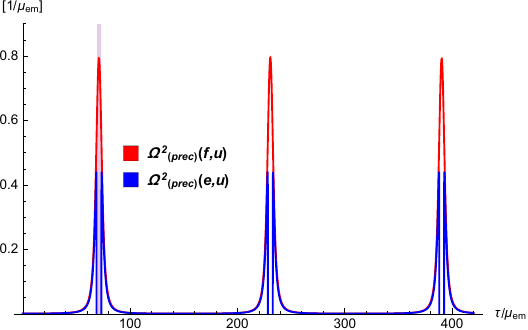}\hfill \\[0pt]
\includegraphics[width=5cm]{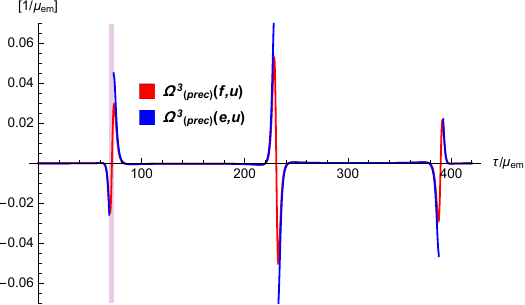}\hfill %
\includegraphics[width=5cm]{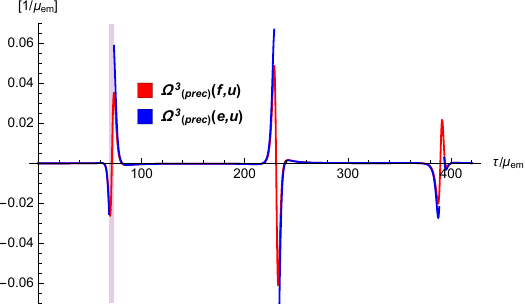}\hfill %
\includegraphics[width=5cm]{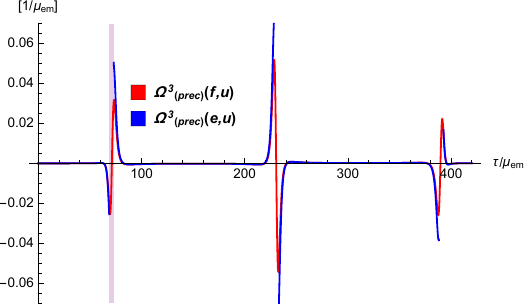}\hfill \\[0pt]
\end{center}
\par
\vskip -0.5cm \caption{(color online). Zoom-whirl orbits are
represented around regular, rotating black holes with
$\protect\gamma =3$ and $a=0.99\protect\mu _{em}$.
The first column shows the orbit around a Bardeen-like black hole ($\protect%
\nu =2$) while the middle and the last around a Hayward-like black hole ($%
\protect\nu =3$). The parameter $q$ is $0.081$ in the first two
columns
while $0.216$ in third one. Applying the notation change $\protect\mu %
\rightarrow \protect\mu _{em}$, the initial values are chosen the
same as in the second column of Figure \protect\ref{ZW1TD}. The
quantities in each line are the same which are presented in Figure
\protect\ref{ZW3TD}.} \label{ZW2regular}
\end{figure}
\end{widetext}

\section{Conclusions\label{Conclusion}}

We have considered numerically the evolution of a spinning test body
governed by the MPD equations, moving along spherical-like, zoom-whirl and
unbound orbits around a Kerr black hole. When the spacetime curvature and
the spin contributions on the right hand sides of the MPD equations can be
neglected, we recovered the corresponding results of Ref. \cite{Bini2017}
for a spherical orbit. However, for higher spin, an amplitude modulation
occured in the harmonic evolution of the spin precessional angular velocity
caused by the spin-curvature coupling. This amplitude modulation also
occured in the $\theta $ Boyer-lindquist coordinate component of the spin
vector.

The existence of zoom-whirl orbits are confirmed by using the MPD dynamics.
The considered zoom-whirl and unbound orbits of spinning body passed over
the ergosphere, where the PN approximation cannot be applied. In all cases
the numerical investigations showed that the spin precessional angular
velocity highly increased near and inside the ergosphere. Thus the direction
of the spin vector is significantly variated during the evolutionary phase
inside the ergosphere. The initial values were chosen such that the test
body moved in the equatorial plane when the spin-curvature coupling is
neglected. Hence, the effect of this coupling occured as a deviation of the
orbit from the equatorial plane. In order to investigate non-trivial spin
evolution, the initial spin direction was chosen to be perpendicular to the
rotation axis of the central black hole. Then, the spin vector evolved in
the equatorial plane of the boosted SO and ZAMO frames when the
spin-curvature coupling is neglected. The deviation of the spin vector from
this equatorial plane was also the effect of the spin-curvature coupling.
Additional effects of the spin-curvature coupling was observed in the
evolutions of the spin precessional angular velocity and of the
Boyer-Lindquist coordinate components of the spin vector.

Zoom-whirl orbits and spin precession including the spin-curvature coupling
were also considered in regular spacetimes containing a central rotating
black hole. Significant differences were observed in the way of deviation of
the orbit from the equatorial plane which were sensitive for the parameters
of the regular black hole. Small deviations were found in the spin
precession angular velocity, which add up over the evolutions. Hence, the
direction of the final spin vector can be very different for different
parameters of the regular black hole.

Finally, we mention that the numeric investigation presented here could be
generalized in the following way. Besides the spin-curvature coupling
another effects would occur if the backreaction of the body to the metric
was not neglected. This backreaction appears as a self-force in the equation
of motion \cite{SF1,SF2,SF3,SF4}, and also causes a deviation from the
geodesic orbit like the spin-curvature coupling.

\medskip \textbf{Acknowledgements}

The work of B. M. was supported by the J\'anos Bolyai Research Scholarship
of the Hungarian Academy of Sciences. The work of Z. K. was supported by the
J\'anos Bolyai Research Scholarship of the Hungarian Academy of Sciences, by
the UNKP-18-4 New National Excellence Program of the Ministry of Human
Capacities and by the Hungarian National Research Development and Innovation
Office (NKFI) in the form of the grant 123996.

\section{Conflict of Interest}

The authors declare no conflict of interest.

\section{APPENDIX A: The relation between the frames $E_{\mathbf{\protect%
\alpha }}\left( e,U\right) $ and $E_{\mathbf{\protect\alpha }}\left(
f,U\right) $\label{comFRametraf}}

\begin{widetext}

The frame vectors $E_{\mathbf{\alpha }}\left( e,U\right) $ derived from the
SO's frame are the following linear combination of $E_{\mathbf{\alpha }%
}\left( f,U\right) $:%
\begin{equation*}
E_{\mathbf{1}}\left( e,U\right) =E_{\mathbf{1}}\left( f,U\right) +\frac{%
\Gamma _{\left( Z\right) }w_{\left( Z\right) }^{\mathbf{1}}}{1+\Gamma
_{\left( S\right) }}\left[ \frac{a\mathcal{B}\sin \theta }{\sqrt{%
-g_{tt}\Sigma \mathcal{A}}}E_{\mathbf{3}}\left( f,U\right) +\left( 1-\sqrt{%
\frac{\Sigma \Delta }{-g_{tt}\mathcal{A}}}\right) \frac{\Gamma _{\left(
Z\right) }\mathbf{w}_{\left( Z\right) }}{1+\Gamma _{\left( Z\right) }}\right]
,
\end{equation*}%
\begin{equation*}
E_{\mathbf{2}}\left( e,U\right) =E_{\mathbf{2}}\left( f,U\right) +\frac{%
\Gamma _{\left( Z\right) }w_{\left( Z\right) }^{\mathbf{2}}}{1+\Gamma
_{\left( S\right) }}\left[ \frac{a\mathcal{B}\sin \theta }{\sqrt{%
-g_{tt}\Sigma \mathcal{A}}}E_{\mathbf{3}}\left( f,U\right) +\left( 1-\sqrt{%
\frac{\Sigma \Delta }{-g_{tt}\mathcal{A}}}\right) \frac{\Gamma _{\left(
Z\right) }\mathbf{w}_{\left( Z\right) }}{1+\Gamma _{\left( Z\right) }}\right]
,
\end{equation*}%
\begin{equation}
E_{\mathbf{3}}\left( e,U\right) \!=\!\left( \sqrt{\frac{\Sigma \Delta }{%
-g_{tt}\mathcal{A}}}+\Gamma _{\left( Z\right) }\right) \frac{E_{\mathbf{3}%
}\left( f,U\right) }{1+\Gamma _{\left( S\right) }}-\frac{\Gamma _{\left(
Z\right) }\mathbf{w}_{\left( Z\right) }}{1+\Gamma _{\left( S\right) }}\left[
\left( 1-\sqrt{\frac{\Sigma \Delta }{-g_{tt}\mathcal{A}}}\right) \frac{%
\Gamma _{\left( Z\right) }w_{\left( Z\right) }^{\mathbf{3}}}{1+\Gamma
_{\left( Z\right) }}+\frac{a\mathcal{B}\sin \theta }{\sqrt{-g_{tt}\Sigma
\mathcal{A}}}\right] .
\end{equation}

The inverse relations are%
\begin{equation*}
E_{\mathbf{1}}\left( f,U\right) =E_{\mathbf{1}}\left( e,U\right) -\frac{%
\Gamma _{\left( S\right) }w_{\left( S\right) }^{\mathbf{1}}}{1+\Gamma
_{\left( Z\right) }}\left[ \frac{a\mathcal{B}\sin \theta }{\sqrt{%
-g_{tt}\Sigma \mathcal{A}}}E_{\mathbf{3}}\left( e,U\right) -\left( 1-\sqrt{%
\frac{\Sigma \Delta }{-g_{tt}\mathcal{A}}}\right) \frac{\Gamma _{\left(
S\right) }\mathbf{w}_{\left( S\right) }}{1+\Gamma _{\left( S\right) }}\right]
,
\end{equation*}%
\begin{equation*}
E_{\mathbf{2}}\left( f,U\right) =E_{\mathbf{2}}\left( e,U\right) -\frac{%
\Gamma _{\left( S\right) }w_{\left( S\right) }^{\mathbf{2}}}{1+\Gamma
_{\left( Z\right) }}\left[ \frac{a\mathcal{B}\sin \theta }{\sqrt{%
-g_{tt}\Sigma \mathcal{A}}}E_{\mathbf{3}}\left( e,U\right) -\left( 1-\sqrt{%
\frac{\Sigma \Delta }{-g_{tt}\mathcal{A}}}\right) \frac{\Gamma _{\left(
S\right) }\mathbf{w}_{\left( S\right) }}{1+\Gamma _{\left( S\right) }}\right]
,
\end{equation*}%
\begin{equation}
E_{\mathbf{3}}\left( f,U\right) \!=\!\left( \sqrt{\frac{\Sigma \Delta }{%
-g_{tt}\mathcal{A}}}+\Gamma _{\left( S\right) }\right) \frac{E_{\mathbf{3}%
}\left( e,U\right) }{1+\Gamma _{\left( Z\right) }}-\frac{\Gamma _{\left(
S\right) }\mathbf{w}_{\left( S\right) }}{1+\Gamma _{\left( Z\right) }}\left[
\left( 1-\sqrt{\frac{\Sigma \Delta }{-g_{tt}\mathcal{A}}}\right) \frac{%
\Gamma _{\left( S\right) }w_{\left( S\right) }^{\mathbf{3}}}{1+\Gamma
_{\left( S\right) }}-\frac{a\mathcal{B}\sin \theta }{\sqrt{-g_{tt}\Sigma
\mathcal{A}}}\right] .  \label{Rdef}
\end{equation}

The frame components of any vector field
\begin{equation}
\mathbf{V}=\overset{(e)}{V^{\mathbf{\alpha }}}E_{\mathbf{\alpha }}\left(
e\right) =\overset{(f)}{V^{\mathbf{\alpha }}}E_{\mathbf{\alpha }}\left(
f\right) ,
\end{equation}%
obey the following transformation rule

\begin{equation*}
\overset{(e)}{V^{\mathbf{1}}}=\overset{(f)}{V^{\mathbf{1}}}+\left[ \left( 1-%
\sqrt{\frac{\Sigma \Delta }{-g_{tt}\mathcal{A}}}\right) \frac{\Gamma
_{\left( Z\right) }\mathbf{w}_{\left( Z\right) }\cdot \mathbf{V}}{1+\Gamma
_{\left( Z\right) }}+\frac{a\mathcal{B}\sin \theta }{\sqrt{-g_{tt}\Sigma
\mathcal{A}}}\overset{(f)}{V^{\mathbf{3}}}\right] \frac{\Gamma _{\left(
Z\right) }w_{\left( Z\right) }^{\mathbf{1}}}{1+\Gamma _{\left( S\right) }},
\end{equation*}%
\begin{equation*}
\overset{(e)}{V^{\mathbf{2}}}=\overset{(f)}{V^{\mathbf{2}}}+\left[ \left( 1-%
\sqrt{\frac{\Sigma \Delta }{-g_{tt}\mathcal{A}}}\right) \frac{\Gamma
_{\left( Z\right) }\mathbf{w}_{\left( Z\right) }\cdot \mathbf{V}}{1+\Gamma
_{\left( Z\right) }}+\frac{a\mathcal{B}\sin \theta }{\sqrt{-g_{tt}\Sigma
\mathcal{A}}}\overset{(f)}{V^{\mathbf{3}}}\right] \frac{\Gamma _{\left(
Z\right) }w_{\left( Z\right) }^{\mathbf{2}}}{1+\Gamma _{\left( S\right) }},
\end{equation*}%
\begin{equation}
\overset{(e)}{V^{\mathbf{3}}}=\left( \sqrt{\frac{\Sigma \Delta }{-g_{tt}%
\mathcal{A}}}+\Gamma _{\left( Z\right) }\right) \frac{\overset{(f)}{V^{%
\mathbf{3}}}}{1+\Gamma _{\left( S\right) }}-\frac{\Gamma _{\left( Z\right) }%
\mathbf{w}_{\left( Z\right) }\cdot \mathbf{V}}{1+\Gamma _{\left( S\right) }}%
\left[ \frac{a\mathcal{B}\sin \theta }{\sqrt{-g_{tt}\Sigma \mathcal{A}}}%
+\left( 1-\sqrt{\frac{\Sigma \Delta }{-g_{tt}\mathcal{A}}}\right) \frac{%
\Gamma _{\left( Z\right) }w_{\left( Z\right) }^{\mathbf{3}}}{1+\Gamma
_{\left( Z\right) }}\right] ,
\end{equation}%
with $\mathbf{w}_{\left( Z\right) }$ introduced in Equation (\ref{Wz}).

The inverse relations are%
\begin{equation*}
\overset{(f)}{V^{\mathbf{1}}}=\overset{(e)}{V^{\mathbf{1}}}+\left[ \left( 1-%
\sqrt{\frac{\Sigma \Delta }{-g_{tt}\mathcal{A}}}\right) \frac{\Gamma
_{\left( S\right) }\mathbf{w}_{\left( S\right) }\cdot \mathbf{V}}{1+\Gamma
_{\left( S\right) }}-\frac{a\mathcal{B}\sin \theta }{\sqrt{-g_{tt}\Sigma
\mathcal{A}}}\overset{(e)}{V^{\mathbf{3}}}\right] \frac{\Gamma _{\left(
S\right) }w_{\left( S\right) }^{\mathbf{1}}}{1+\Gamma _{\left( Z\right) }},
\end{equation*}%
\begin{equation*}
\overset{(f)}{V^{\mathbf{2}}}=\overset{(e)}{V^{\mathbf{2}}}+\left[ \left( 1-%
\sqrt{\frac{\Sigma \Delta }{-g_{tt}\mathcal{A}}}\right) \frac{\Gamma
_{\left( S\right) }\mathbf{w}_{\left( S\right) }\cdot \mathbf{V}}{1+\Gamma
_{\left( S\right) }}-\frac{a\mathcal{B}\sin \theta }{\sqrt{-g_{tt}\Sigma
\mathcal{A}}}\overset{(e)}{V^{\mathbf{3}}}\right] \frac{\Gamma _{\left(
S\right) }w_{\left( S\right) }^{\mathbf{2}}}{1+\Gamma _{\left( Z\right) }},
\end{equation*}%
\begin{equation}
\overset{(f)}{V^{\mathbf{3}}}=\!\left( \sqrt{\frac{\Sigma \Delta }{-g_{tt}%
\mathcal{A}}}+\Gamma _{\left( S\right) }\right) \frac{\overset{(e)}{V^{%
\mathbf{3}}}}{1+\Gamma _{\left( Z\right) }}+\frac{\Gamma _{\left( S\right) }%
\mathbf{w}_{\left( S\right) }\cdot \mathbf{V}}{1+\Gamma _{\left( Z\right) }}%
\left[ \frac{a\mathcal{B}\sin \theta }{\sqrt{-g_{tt}\Sigma \mathcal{A}}}%
-\left( 1-\sqrt{\frac{\Sigma \Delta }{-g_{tt}\mathcal{A}}}\right) \frac{%
\Gamma _{\left( S\right) }w_{\left( S\right) }^{\mathbf{3}}}{1+\Gamma
_{\left( S\right) }}\right] ,
\end{equation}%
with $\mathbf{w}_{\left( S\right) }$ introduced in Equation (\ref{Ws}).

\end{widetext}

\begin{widetext}

\begin{figure}[H]
\begin{center}
\includegraphics[width=5cm]{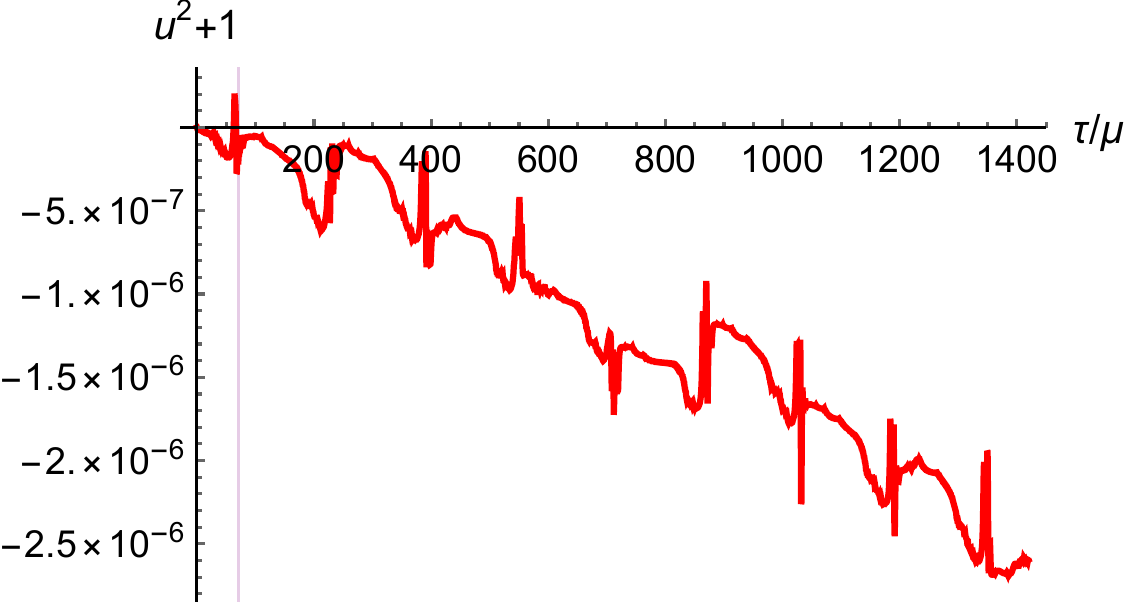}\hfill %
\includegraphics[width=4cm]{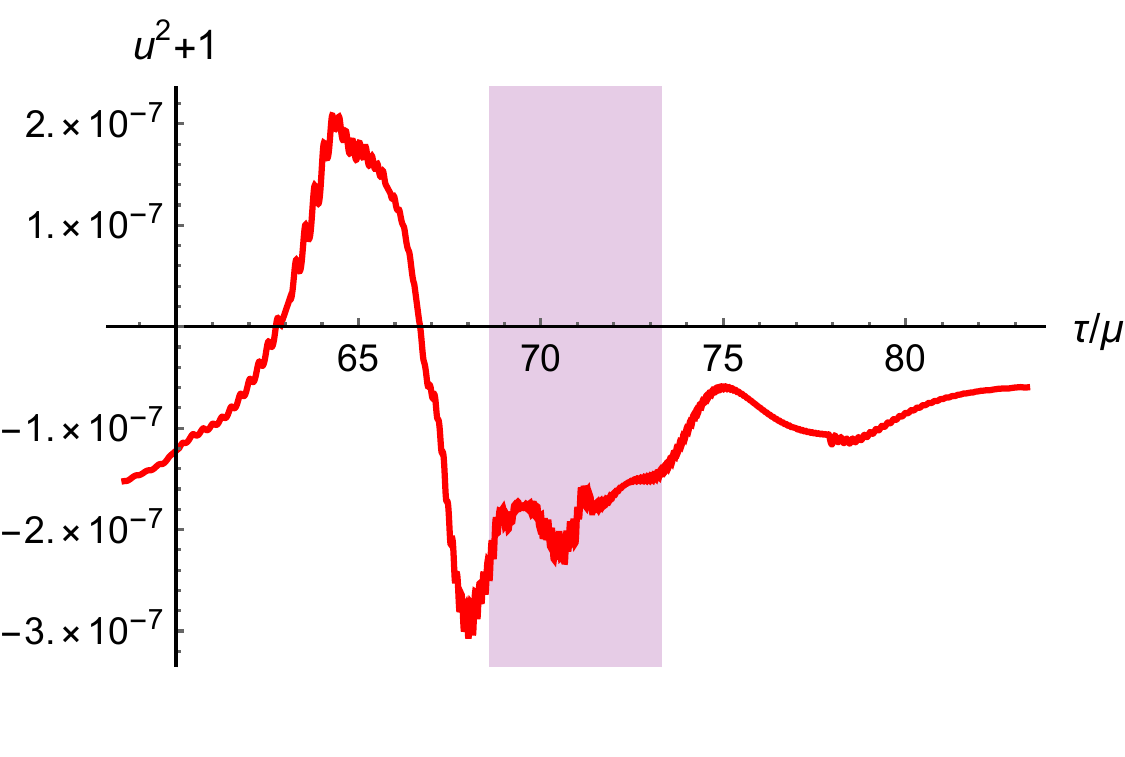}\hfill %
\includegraphics[width=4cm]{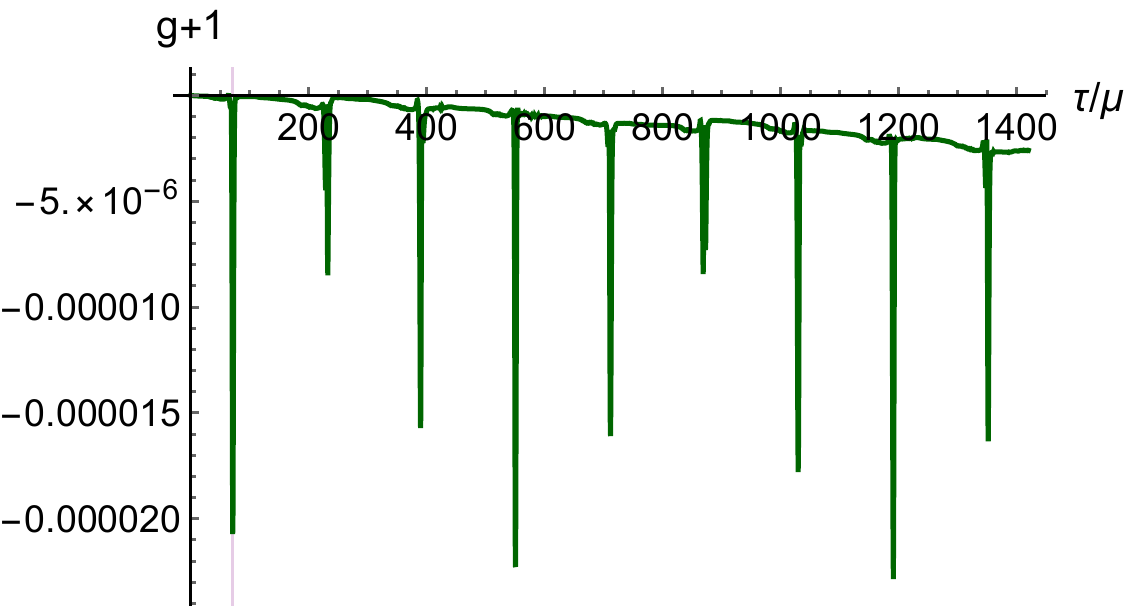}\hfill %
\includegraphics[width=4cm]{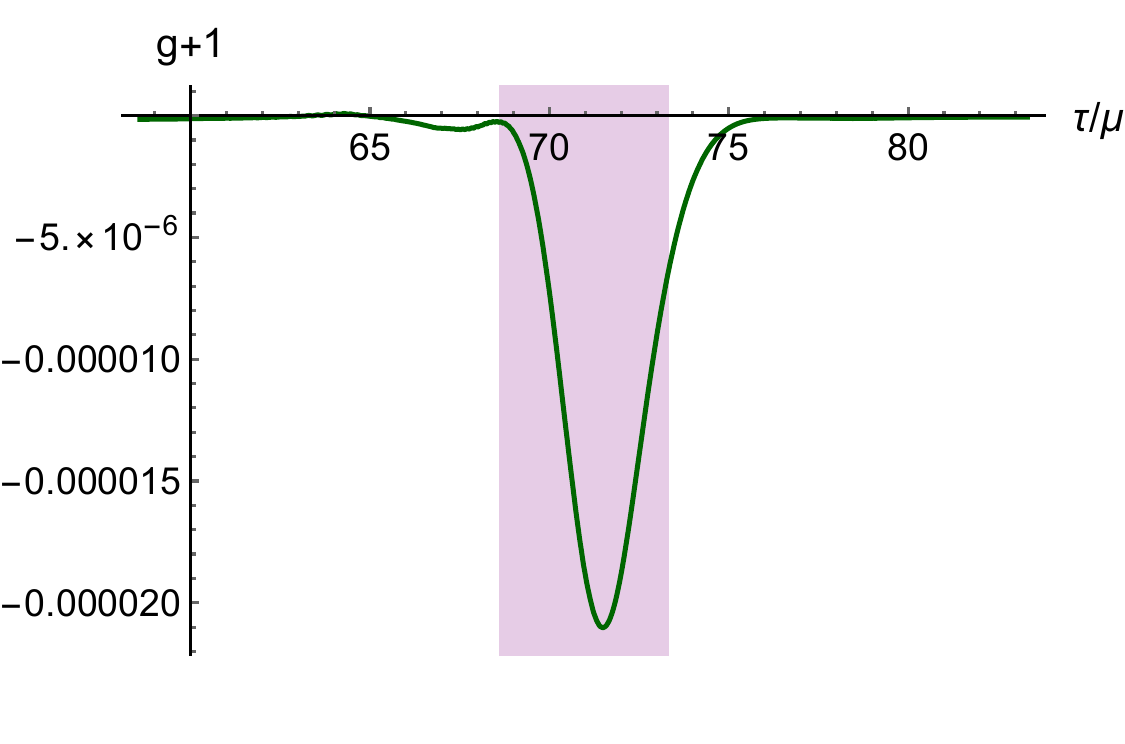}
\end{center}
\caption{(color online). The evolutions of $u^{2}=u_{a}u^{a}$ and
$g$ on longer and shorter timescales for a zoom-whirl orbit
presented on the left hand sides of Figures \protect\ref{ZW1TD} and
\protect\ref{ZW1TDb}.} \label{c1}
\end{figure}

\begin{figure}[H]
\begin{center}
\includegraphics[width=5cm]{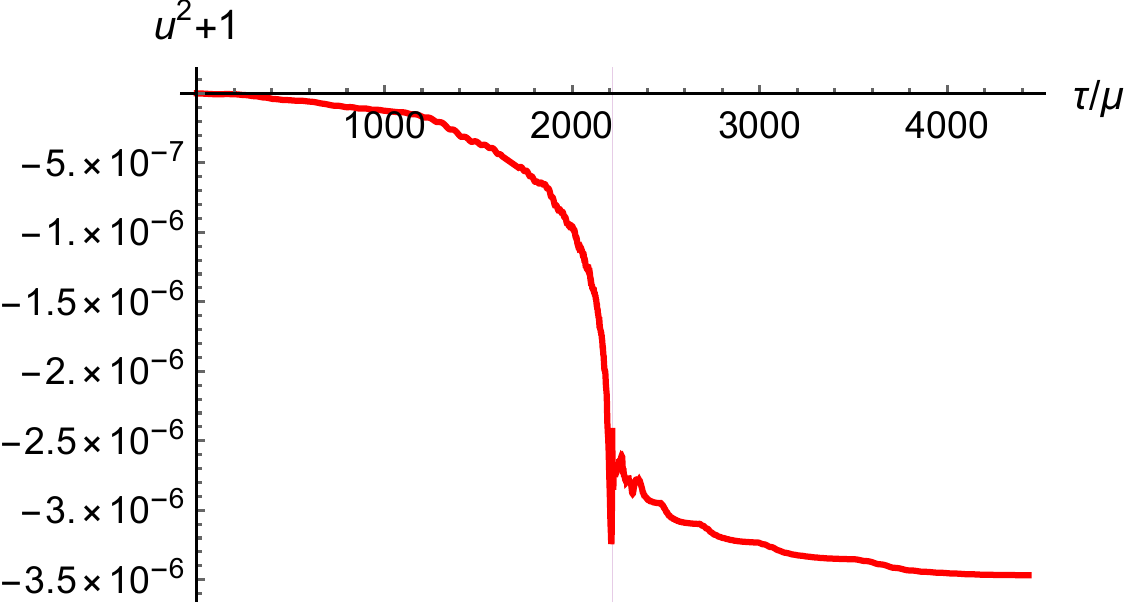}\hfill %
\includegraphics[width=4cm]{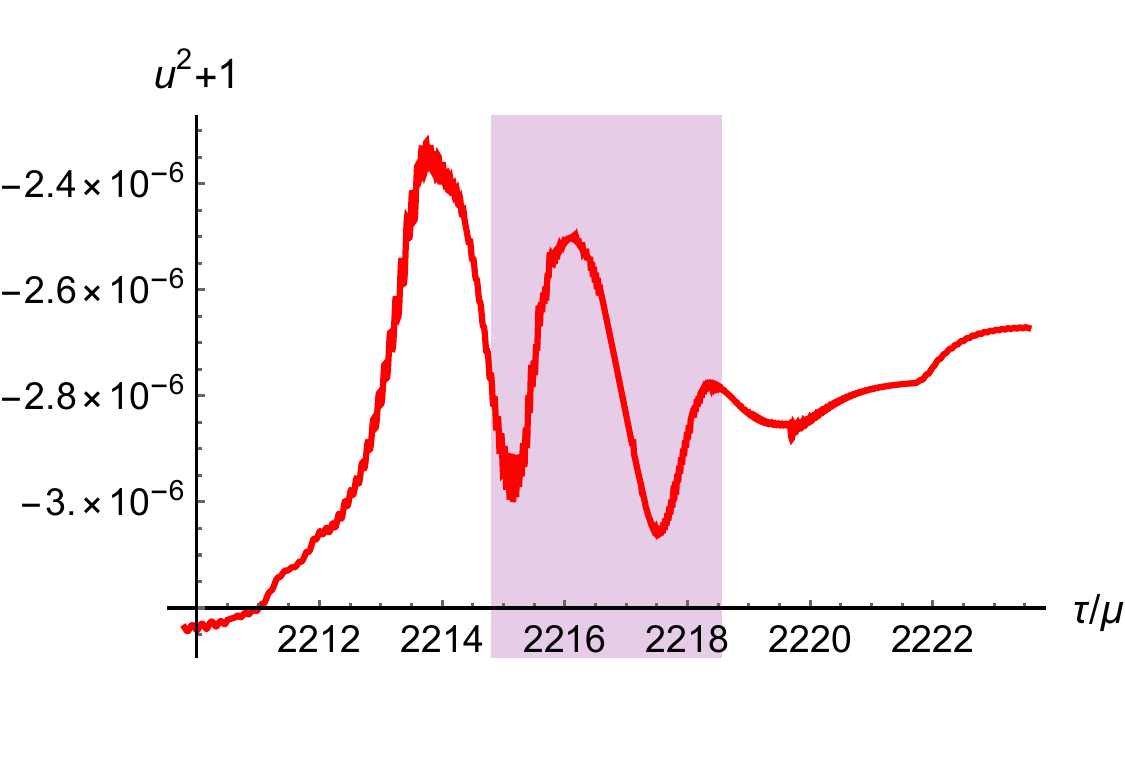}\hfill %
\includegraphics[width=4cm]{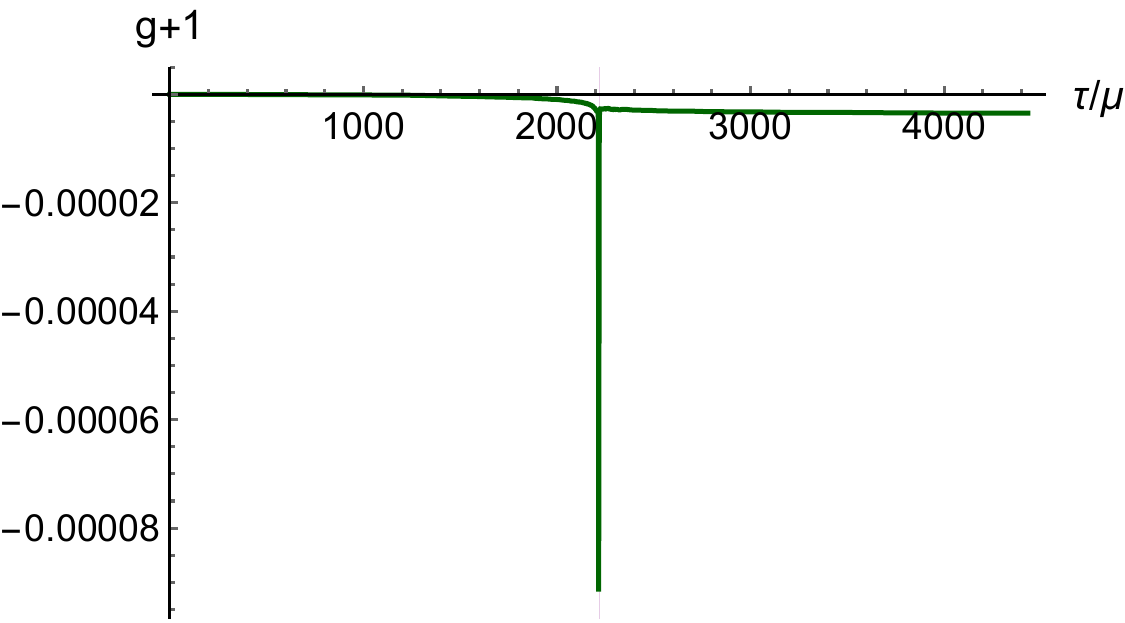}\hfill %
\includegraphics[width=4cm]{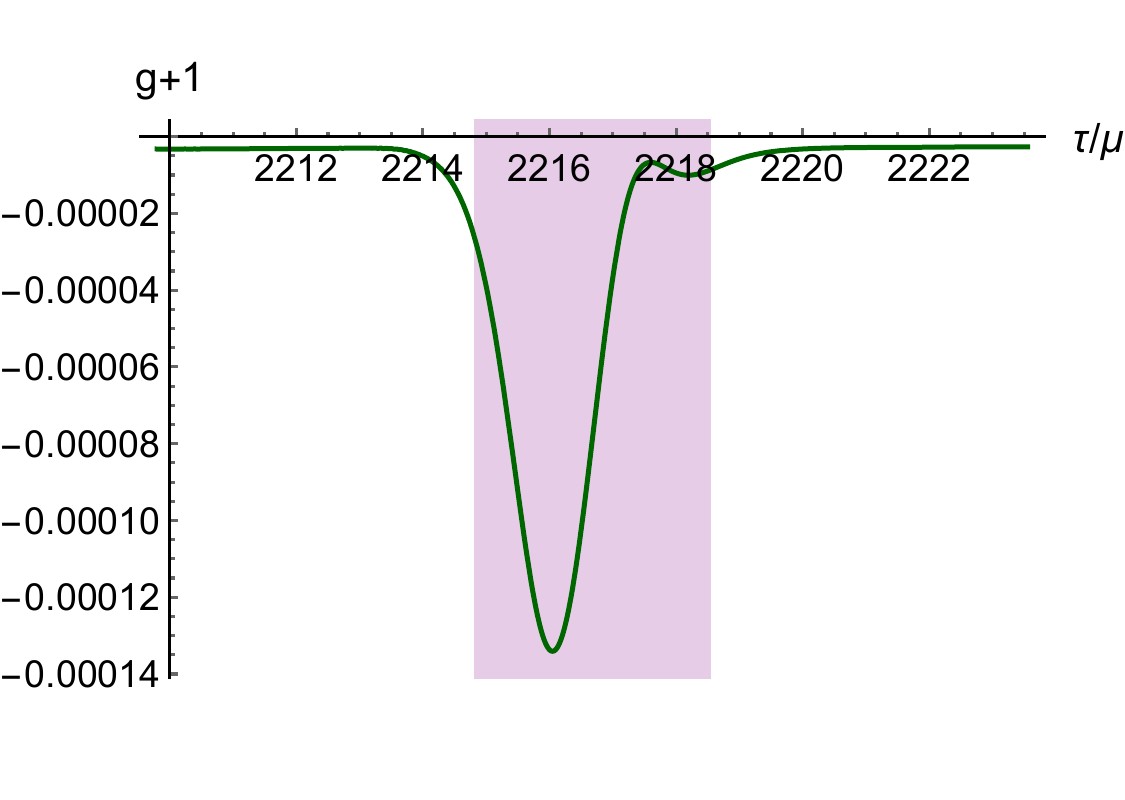}
\end{center}
\caption{(color online). The evolutions of $u^{2}=u_{a}u^{a}$ and
$g$ on longer and shorter timescales are shown for an unbound orbit
presented on the left hand sides of Figures \protect\ref{HB1TD} and
\protect\ref{HB1TDb}.} \label{c2}
\end{figure}

\end{widetext}

\section{APPENDIX B: Checking the validity of the MPD equations \label{para}}

The contraction of the inverse of the velocity-momentum relation (\ref{u})
with $u_{a}$ gives that the sign of $p_{a}u^{a}$ is determined by the
quantity:
\begin{equation}
g=u_{a}u^{a}-\frac{1}{2M^{2}}u^{b}R_{ebcd}S^{cd}S^{ae}u_{a},  \label{g}
\end{equation}%
which corresponds to $\dot{x}\tilde{T}\dot{x}=\dot{x}G\dot{x}$ in Ref. \cite%
{DR3}. Both the functions $g$ \ and $u_{a}u^{a}$ are shown on Figures \ref%
{c1} and \ref{c2} for two cases when the test body follows a
zoom-whirl and an unbound orbit, respectively. In both cases $g$
takes values close to -1 during the whole evolution and
$u_{a}u^{a}=-1$. Hence, the MPD equations are applied where they are
valid. Similar is hold along the all trajectories presented in the
article.

\end{document}